\documentclass[aps,prd,twocolumn,preprintnumbers,showpacs,nofootinbib,amssymb]{revtex4}
\usepackage{graphicx}
\usepackage{amsmath}
\usepackage{amssymb}
\usepackage{amsfonts}
\usepackage{bm}
\usepackage{color}

\def\be{\begin{equation}}
\def\ee{\end{equation}}
\def\bea{\begin{eqnarray}}
\def\eea{\end{eqnarray}}

\begin{document}

\title{Polytropes, logotropes, the universal value of the surface density
of dark matter halos, and the value of the cosmological constant}
\author{Pierre-Henri Chavanis}
%\email{chavanis@irsamc.ups-tlse.fr}
\affiliation{Laboratoire de
Physique Th\'eorique, Universit\'e de Toulouse, CNRS, UPS, France}

\begin{abstract} 

We discuss the connection between logotropes and polytropes in astrophysics and
cosmology. The logotropic equation of state $P=A\ln(\rho/\rho_P)$ may be seen as
a degenerate form of the polytropic equation of state $P=K\rho^{\gamma}$ in the
limit $\gamma\rightarrow 0$, $K\rightarrow\infty$ with $A=K\gamma$ fixed, after
removing an infinite constant. The logotropic distribution function corresponds
to the polytropic distribution function of index $\gamma=0$ (i.e. $n=-1$) for
which the density is finite but the pressure diverges logarithmically. We show
that the polytropic and
logotropic distribution functions can be obtained  in the nondegenerate limit of
the Lynden-Bell theory of violent relaxation for a particular distribution of
phase levels given by the $\chi$-squared distribution. This provides a
justification of the Tsallis entropy  from the Lynden-Bell entropy. The
logotropic distribution function presents a power-law energy tail decreasing as
$\epsilon^{-5/2}$. 
Interestingly, this ``universal'' power-law tail is predicted by recent kinetic
theories of collisionless relaxation based on the coarse-grained Vlasov equation
and on the secular dressed diffusion (SDD) equation. When coupled to gravity,
the associated density profile decreases as $r^{-1}$. This may explain the
universal surface density of dark
matter halos $\Sigma_0=\rho_0 r_h=0.01955\, c^2\sqrt{\Lambda}/G=133\,
M_{\odot}/{\rm pc}^2$ determined from the empirical Burkert profile, or account
for an effective Navarro-Frenk-White (NFW) density cusp $\rho\sim \rho_s r_s/r$
with
$\rho_s r_s=0.00472\, c^2\sqrt{\Lambda}/G=32.0\,
M_{\odot}/{\rm pc}^2$. This
also accounts for the universal gravitational acceleration
$g=0.0291\, c^2\sqrt{\Lambda}=2.76\times 10^{-11}\, {\rm
m\, s^{-2}}$ felt by a test particle and for the Tully-Fisher
relation $v_h^4/M_h=0.0291\, G c^2\sqrt{\Lambda}=3.66\times
10^{-3}\,
{\rm km}^{4}{\rm s}^{-4}M_{\odot}^{-1}.$ Remarkably, these
predictions are obtained from the
logotropic
model without adjustable parameter. The logotropic model can thus provide
an alternative to the modification of Newtonian dynamics (MOND) theory.
We recall how the logotropic model leads to a very accurate
(possibly exact) expression of the cosmological constant
$\Lambda={G^2m_e^6}/{\alpha^6\hbar^4}=1.36\times
10^{-52}\, {\rm m^{-2}}$ in terms of the mass of the electron 
and the fundamental constants of physics [P.H. Chavanis, Phys. Dark Univ. {\bf
44}, 101420 (2024)]. This refined Eddington relation, which expresses the
commensurability between the surface density of the electron and the surface
density of the Universe, may be related to the holographic principle.

\end{abstract}


\maketitle

\section{Introduction}

Models of stars described by the isothermal equation of state $P=\rho k_B T/m$ or by the  polytropic equation of state $P=K\rho^{\gamma}$ with $\gamma=1+1/n$ were developed in astrophysics since the 19th century. The polytropic equation of state was introduced by Lord Kelvin (W. Thomson) \cite{kelvin1,kelvin2} in considerations relating to the convective equilibrium of fluids under the influence of gravity. It was also used by Lane \cite{lane} in his theoretical investigations on the temperature of the Sun, assuming that the mass of the Sun is in convective equilibrium.  The isothermal equation of state was used by  Z\"ollner \cite{zollner}, Thiesen \cite{thiesen}, Ritter \cite{ritteriso}, Lord Kelvin \cite{kelvin2},  Hill \cite{hill}  and Darwin \cite{darwin} to describe the equilibrium of a gas under its own gravitation. This self-gravitating gas could represent a nebula, a star like the Sun, a gaseous planet, a swarm of meteorites or comets, or even the Earth.

The results of these early investigations are collected and expounded in the books of Emden \cite{emden} and Chandrasekhar \cite{chandrabook}. Polytropic stars are described by the Lane-Emden equation. Some analytical solutions were discovered by Ritter \cite{ritter} for $n=1$ (it corresponds to Laplace's \cite{laplace} celebrated law of density for the Earth interior as discussed in footnote 10 of \cite{jeansapp}) and Schuster \cite{schuster} for $n=5$. Isothermal stars are described by the Emden equation, which can be viewed as a dimensionless form of the Boltzmann-Poisson equation.\footnote{The Emden equation already appeared in Eq. (512) of Ritter \cite{ritteriso}, in Eq. (14) of  Lord Kelvin \cite{kelvin2}, and in  Eq. (14) of Darwin \cite{darwin}.}  The density profile of self-gravitating isothermal spheres decreases at large distances as $r^{-2}$ with damped oscillations superimposed \cite{emden,chandrabook}.\footnote{The singular solution of the isothermal sphere $\rho=k_B T/(2\pi G m  r^{2})$ was found by Z\"ollner \cite{zollner} and  Ritter \cite{ritteriso}. It was also reported in the paper of Lord Kelvin \cite{kelvin2} whose interest in the problem originated in a question set by Tait in an examination paper. The first oscillation about the $r^{-2}$ asymptotic behavior of the density was found by Ritter \cite{ritteriso} and Darwin \cite{darwin}. The infinite series of damped oscillations leading to a spiral in certain representations were found by Thiesen \cite{thiesen} and  Hill \cite{hill} and studied in detail by Emden \cite{emden} and Chandrasekhar \cite{chandrabook}. } Similar results were obtained for polytropic spheres of index $n>5$ \cite{emden,chandrabook}.

The isothermal and polytropic equations of state were  used in the beginning of the 20th century by von Zeipel \cite{zeipel} and Plummer \cite{plummer} to describe clusters of stars (stellar systems and globular clusters). Inspired by an idea of Lord Kelvin, von Zeipel \cite{zeipel} modeled the density distribution of stars in a globular cluster by analogy with the density profile of a self-gravitating isothermal gas and found a good agreement with observations in the central parts of the cluster.  On the other hand, Plummer \cite{plummer}  modeled the density distribution of stars in a globular cluster by analogy with the density profile of a self-gravitating polytropic gas in convective equilibrium. He noted that the polytrope of index $n=5$, whose spatial density decreases at large distances as $r^{-5}$, provides a good agreement with observations in the outer parts of the cluster. This is now called the ``Plummer distribution''. At the end of his paper,
Plummer \cite{plummer} proposed to model the density of stars in a globular
cluster by an isothermal distribution in the center and by a
polytropic distribution of index $n = 5$ in the envelope. In other
words, he assumed that globular clusters have a central core
in isothermal equilibrium and an outer envelope in convective
equilibrium.

However, Eddington \cite{edd1913,edd1915} emphasized that stellar systems cannot be considered as gaseous systems with a short mean free path as assumed by von Zeipel \cite{zeipel} and Plummer \cite{plummer}. On the contrary, they are rather in the collisionless limit in which the stars have a long mean free path, larger than the size of the system. They are thus described by a distribution function (DF) $f({\bf r},{\bf v},t)$  which is governed by the collisionless Boltzmann equation coupled to the Poisson equation. This equation was explicitly introduced by Jeans \cite{jeansT}.\footnote{The same equation was later introduced by Vlasov \cite{vlasov} in plasma physics. It is usually called the Vlasov  equation. See H\'enon \cite{henonV} for a discussion about the name that should be given to this equation.} Its stationary solutions are functions of the integrals of the motion as specified by the Jeans theorem \cite{jeansT}. For spherically symmetric and isotropic stellar systems, the DF only depends on the individual energy $\epsilon=v^2/2+\Phi$ of the stars.  Jeans \cite{jeansT}  pointed out that the Maxwell-Boltzmann distribution is a particular solution of the collisionless Boltzmann equation associated with the isothermal equation of state.\footnote{Eddington \cite{edd1913,edd1915} previously derived a self-consistent quasisteady state of collisionless stellar systems  with a Schwarzschild velocity distribution containing the Maxwell distribution as a special case.} Jeans \cite{jeansiso} mentioned that the effect of “collisions” (encounters) between stars would tend
to establish an isothermal (Maxwell-Boltzmann) distribution. Eddington \cite{edd1916} reconsidered the results of von Zeipel \cite{zeipel} and Plummer \cite{plummer} from the point of view of collisionless stellar systems instead of collisional gases. The connection is made through the so-called Eddington formula which relates the DF $f(\epsilon)$ of a spherically symmetric and isotropic stellar system to the density $\rho(\Phi)$  of the corresponding barotropic gas. Eddington \cite{edd1916} also introduced a special class of stationary solutions of the Vlasov equation with a DF $f=A(\epsilon_m\mp\epsilon)^{n-3/2}$, called stellar polytropes, which are associated with polytropic stars. Indeed, by computing their density and pressure as given by kinetic theory he showed that stellar polytropes can be formally associated with a barotropic gas with a polytropic equation of state $P=K\rho^{1+1/n}$.

Later, Ipser \cite{ipser} showed that the polytropic DF could be obtained by 
extremizing a certain functional $S=-\int f^{1+1/(n-3/2)}\, d{\bf r}d{\bf v}$ 
at fixed mass and energy and that a maximum of this functional is dynamically
stable with respect to the Vlasov-Poisson equations. He also generalized these
results to a larger class of DFs of the form $f=f(\epsilon)$ with
$f'(\epsilon)<0$, which are associated with a functional $S=-\int C(f)\, d{\bf
r}d{\bf v}$, where $C$ is an arbitrary convex function.

Similar results were obtained more recently in the context 
of Tsallis \cite{tsallis} generalized thermodynamics. Plastino and Plastino
\cite{pp} noticed that stellar polytropes  can be obtained by extremizing the
Tsallis entropy $S=-\frac{1}{q-1}\int (f^q-f)\, d{\bf r}d{\bf v}$ with
$q=1+1/(n-3/2)$ at fixed mass and energy (isothermal galaxies described by the
Boltzmann entropy are recovered in the limit $q\rightarrow 1$ or $n\rightarrow
+\infty$). They mentioned that stellar polytropes with an index $3/2\le n\le 5$
have a finite mass unlike isothermal galaxies, and claimed that generalized
thermodynamics solves the problems of standard thermodynamics when applied to
gravity.\footnote{We do not agree with that claim. The absence of Boltzmann
equilibrium state for stellar systems simply reflects the physical fact that
these systems have the tendency to evaporate and undergo core collapse.
Therefore, the statistical mechanics of self-gravitating systems is essentially
an out-of-equilibrium problem which can be correctly dealt with
kinetic theory \cite{aakin}.} Taruya and Sakagami \cite{ts1,ts2,ts3} and
Chavanis and Sire \cite{aaPOLY,grand,anomalous,aastab} studied the dynamical and
generalized thermodynamical stability of polytropic stars and stellar
polytropes. In the context of collisionless stellar systems described by the
Vlasov-Poisson equations, Tsallis distributions correspond to the stellar
polytropes of Eddington \cite{edd1916} and Tsallis entropy corresponds to the
functional introduced by Ipser \cite{ipser}.\footnote{The Tsallis functional is
slightly different from Ipser's functional (they essentially differ by a
constant) in the sense that, using l'H\^opital's rule, the Tsallis functional
reduces to the Boltzmann functional for $q\rightarrow 1$ or $n\rightarrow
+\infty$ while Ipser's functional is indefinite in that case (it is defined only
for $n<+\infty$).} In this sense, the maximization of the Tsallis entropy at
fixed mass and energy provides a sufficient condition of dynamical stability
with respect to the Vlasov-Poisson equations, not a condition of generalized
thermodynamical stability.

This ``dynamical'' interpretation of the Tsallis functional has been defended by
Chavanis and Sire  \cite{aaPOLY,grand,anomalous,aastab} while Plastino and
Plastino \cite{pp} and Taruya and Sakagami  \cite{ts1,ts2,ts3}  have defended a
``generalized thermodynamical'' interpretation. According to Chavanis and Sire
\cite{aaPOLY,grand,anomalous,aastab}, for collisionless systems with long-range
interactions,  the generalized thermodynamical formalism is {\it effective}, but
it is useful to study the dynamical stability of the system with respect to the
Vlasov-Poisson equations through a {\it thermodynamical analogy}. Generalized
thermodynamics with a true thermodynamical interpretation can be relevant in
other  contexts, different from collisionless systems with long-range
interactions, as discussed in Sec. IV of \cite{anomalous} and in
\cite{csTsallis,assise}.\footnote{Actually, in this paper, we shall provide a
new thermodynamical interpretation of the Tsallis entropy for collisionless
systems with long range interactions on the coarse-grained scale.  Indeed, we
will show that it can be obtained  in the nondegenerate limit of the Lynden-Bell
\cite{lb} theory of violent relaxation for a particular distribution 
of phase levels given by the $\chi$-squared distribution.} For example, 
various interpretations of Tsallis entropy and Tsallis generalized
thermodynamics have been given in relation to collisional systems with
long-range interactions \cite{tsprl,ccgm}, nonlinear kinetic equations
\cite{kaniadakis,kingen}, nonlinear Fokker-Planck (NFP) equations
\cite{frank,gen,anomalous,nfp,ggpp}, and the time dependent density functional
theory (TDFT) in the physics of simple
 liquids \cite{back}.\footnote{These studies involve kinetic equations such as the usual \cite{tsprl,ccgm} or polytropic \cite{kaniadakis,kingen} Boltzmann, Landau and Lenard-Balescu equations in the microcanonical ensemble (where the energy is conserved) or such as the polytropic Kramers and Smoluchowski equations \cite{frank,gen,anomalous,nfp,ggpp,back} in the canonical ensemble (where  the system is in contact with a thermal bath). In the collisional case, self-gravitating polytropic systems with index $n\ge 3$ can experience a collapse in the canonical ensemble \cite{aaPOLY,anomalous} (like isothermal stars and  self-gravitating Brownian particles \cite{aaISO,sc2002}) while self-gravitating polytropic systems with index $n\ge 5$ can experience a form of gravothermal catastrophe in the microcanonical ensemble \cite{anomalous} (like globular clusters, isothermal stellar systems and self-gravitating Brownian particles \cite{antonov,lbw,sc2002}).}
 In the last case, Tsallis free energy can be interpreted as a particular form
of excess free energy taking into account correlations and microscopic
(small-scale) constraints between the particles. See \cite{tsallisbook} for an
extensive list of references related to Tsallis generalized thermodynamics and
its applications to numerous domains of physics.\footnote{The discussion about
``generalized entropies'' may be sometimes confusing because the interpretation
depends on the situation contemplated and the notions of ``dynamics'' and
``thermodynamics'' are sometimes intermingled. There is no universal
interpretation of generalized entropies. Their interpretation 
has to be given case by case.}

The logotropic equation of state was introduced in the context of stellar structure by McLaughlin and Pudritz \cite{mlp} to describe gravitational collapse and star formation in nonisothermal spheres. Later, Chavanis and Sire \cite{cslogo} studied logotropic gaseous spheres in the context of Tsallis generalized thermodynamics.\footnote{We met the logotropic equation of state after a referee of our paper \cite{anomalous} suggested that we should extend our study to logotropes.}  They showed that logotropes are related to polytropes of index $\gamma=0$ (i.e. $n=-1$). More precisely, the logotropic equation of state $P=A\ln(\rho/\rho_P)$ may be seen as a degenerate form of the polytropic equation of state $P=K\rho^{\gamma}$ in the limit $\gamma\rightarrow 0$, $K\rightarrow\infty$ with $A=K\gamma$ fixed after removing an infinite constant. This limit gives a sense to the Lane-Emden equation of index $n=-1$. Chavanis and Sire \cite{cslogo} studied the dynamical and generalized thermodynamical stability of 
self-gravitating   logotropic gaseous spheres confined within a box. They also considered their dynamical evolution (evaporation and collapse) governed by the logotropic Smoluchowski-Poisson equations. 

Ten years later, we met again the logotropes in an unexpected area \cite{epjp,lettre,jcap,preouf1,action,logosf,preouf2,ouf,gbi}. 
It is an observational evidence that dark matter (DM) halos have a universal surface density \cite{kormendy,spano,donato}
\begin{eqnarray}
\Sigma_0^{\rm obs}=\rho_0 r_h=141_{-52}^{+83}\, M_{\odot}/{\rm pc}^2.
\label{univ}
\end{eqnarray}
By looking for the equation of state that has such a property we were naturally led to the logotropic equation of state. Then, by applying the logotropic equation of state both at the galactic level (``small'' scales) and in cosmology (``large'' scales) in the spirit of a unified dark matter and dark energy (UDME) model, we could obtain a prediction of the surface density of DM halos
\begin{eqnarray}
\Sigma_0=0.01955\, \frac{c^2\sqrt{\Lambda}}{G}=133\, M_{\odot}/{\rm pc}^2
\end{eqnarray}
in terms of the cosmological constant  $\Lambda$  without adjustable parameter.
With the measured value of the
cosmological constant, $\Lambda_{\rm obs}=1.11\times 10^{-52}\, {\rm m^{-2}}$,
this expression turns out to be remarkably consistent with the observational
value of the universal surface density from Eq. (\ref{univ}). This is either a
very important result or a striking  coincidence. On the other hand, by
remarking that the surface density of DM halos is of the same order of magnitude
as the surface density of the electron, we obtained a qualitative relation
between the cosmological constant and the mass of the electron. In this manner,
we recovered (without knowing it initially)  the Eddington relation
\cite{eddington1931lambda}. Then, by developing some arguments (see Sec.
\ref{sec_value} and Appendix \ref{sec_just} for a summary of our peregrination),
we obtained a refined Eddington relation
\begin{eqnarray}
\Lambda=\frac{G^2m_e^6}{\alpha^6\hbar^4}=1.36\times 10^{-52}\, {\rm m^{-2}},
\end{eqnarray}
which provides a very accurate prediction of the cosmological constant. This relation turns out to be in very good agreement with the observational value  $\Lambda_{\rm obs}=1.11\times 10^{-52}\, {\rm m^{-2}}$.  Again, this is either a very important result or a striking coincidence.

In the present paper, we discuss in more detail the connection between logotropes and polytropes. In particular, we show that the density $\rho(\Phi)$ associated with the logotropic equation of state can be obtained from the polytropic DF $f=A(\epsilon_m+\epsilon)^{-5/2}$ of index $\gamma=0$ (i.e. $n=-1$). We will therefore call it the logotropic DF.  This DF decreases with the velocity as $v^{-5}$  so that  the pressure diverges logarithmically. This logarithmic behavior is of course characteristic of the logotropic equation of state.

Recently, two groups of researchers obtained very interesting results in relation to the collisionless relaxation of systems with long-range interactions (like plasmas and self-gravitating systems). Ewart {\it et al.} \cite{ewart1,ewart2}  showed that the Lynden-Bell \cite{lb} theory of violent relaxation, when formulated in the multispecies case, leads to a universal DF with a power-law tail scaling as $\overline{f}\sim v^{-5}$.\footnote{In a previous paper \cite{superstat}, we have studied the 
multispecies Lynden-Bell \cite{lb} theory of violent relaxation and showed that
it can lead to coarse-grained DFs with power-law tails in certain cases. We also
made the connection with Tsallis distributions (polytropes) and argued that the 
multispecies Lynden-Bell distribution may be viewed as a form of
superstatistics. We showed that the polytropic distributions can be obtained  in
the nondegenerate limit of the Lynden-Bell theory of violent relaxation for a
particular distribution of phase levels given by the $\chi$-squared
distribution. These DFs have power-law tails for which the result of Ewart {\it
et al.} \cite{ewart1,ewart2} is a particular case. In our point of view, a
$v^{-5}$ tail may not be universal (other non-Boltzmannian distributions may
arise in practice \cite{superstat}) but a DF with a $v^{-5}$ tail appears in
certain situations of physical interest \cite{ewart2}.} On the other hand, Banik
{\it et al.}  \cite{banik1,banik2}  showed that the secular dressed diffusion
(SDD) equation \cite{epjp1,sdduniverse} describing collisionless systems with
long-range interactions stochastically forced by an external medium naturally
generates a DF with a power-law tail decreasing as $f\sim v^{-5}$ in the
homogeneous case and as $f\sim \epsilon^{-5/2}$ in the inhomogeneous case. These
two results are strikingly similar while being based on very different physical
assumptions. In the present paper, we remark that the universal power-law tails
predicted by Ewart {\it et al.}  \cite{ewart1,ewart2} and Banik {\it et al.} 
\cite{banik1,banik2} correspond to the power-law tail $f\sim \epsilon^{-5/2}$ of
the logotropic distribution. This gives further support to the logotropic model
introduced in our previous works 
\cite{epjp,lettre,jcap,preouf1,action,logosf,preouf2,ouf,gbi}.  This is one more time a very interesting result or a striking coincidence. When coupled to gravity, the logotropic density profile decreases as $\rho\sim r^{-1}$.  As discussed in \cite{epjp,lettre,jcap,preouf1,action,logosf,preouf2,ouf,gbi}  this may either explain the density cusp of the NFW distribution of DM halos (see Appendix 1 of \cite{gbi} and \cite{banik2}) or account for the universal surface density of DM halos \cite{epjp,lettre,jcap,preouf1,action,logosf,preouf2,ouf,gbi}.

This paper is organized as follows.  In Sec. \ref{sec_baro}, we discuss 
the basic properties of gaseous stars described by the Euler-Poisson equations
with a polytropic and a logotropic equation of state. In Sec. \ref{sec_gal}, we
discuss the basic properties of stellar polytropes and stellar logotropes that
are particular stationary solutions of the Vlasov-Poisson
equations.\footnote{Throughout the paper, for convenience, we call 
``gaseous star'' a self-gravitating fluid system described by the Euler-Poisson
equations and ``stellar system'' a collisionless self-gravitating system
described by the Vlasov-Poisson equations. Of course, our discussion also
applies to systems  other than stars and galaxies, e.g., DM and plasmas.} In
Sec. \ref{sec_ew}, following our previous work \cite{superstat}, we show how
polytropic DFs arise from the multispecies Lynden-Bell theory of violent
relaxation in certain cases. In Secs. \ref{sec_ew} and \ref{sec_banik},  we
recall the arguments of Ewart {\it et al.}  \cite{ewart1,ewart2} and Banik {\it
et al.}  \cite{banik1,banik2} showing how universal power-law tails decreasing
as $f\sim\epsilon^{-5/2}$ emerge naturally  from the Lynden-Bell theory of
violent relaxation (Sec. \ref{sec_ew}) of from the SDD equation (Sec.
\ref{sec_banik}).  We then note that the logotropic DF, corresponding to a
polytrope of index $n=-1$, presents a power-law energy tail precisely decreasing
as $f\sim \epsilon^{-5/2}$. Therefore, the logotropic DF seems to be singled out
by kinetic theories of collisionless relaxation based on the coarse-grained
Vlasov equation or on the SDD equation. In Sec. \ref{sec_logodm}, we apply the
logotropic equation of state to DM halos and detail the main properties of this
model. In particular, we emphasize the $\rho\sim r^{-1}$ decay of the density
profile of DM halos and their universal surface density. Finally, in Sec.
\ref{sec_value}, we combine these general results with other considerations to
predict the value of the cosmological constant in terms of the mass of the
electron. The Appendices provide additional results, which complete and refine
the discussion of the main text.

\section{Self-gravitating systems described by an equation of state: barotropic stars}
\label{sec_baro}

\subsection{Fundamental condition of hydrostatic equilibrium}

We consider a self-gravitating gaseous system (star) described by the Euler-Poisson equations
\begin{eqnarray}
\frac{\partial\rho}{\partial t}+\nabla\cdot (\rho {\bf u})=0,
\label{g1}
\end{eqnarray}
\begin{eqnarray}
\frac{\partial {\bf u}}{\partial t}+{\bf u}\cdot \nabla {\bf u}=-\frac{1}{\rho}\nabla P-\nabla\Phi,
\label{g2}
\end{eqnarray}
\begin{eqnarray}
\Delta\Phi=S_d G\rho,
\label{g3}
\end{eqnarray} 
where $S_d=2\pi^{d/2}/\Gamma(d/2)$ denotes the surface of a hypersphere of unit radius in a $d$-dimensional space (for the sake of generality we work in a space of dimension $d$). This system of equations can be closed by prescribing a barotropic equation of state $P=P(\rho)$. 

The
equilibrium state of the Euler-Poisson equations results from the balance between
the gravitational attraction and the repulsion due to the pressure
force. It is described by the 
equation of
hydrostatic equilibrium
\begin{eqnarray}
\nabla P+\rho\nabla\Phi={\bf 0}
\label{g4}
\end{eqnarray}
coupled to the Poisson equation (\ref{g3}). These equations  can be combined into a single differential equation
\begin{eqnarray}
-\nabla\cdot \left(\frac{\nabla
P}{\rho}\right )=S_d G\rho,
\label{g5}
\end{eqnarray}
which determines the density
profile $\rho({\bf r})$. This is the fundamental differential equation of hydrostatic equilibrium.

The energy functional associated with the Euler-Poisson equations reads
\begin{equation}
E_{\rm tot}=\frac{1}{2}\int \rho {\bf u}^2\, d{\bf r}+\int\rho\int^{\rho}\frac{P(\rho')}{{\rho'}^2}\, d\rho'd{\bf r}+\frac{1}{2}\int\rho\Phi\, d{\bf r}.
\label{g6}
\end{equation}
This is the sum of the macroscopic kinetic energy $\Theta_c$, the internal energy $U$ and the gravitational energy $W$. It can be shown that the Euler-Poisson equations conserve the total mass $M=\int \rho\, d{\bf r}$ and the total energy $E_{\rm tot}$.

Using general arguments based on these conservation laws, one can show that an equilibrium state of the Euler-Poisson equations is an extremum of energy at fixed mass. Furthermore, an equilibrium state of the Euler-Poisson equations is dynamically stable if, and only if, it is a minimum of energy at fixed mass \cite{aastab}. Writing the first variations as $\delta E_{\rm tot}-\mu\delta M=0$, where $\mu$ is a Lagrange multiplier,  we get ${\bf u}={\bf 0}$ and 
\begin{equation}
\int \frac{P'(\rho')}{\rho'}\, d\rho'+\Phi=\mu.
\label{g7}
\end{equation}
Here $\mu$ is similar to a chemical potential. Taking the gradient of this relation, we obtain the condition of hydrostatic equilibrium (\ref{g4}). Conversely, integrating the condition of hydrostatic equilibrium  (\ref{g4}) with the equation of state $P=P(\rho)$ we see that the density $\rho$ is a function $\rho=\rho(\Phi)$ of the gravitational potential $\Phi$ determined by Eq. (\ref{g7}). From Eq. (\ref{g4}) or Eq. (\ref{g7}), we obtain 
\begin{equation}
\frac{P'(\rho)}{\rho}=-\frac{1}{\rho'(\Phi)}.
\label{g7b}
\end{equation}
The squared speed of sound is $c_s^2=P'(\rho)$. In the usual case where $P'(\rho)>0$, the density is a decreasing function of the gravitational potential: $\rho'(\Phi)<0$.  The dynamical stability of these solutions is discussed in Sec. \ref{sec_stabg} below.

\subsection{Polytropic equation of state}

For the polytropic equation of state
\begin{equation}
P=K\rho^{\gamma},
\label{g8}
\end{equation}
where $K$ is the polytropic constant and $\gamma=1+1/n$ is the polytropic index, the energy functional takes the form
\begin{equation}
E_{\rm tot}=\frac{1}{2}\int \rho {\bf u}^2\, d{\bf r}+\frac{K}{\gamma-1}\int (\rho^{\gamma}-\rho)\, d{\bf r}+\frac{1}{2}\int\rho\Phi\, d{\bf r}.
\label{g9}
\end{equation}
It can be interpreted as a Tsallis free energy in physical space $E_{\rm tot}=\Theta_c+W-KS_{\gamma}$, where $\Theta_c+W$ is the kinetic $+$ potential energy, $K$ plays the role of a generalized (polytropic) temperature,\footnote{The polytropic temperature $K$, which is uniform,  is different from the kinetic temperature. The kinetic temperature defined by $T({\bf r})=P/\rho$ is equal to $T({\bf r})=K\rho^{\gamma-1}$ for polytropes (when $K>0$). It is position-dependent. Its gradient $\nabla T/T=(\gamma-1)\nabla\rho/\rho$, or $\nabla T=-[(\gamma-1)/\gamma]\nabla\Phi$ for a state of hydrostatic equilibrium, is determined by the polytropic index $\gamma$ (see Ref. \cite{csTsallis} for a discussion on the different notions of temperature in polytropic fluids).} and $S_\gamma=-\frac{1}{\gamma-1}\int (\rho^{\gamma}-\rho)\, d{\bf r}$ is the Tsallis entropy \cite{aaPOLY,anomalous,csTsallis}.\footnote{For $\gamma=2$ (i.e. $n=1$), yielding a linear relationship $\rho=C(\epsilon_m-\Phi)_+$ between the density and the gravitational potential [see Eq. (\ref{g11}) below], the Tsallis entropy $S_2=-\int (\rho^{2}-\rho)\, d{\bf r}$ is similar to minus the enstrophy $\Gamma_2=\int \omega^2\, d{\bf r}$ in 2D hydrodynamics \cite{varp}. The same comment applies to stellar polytropes in Sec. \ref{sec_gal}. In that case, the maximum entropy state is equivalent to the minimum enstrophy state.} Since $M$ is conserved we also have, up to an additive constant, $U=\frac{1}{\gamma-1}\int P\, d{\bf r}=\frac{K}{\gamma-1}\int \rho^{\gamma}\, d{\bf r}$ and
\begin{equation}
E_{\rm tot}=\frac{1}{2}\int \rho {\bf u}^2\, d{\bf r}+\frac{1}{\gamma-1}\int P\, d{\bf r}+\frac{1}{2}\int\rho\Phi\, d{\bf r}.
\label{g10}
\end{equation}
The polytropic equilibrium density profile can be obtained by extremizing the energy at fixed mass. Writing $\delta E_{\rm tot}-\mu\delta M=0$, we get
\begin{equation}
\rho=\left \lbrack \frac{1}{\gamma}+\frac{\gamma-1}{K\gamma}(\mu-\Phi)\right\rbrack_+^{1/(\gamma-1)}.
\label{g11}
\end{equation}
It can be interpreted as a Tsallis distribution in physical space. 

For the polytropic equation of state (\ref{g8}), the squared speed of sound is  $c_s^2=P'(\rho)=K\gamma\rho^{\gamma-1}$. We shall assume in the following that $c_s^2>0$, which is a necessary condition of dynamical stability \cite{wdsD} (the case $c_s^2<0$ is discussed in Appendix \ref{sec_cs2neg}).\footnote{Since $\Phi(r)$ increases with the distance [$d\Phi/dr=GM(r)/r^2\ge 0$], the condition of hydrostatic equilibrium from Eq. (\ref{g4}) implies that $P(r)$ decreases with the distance. When $c_s^2>0$ the density decreases with the distance and when $c_s^2<0$ the density increases with the distance.} Therefore, we assume $K\gamma>0$ or, equivalently,  $K(n+1)/n>0$. We must distinguish two cases:

(i) $n>0$: in that case $K>0$ and Eq. (\ref{g11}) can be rewritten as
\begin{equation}
\rho=C\left (\epsilon_m-\Phi \right )_+^{n}
\label{g11a}
\end{equation}
with $C=1/\lbrack K(1+n)\rbrack^n$ and $\epsilon_m=Kn+\mu$ (this notation will take more sense in Sec. \ref{sec_gal}). The pressure is always positive.

(ii) $n<0$: in that case $K(n+1)<0$ and Eq. (\ref{g11}) can be rewritten as
\begin{equation}
\rho=C\left (\epsilon_m+\Phi \right )^{n}
\label{g11b}
\end{equation}
with $C=1/|K(1+n)|^n$ and $\epsilon_m=-Kn-\mu$. For $n<-1$, the pressure is positive ($K>0$). For $-1<n<0$, the pressure is negative ($K<0$) but it has a repulsive effect.

The index $n=0$ corresponds to a homogeneous sphere.

The index  $n=-1$ (i.e. $\gamma=0$) with $K$ finite corresponds to a constant pressure ($P=K$). There is no equilibrium state in that case since the gradient of the pressure vanishes identically. Still, the Lane-Emden equation [see Eq. (\ref{g14}) below] is mathematically well-defined for $n=-1$.  We will see in Sec. \ref{sec_logog} that the index $n=-1$ with $K\rightarrow \infty$ is connected to the logotropic equation of state.

For $n\rightarrow \pm\infty$, we obtain the isothermal equation of state $P=\rho k_B T/m$, which leads to the Boltzmann distribution $\rho=A e^{-\beta m \Phi}$. It is associated with the Boltzmann entropy $S=-k_B \int \frac{\rho}{m}\ln\rho\, d{\bf r}$, which is the limit of the Tsallis entropy for $\gamma\rightarrow 1$. In that case,  $E_{\rm tot}=\Theta_c+W-TS$.

\subsubsection{Lane-Emden equation}

For the polytropic equation of state (\ref{g8}), the fundamental differential equation of hydrostatic equilibrium (\ref{g5}) becomes
\begin{eqnarray}
-\frac{K\gamma}{\gamma-1}\Delta\rho^{\gamma-1}=S_d G\rho.
\label{g12}
\end{eqnarray}
Assuming that the system is spherically symmetric and making the change of variables 
\begin{eqnarray}
\rho=\rho_0\theta^n,\qquad \xi=\left\lbrack \frac{S_d G\rho_0^{1-1/n}}{|K(n+1)|}\right\rbrack^{1/2}r=r/r_0,
\label{g13}
\end{eqnarray}
where $\rho_0$ is the central density and $r_0$ is the polytropic core radius,  we can write Eq. (\ref{g12}) under the form 
\begin{equation}
\label{g14}
\frac{1}{\xi^{d-1}}\frac{d}{d\xi}\left (\xi^{d-1}
\frac{d\theta}{d\xi}\right )=-\epsilon\theta^n
\end{equation}
with $\theta(0)=1$ and $\theta'(0)=0$. We have $\epsilon=+1$ in case (i) and $\epsilon=-1$ in case (ii). The kinetic temperature $T=P/\rho=K\rho^{1/n}=K\rho_0^{1/n}\theta$ is proportional to $\theta$ for $n>0$ and for $n<-1$, i.e.,  $T\propto \theta$ when  $K>0$. Eq. (\ref{g14}) with $\epsilon=+1$ is the celebrated Lane-Emden equation \cite{emden,chandrabook,anomalous}.\footnote{Emden \cite{emden}, Chandrasekhar \cite{chandrabook} and Chavanis and Sire \cite{anomalous} did not consider the case of polytropes of index $n<0$. The case $\epsilon=-1$ can be formally obtained from the case $\epsilon=1$ by making the substitution $\xi\rightarrow i\xi$.} The normalized density profile $(\rho/\rho_0)(r/r_0)$ is universal as a consequence of the homology invariance of
the solutions of the Lane-Emden equation \cite{emden,chandrabook,anomalous}. In the following, we summarize and extend the results of \cite{emden,chandrabook,anomalous} to arbitrary dimension and arbitrary index.

\subsubsection{Taylor expansion}

For $\xi\rightarrow 0$, we have the Taylor expansion
\begin{equation}
\label{g15}
\theta=1-\frac{\epsilon}{2d}\xi^2+\frac{n}{8d(d+2)}\xi^4+...
\end{equation}
The function $\theta(\xi)$ decreases when $\epsilon=+1$ [case (i)] and increases
when $\epsilon=-1$ [case (ii)]. The density $\rho\propto \theta^n$ and the
pressure $P\propto \pm\theta^{n+1}$  always decrease with the distance 
(see footnote 15). Close to the center, the regular density profile behaves as
\begin{equation}
\label{g15prof}
\rho=\rho_0\left ( 1-\frac{|n|}{2d}\frac{r^2}{r_0^2}+...\right ),
\end{equation}
indicating the presence of a core. The density profile of the core may be approximated by
\begin{equation}
\label{g15profa}
\rho_{\rm core}=\frac{\rho_0}{\left ( 1+\frac{1}{2d}\frac{r^2}{r_0^2}\right )^{|n|}}.
\end{equation}

Case (i): For $0\le n<n_5=(d+2)/(d-2)$ when $d>2$ and for any $n\ge 0$ when $d\le 2$ (see below), the function $\theta(\xi)$ vanishes at a finite distance $\xi_1$ (i.e. $\theta_1=\theta(\xi_1)=0$).\footnote{Using the technique of Pad\'e approximants, Pascual \cite{pascual} obtained approximate analytical solutions of the Lane-Emden equation valid for any value of $n$ in $d=3$ and found that $\xi_1\simeq 15.0(5-n)^{-9/8}$ with an error smaller than $1\%$ for $0\le n \le 4$.} Close to $\xi_1$, we have
\begin{eqnarray}
\label{complete}
\theta=-\xi_1\theta'_1\Biggl\lbrack\frac{\xi_1-\xi}{\xi_1}+\frac{d-1}{2}\left (\frac{\xi_1-\xi}{\xi_1}\right )^2\nonumber\\
+\frac{d(d-1)}{6}\left (\frac{\xi_1-\xi}{\xi_1}\right )^3+...\Biggr\rbrack,
\end{eqnarray}
where $\theta'_1=\theta'(\xi_1)$.  In that case, the density $\rho\propto \theta^n$ and the pressure $P\propto \theta^{n+1}$ of the system vanish at some finite radius $R=r_0\xi_1$  (i.e. $\rho(R)=P(R)=0$). The polytropes have a compact density profile. We will say that they are are complete. For $r\rightarrow R^-$, their density vanishes like
\begin{eqnarray}
\label{bru}
\rho\sim \frac{\rho_0}{r_0^n}(-\theta'_1)^n(R-r)^n.
\end{eqnarray}
The mass-radius relation of complete polytropes in $d$ dimensions is \cite{emden,chandrabook,anomalous}
\begin{eqnarray}
\label{completemr}
M^{(n-1)/n}R^{(d-2)(n_3-n)/n}=\frac{K(1+n)}{GS_d^{1/n}}\omega_n^{(n-1)/n},
\end{eqnarray}
where $n_3=d/(d-2)$ and $\omega_n=-\xi_1^{(n+1)/(n-1)}\theta'_1$. 

Case (ii): The function $\theta(\xi)$ increases and tends to $+\infty$ at infinity. The density $\rho\propto \theta^n$  decreases and tends to zero at infinity. The pressure $P\propto \pm\theta^{n+1}$ decreases and tends to $0$ (when $n<-1$) or to $-\infty$ (when $-1<n<0$) at infinity. In $d\ge 2$, the mass in infinite. In $d<2$ the mass is finite when $n<n_3$ and infinite when $n_3<n<0$ (see below).

\subsubsection{Singular solution}

For certain indices $n$ (see below), the Lane-Emden equation (\ref{g14}) admits a power-law solution $\theta_s\propto \xi^{-2/(n-1)}$ leading to a singular density profile  $\rho_s\propto \theta_s^n\propto r^{-2n/(n-1)}$.

In case (i) the singular solution is
\begin{equation}
\label{g16}
\theta_s=\left\lbrace \frac{2\left\lbrack (d-2)n-d\right\rbrack}{(n-1)^2}\right \rbrace^{1/(n-1)}\xi^{-2/(n-1)}
\end{equation}
provided that $(d-2)n-d>0$. The singular solution exists for $n>n_3=d/(d-2)$ in dimensions $d>2$. The corresponding singular density profile $\rho_s\propto r^{-2n/(n-1)}$ is integrable for $r\rightarrow 0$ but not for $r\rightarrow +\infty$.

In case (ii) the singular solution is
\begin{equation}
\label{g17}
\theta_s=\left\lbrace \frac{2\left\lbrack d-(d-2)n\right\rbrack}{(n-1)^2}\right \rbrace^{1/(n-1)}\xi^{2/(1-n)}
\end{equation}
provided that $d-(d-2)n>0$. The singular solution exists for any $n<0$ in dimensions $d\ge 2$ and for $n_3=d/(d-2)<n<0$ in dimensions $d<2$. In particular, when $d=1$, the singular solution exists only for $-1<n<0$. The corresponding singular density profile $\rho_s\propto r^{-2n/(n-1)}$ is integrable for $r\rightarrow 0$ but not for $r\rightarrow +\infty$.

For $n=-1$ the singular solution is 
\begin{equation}
\label{g18}
\theta_s=\frac{\xi}{\sqrt{d-1}}
\end{equation}
provided that $d>1$. The dimension $d=1$ is treated in Sec. \ref{sec_dimun}.

\subsubsection{Asymptotic behavior for $r\rightarrow +\infty$}
\label{sec_abi}

We can show that, in general,  the regular unbounded solutions of the Lane-Emden equation (with a finite central density $\rho_0$) behave asymptotically like the singular solution. In particular, the density decreases as $\rho \propto r^{-2n/(n-1)}$ for $r\rightarrow +\infty$. We note that the exponent does not depend on the dimension of space $d$.

To prove this result, we make the change of variables $t=\ln\xi$ and $\theta=\xi^{-2/(n-1)}z$ \cite{emden,chandrabook,anomalous}, which transforms the Lane-Emden equation (\ref{g14}) into the differential equation
\begin{equation}
\label{g19}
\frac{d^2z}{dt^2}+\frac{(d-2)n-(d+2)}{n-1}\frac{dz}{dt}=-\epsilon z^n-2\frac{d+(2-d)n}{(n-1)^2}z.
\end{equation}
This equation is similar to the damped equation of motion of a fictive particle of unit mass in a potential
\begin{equation}
\label{g20}
V(z)=\epsilon \frac{z^{n+1}}{n+1}+\frac{d+(2-d)n}{(n-1)^2}z^2,
\end{equation}
where $z$ plays the role of the position and $t$ the role of time. 
The extremum of the potential is located at
\begin{equation}
\label{g21}
z_s=\left\lbrace \epsilon \frac{2\left\lbrack (d-2)n-d\right\rbrack}{(n-1)^2}\right \rbrace^{1/(n-1)}.
\end{equation}
Since
\begin{equation}
\label{g22}
V''(z_s)=\frac{2\left\lbrack (d-2)n-d\right\rbrack}{n-1},
\end{equation}
we see that when the conditions below Eqs. (\ref{g16}) and (\ref{g17}) are fulfilled, $z_s$ is always a minimum of $V(z)$ i.e. $V''(z_s)>0$. Furthermore, in case (i), the coefficient in front of $dz/dt$ is positive (implying damping)  if, and only if, $n>n_5=(d+2)/(d-2)$. In case (ii) this coefficient is always positive. When these conditions are fulfilled, the fictive particle reaches the minimum of the potential $z_s$ for $t\rightarrow +\infty$. This implies that the regular solution of the Lane-Emden equation behaves like the singular solution for $\xi\rightarrow +\infty$.

Combining the previous results, we can make the following claims.

In case (i) the regular polytropes are unbounded (incomplete) in dimensions $d>2$ for $n>n_5$. Their density behaves as $\rho \propto r^{-2n/(n-1)}$ for $r\rightarrow +\infty$  and their total mass is infinite. The regular polytropes are complete in dimensions $d>2$ for $n<n_5$ even though there exists an unbounded  singular solution for $n_3<n\le n_5$ (the index $n=n_5$ leading to an unbounded solution with a finite mass is treated specifically in Sec. \ref{sec_schuster}). The regular polytropes are complete in dimensions $d\le 2$.

In case (ii) the regular polytropes are unbounded in dimensions $d\ge 2$ for any $n<0$. The regular polytropes are unbounded in dimensions $d<2$ for $n_3=d/(d-2)<n<0$.  In particular, the regular polytropes are unbounded in dimension $d=1$ for $n_3=-1<n<0$. Their density behaves as $\rho \propto r^{-2n/(n-1)}$ for $r\rightarrow +\infty$  and their total mass is infinite. The regular polytropes are probably also unbounded in dimensions $d<2$ for $n\le n_3$ but they have a finite mass. Indeed, we find unbounded solutions with a finite mass for $n=n_5\le n_3$ when $d<2$  in Sec. \ref{sec_schuster} and for $n<n_3=-1$ when $d=1$ in Sec. \ref{sec_dimun} (the index $n=-1$ is treated specifically in Sec.  \ref{sec_logog} and has an infinite mass).

We can study the correction to the asymptotic behavior of the density profile by writing $z=z_s+z'$ and linearizing Eq. (\ref{g19}) for $z'\ll z_s$, giving 
\begin{equation}
\label{g23}
\frac{d^2z'}{dt^2}+\frac{(d-2)n-(d+2)}{n-1}\frac{dz'}{dt}+\frac{2\lbrack (d-2)n-d\rbrack}{n-1}z'=0.
\end{equation}
This equation is valid in the two cases (i) and (ii). Proceeding as in \cite{emden,chandrabook,anomalous}  we can establish the following results.

Case (i): In dimensions $d>10$, the density profile displays damped oscillations around the asymptotic solution for $n_5<n<n_{-}$ and an overdamped convergence (without oscillations) towards the  asymptotic solution for $n\ge n_{-}$, where $n_-$ is defined (together with $n_+$ used below) by
\begin{equation}
\label{g24}
n_{\pm}=\frac{-d^2+8d-4\pm 8\sqrt{d-1}}{(d-2)(10-d)}.
\end{equation}
In dimensions $2<d\le 10$,  the density profile displays damped oscillations around the asymptotic solution for $n>n_5$.

Case (ii): In dimensions $d\ge 10$, the density profile displays an overdamped convergence (without oscillations) towards the  asymptotic solution for any $n<0$. In dimensions $2<d<10$, the density profile displays damped oscillations around the asymptotic solution for $n<n_{-}$ and an overdamped convergence (without oscillations)  towards the  asymptotic solution for $n_{-}\le n<0$. In dimension $d=2$,  the density profile displays damped oscillations around the asymptotic solution for any $n<0$. In dimensions $1<d<2$, the density profile displays an overdamped convergence (without oscillations) towards the  asymptotic solution for $n_3<n\le n_+$, damped oscillations around the asymptotic solution for $n_{+}<n<n_{-}$, and an overdamped convergence  (without oscillations) towards the  asymptotic solution for $n_{-}\le n<0$. In dimension $d=1$,  the density profile displays an overdamped convergence (without oscillations)  towards the  asymptotic solution for  $n_3=-1<n<0$. In dimensions $d<1$,  the density profile displays an overdamped convergence (without oscillations) towards the  asymptotic solution for $n_3<n<0$.

{\it Remark:} We have seen that, in certain situations, the density profile of self-gravitating polytropes decreases at large distances as $r^{-2n/(n-1)}$ with damped oscillations superimposed (see \cite{emden,chandrabook,anomalous} for more details). When plotted in terms of certain homology variables (e.g. the Milne \cite{milnevariables1,milnevariables2} variables $u$ and $v$) the solution of the Lane-Emden equation forms a spiral in the $(u,v)$ plane \cite{emden,chandrabook,anomalous}. A similar behavior is obtained in terms of thermodynamic variables. Indeed, the  caloric curves of box-confined polytropes (including the isothermal sphere) have the form of a spiral \cite{anomalous}.

\subsubsection{$d$-dimensional Schuster solution}
\label{sec_schuster}

For the index $n_5=(d+2)/(d-2)$, the Lane-Emden equation (\ref{g14}) admits an analytical solution
\begin{equation}
\label{g25}
\theta_5=\frac{1}{\left\lbrack 1+\frac{\epsilon}{d(d-2)}\xi^2\right \rbrack^{(d-2)/2}}.
\end{equation}
The corresponding density profile is
\begin{equation}
\label{g25prof}
\rho_5(r)=\frac{\rho_0 }{\left\lbrack 1+\frac{S_d G\rho_0^{\frac{4}{d+2}}}{2d^2K}r^2\right \rbrack^{(d+2)/2}}.
\end{equation}
In case (i) this solution is valid in dimensions $d>2$ and we have $1<n_5<+\infty$. In case (ii) this solution is valid in dimensions $d<2$ and we have $-\infty<n_5<-1$.\footnote{The case of logotropes $n=-1$ corresponds to $d\rightarrow 0$.} There is no solution with $-1<n_5<1$ (for $d>0$).  For $\xi\rightarrow +\infty$, we have $\theta_5\sim \xi^{-(d-2)}$, implying $\rho_5\propto \theta_5^{(d+2)/(d-2)} \propto r^{-(d+2)}$. This asymptotic behavior is different from the asymptotic behavior of the unbounded solutions determined in Sec. \ref{sec_abi}. The mass of the configuration is always finite.\footnote{By contrast, the moment of inertia $I=\int \rho r^2\, d{\bf r}$ diverges logarithmically.} The mass-central density relation reads
\begin{equation}
\label{g25profw}
M=\frac{S_d}{d}\left (\frac{2Kd^2}{S_d G}\right )^{d/2}\frac{1}{\rho_0^{\frac{d-2}{d+2}}}.
\end{equation}

In dimension $d=3$, this analytical solution has been found by Schuster \cite{schuster} in the context of gaseous stars with polytropic index $n=5$. It was used by Plummer \cite{plummer} as a model of galaxy (see Appendix A of \cite{acpre} for some historical details). In dimension $d=2$, the solution from Eq. (\ref{g25}) is singular but it can be related to the profile $\psi(\xi)$ of an isothermal cylinder ($n\rightarrow +\infty)$ which is the solution of the Emden equation \cite{sc2002,anomalous,tcdim2,virialrot}. This solution exists for a unique value of the temperature $K\equiv k_B T/m=GM/4$, or for a unique value of the mass $M=4k_B T/Gm$, independent of the central density, and the corresponding density profile is
\begin{equation}
\label{g25profiso}
\rho_5(r)=\frac{\rho_0 }{\left ( 1+\frac{\pi G\rho_0}{4K}r^2\right )^{2}},
\end{equation}
in agreement with Eqs. (\ref{g25prof}) and (\ref{g25profw}) for $d=2$.
In dimension $d=1$, we have $n_5=-3$ and 
\begin{equation}
\label{g26}
\theta_5=\left (1+\xi^2\right )^{1/2},\quad \rho_5(r)=\frac{\rho_0 }{\left ( 1+\frac{G\rho_0^{{4}/{3}}}{K}r^2\right )^{3/2}}.
\end{equation}
This density profile has been used recently in \cite{roule}.

We recall the equilibrium virial theorem for a gaseous star \cite{emden,chandrabook,anomalous}
\begin{equation}
\label{vir1}
2E_{\rm kin}+(d-2)W=0\qquad (d\neq 2),
\end{equation}
\begin{equation}
\label{vir2}
2E_{\rm kin}-\frac{GM^2}{2}=0\qquad (d=2),
\end{equation}
where $E_{\rm kin}=\frac{1}{2}\int fv^2\, d{\bf r}d{\bf v}=\frac{d}{2}\int P\, d{\bf r}$ is the kinetic energy (with $P=\frac{1}{d}\int fv^2\, d{\bf v}$) and $W=\frac{1}{2}\int\rho\Phi\, d{\bf r}$ is the potential energy. For gaseous polytropes, using Eq. (\ref{g10}), we get  $U=n\int P\, d{\bf r}=\frac{2n}{d}E_{\rm kin}$ and $E_{\rm tot}=(2n/d)E_{\rm kin}+W$. Then, using Eqs. (\ref{vir1}) and (\ref{vir2}) we find that  $E_{\rm tot}=(1-n/n_3)W$ ($d\neq 2$) and $E_{\rm tot}=nGM^2/4+W$ ($d=2$) \cite{wdsD}. From these relations, it can be shown that gaseous polytropes are stable if, and only if, $1/n>1/n_3=(d-2)/d$ \cite{wdsD}. For the index $n=n_5$, calculating $E_{\rm kin}=\frac{d}{2}\int P\, d{\bf r}$ with the density profile from Eq. (\ref{g25prof}) and using Eq. (\ref{vir1}), we find for $d\neq 2$ that
\begin{equation}
\label{vir3}
W=-\frac{d}{d-2}KS_d\frac{\sqrt{\pi}\Gamma\left (\frac{d}{2}\right )}{\Gamma\left (\frac{1+d}{2}\right )}\left (\frac{d^2K}{2S_dG}\right )^{d/2}\qquad (d\neq 2),
\end{equation}
and $E_{\rm tot}=-(2/d)W$. We note that the energy is independent of the central density. The isothermal case $d=2$ is treated in \cite{sc2002,anomalous,tcdim2,virialrot} giving
\begin{equation}
\label{vir3b}
W=\frac{GM^2}{4}\left \lbrack 1+\ln\left (\frac{M}{\pi \rho_0}\right )\right\rbrack\qquad (d=2).
\end{equation}
The total energy $E_{\rm tot}=(GM^2/4)\lbrack \ln(M/\pi)-1\rbrack$ is constant
\cite{tcdim2,virialrot}. 
For the other indices $n\neq n_5$, the potential energy $W$ is given by the Betti-Ritter formula  \cite{emden,chandrabook,anomalous} 
\begin{equation}
\label{br1}
W=-\frac{d}{d+2-(d-2)n}\frac{GM^2}{(d-2)R^{d-2}}\qquad (d\neq 2),
\end{equation}
\begin{equation}
\label{br2}
W=-(n+1)\frac{GM^2}{8}+\frac{1}{2}GM^2\ln R\qquad (d=2),
\end{equation}
which is valid for $0\le n<+\infty$ when the density profile has a compact support: $\rho(R)=0$.
In $d=3$, we get $W=-\frac{3}{5-n}\frac{GM^2}{R}$ \cite{betti,ritter332}.

{\it Remark:} The Lane-Emden equation also admits analytical solutions for $n=0$
and $n=1$ in any dimension of space $d$ \cite{emden,chandrabook,anomalous} (see,
e.g., \cite{prd1,frontiers} for specific applications of the index $n=1$ in
relation 
to Bose-Einstein condensate DM).

\subsubsection{Dimension $d=1$}
\label{sec_dimun}

In $d=1$ dimension, the Lane-Emden equation (\ref{g14}) becomes
\begin{equation}
\label{g27}
\frac{d^2\theta}{d\xi^2}=-\epsilon\theta^n,
\end{equation}
with $\theta(0)=1$ and $\theta'(0)=0$. It is similar to the equation of motion of a fictive particle of unit mass in a potential
\begin{equation}
\label{g28}
V(\theta)=\frac{\epsilon}{n+1}\theta^{n+1}\qquad (n\neq -1),
\end{equation}
where $\theta$ plays the role of the position and $\xi$ the role of the time. The first integral of motion, which corresponds to the energy of the fictive particle, is
\begin{equation}
\label{g29}
E=\frac{1}{2}\left (\frac{d\theta}{d\xi}\right )^2+\frac{\epsilon}{n+1}\theta^{n+1},
\end{equation}
where $E$ is a constant. It is determined by the initial condition giving $E=\epsilon/(n+1)$. The solution of Eq. (\ref{g27}) is then given in reversed form by
\begin{equation}
\label{g30}
\xi=\epsilon\int_{\theta}^{1} \frac{dx}{\sqrt{\frac{2\epsilon}{n+1}(1-x^{n+1})}}.
\end{equation}

In case (i) we get
\begin{equation}
\label{g31}
\xi=\int_{\theta}^{1} \frac{dx}{\sqrt{\frac{2}{n+1}(1-x^{n+1})}}.
\end{equation}
The polytropes are complete and the density vanishes ($\theta=0$) at the normalized radius
\begin{equation}
\label{g32}
\xi_1=\left (\frac{n+1}{2}\right )^{1/2}\sqrt{\pi}\frac{\Gamma\left (1+\frac{1}{1+n}\right )}{\Gamma\left (\frac{1}{2}+\frac{1}{1+n}\right )}.
\end{equation}
At that point we have
\begin{equation}
\label{ami}
\theta'_1=-\sqrt{\frac{2}{n+1}}.
\end{equation}
For $n=1$, we obtain $\theta=\cos\xi$, $\xi_1=\pi/2$, and $\theta'_1=-1$.

In case (ii) we get
\begin{equation}
\label{g33}
\xi=\int_{1}^{\theta} \frac{dx}{\sqrt{\frac{2}{n+1}(x^{n+1}-1)}}.
\end{equation}
When $n_3=-1<n<0$, we find that 
\begin{equation}
\label{g34}
\theta\sim \left\lbrack \frac{(1-n)^2}{2(n+1)}\right\rbrack^{1/(1-n)}\xi^{2/(1-n)}\quad (\xi\rightarrow +\infty),
\end{equation}
implying that $\rho\propto r^{2n/(1-n)}$ for $r\rightarrow +\infty$. The total mass diverges. This returns the asymptotic result found in Sec. \ref{sec_abi}. When $n<n_3=-1$, we find that
\begin{equation}
\label{g35}
\theta\sim \sqrt{\frac{-2}{n+1}}\xi \quad (\xi\rightarrow +\infty),
\end{equation}
implying that $\rho\propto r^{-|n|}$ for $r\rightarrow +\infty$.
The system is unbounded but its total mass is finite
(the moment of inertia is finite when $n<-3$). For $n=n_5=-3$, we recover the
Schuster solution from Eq. (\ref{g26}). In that case, the potential energy is
$W=2\pi (K^3/4G)^{1/2}$ [see Eq. (\ref{vir3})]. For $0\le n<+\infty$, the
potential energy is given by $W=GM^2R/(3+n)$ [see Eq. (\ref{br1})]. 

In any dimension, the total mass is given by
\begin{equation}
\label{massepoly}
M=-\epsilon\rho_0 \left\lbrack \frac{|K(n+1)|}{S_d G\rho_0^{1-1/n}}\right\rbrack^{d/2}
S_d\xi_1^{d-1}\theta'_1,
\end{equation}
where $\xi_1$ may be finite or infinite. For $n=n_5$, we recover Eq. (\ref{g25profw}). When $R$ is finite, eliminating $\rho_0$ between Eqs. (\ref{massepoly}) and (\ref{g13}) taken at $\xi=\xi_1$, we obtain the mass-radius relation from Eq. (\ref{completemr}). Applying Eq. (\ref{massepoly}) in $d=1$ and using Eqs. (\ref{ami}) and (\ref{g35}), we obtain for $n\ge 0$ and for $n<n_3=-1$:
\begin{equation}
\label{massepoly1}
M=2\left (\frac{K}{G}\right )^{1/2}\rho_0^{\frac{n+1}{2n}}.
\end{equation}
For $n\rightarrow \infty$, we recover the result $M=2(k_B T/Gm)^{1/2}\rho_0^{1/2}$ of isothermal sheets \cite{tcdim2}. Their density profile is
\begin{equation}
\label{rho1d}
\rho(x)=\frac{\rho_0}{\cosh^2\left (\frac{GMmx}{2k_B T}\right )}.
\end{equation}
Their moment of inertia is $I=\frac{\pi^2}{12}M(2k_B T/GMm)^2$ and their
potential energy, obtained from the equilibrium virial theorem (\ref{vir1}), 
is $W=2E_{\rm kin}=Nk_B T$. The total energy is $E_{\rm tot}=Nk_B T(\ln
\rho_0-1+2\ln 2)$.

For $n=-1$, the potential is
\begin{equation}
\label{g36}
V(\theta)=-\ln\theta.
\end{equation}
The first integral of motion takes the form
\begin{equation}
\label{g37}
E=\frac{1}{2}\left (\frac{d\theta}{d\xi}\right )^2-\ln\theta,
\end{equation}
where $E$ is a constant. It is determined by the initial condition giving $E=0$. The solution of Eq. (\ref{g27}) is then given in reversed form by
\begin{equation}
\label{g38}
\xi=\int_{1}^{\theta} \frac{dx}{\sqrt{2\ln x}}.
\end{equation}
Making the change of variables $\ln x=y^2$, we obtain
\begin{equation}
\label{g39}
\xi=\sqrt{2}\theta F(\sqrt{\ln\theta}),
\end{equation}
where
\begin{equation}
\label{g40}
F(x)=e^{-x^2}\int_0^x e^{y^2}\, dy
\end{equation}
is Dawson's integral. For $\xi\rightarrow +\infty$, we get
\begin{equation}
\theta\sim \xi\sqrt{2\ln\xi}.
\label{g40b}
\end{equation}
This asymptotic behavior is different from the asymptotic behavior of the unbounded solutions determined in Sec. \ref{sec_abi} (see also Sec. \ref{sec_logog}). 

{\it Remark:} Coincidentally, the 1D Lane-Emden equation of index $n=1$ (Laplace), $n=-3$ (Schuster, Plummer) and $n=-1$ (logotropes) also occurs in quantum mechanics in relation to a generalized Schr\"odinger equation \cite{nottale2}.

\subsection{Logotropic equation of state}
\label{sec_logog}

In this section, we study the logotropic equation of state
\begin{eqnarray}
P=A\ln\left (\frac{\rho}{\rho_*}\right ).
\label{g41}
\end{eqnarray}
We assume $A>0$ in order to have $c_s^2=P'(\rho)=A/\rho>0$. The pressure is positive when $\rho>\rho_*$ and negative when $\rho<\rho_*$. We first recall how the logotropic equation of state can 
be obtained from the polytropic equation of state in a certain limit \cite{cslogo,epjp} (see also Appendix A of \cite{logosf}). 

To that purpose we consider a self-gravitating gaseous system satisfying the condition of hydrostatic equilibrium (\ref{g4}). For the polytropic equation of state (\ref{g8}),
this condition can be written as 
\begin{eqnarray}
K\gamma\rho^{\gamma-1}\nabla\rho+\rho\nabla\Phi={\bf 0}.
\label{g42}
\end{eqnarray}
Taking the limit $\gamma\rightarrow 0$ and $K\rightarrow\infty$ with
$A=K\gamma$ fixed, we obtain
\begin{eqnarray}
\frac{A}{\rho}\nabla\rho+\rho\nabla\Phi={\bf 0}.
\label{g43}
\end{eqnarray}
Comparing this equation with Eq. (\ref{g4}), we see that the pressure
involved in this expression corresponds to the logotropic equation of
state (\ref{g41}) where $\rho_*$ is a constant of integration. This shows that the logotropic equation of state can be viewed as a degenerate form of the polytropic equation of state with index $\gamma=0$ (i.e. $n=-1$). Note that the logotropic equation of state differs from a pure polytrope of index $\gamma=0$ (or $n=-1$) and fixed $K$ which has a constant pressure $P=K$.  Therefore, the limit
 $\gamma\rightarrow 0$ and $K\rightarrow \infty$ with $A=K\gamma$ fixed leading
to the logotropic equation of state is very peculiar.

We can obtain the logotropic equation of state
(\ref{g41}) directly
from the polytropic equation of state (\ref{g8}) by
writing
\begin{eqnarray}
P=Ke^{\gamma\ln \rho}
\label{g44}
\end{eqnarray}
and expanding the right hand side  for $\gamma\rightarrow 0$. This yields
\begin{eqnarray}
P=K(1+\gamma\ln\rho+...).
\label{g45}
\end{eqnarray}
In the limit $\gamma\rightarrow 0$ and $K\rightarrow \infty$ with
$A=K\gamma$ fixed, we get
\begin{eqnarray}
P=A\ln\rho+K+...
\label{g46}
\end{eqnarray}
The drawback with this calculation is that it yields an infinite constant
($K\rightarrow +\infty$) in addition to the logotropic equation of state $A\ln\rho$. 
Therefore, this procedure is not well-justified mathematically. The infinite
constant has to be removed by hand.\footnote{This procedure may be related to a
form of renormalization (see, e.g., \cite{diracrenorm}).}  By contrast, the
calculation based on the condition of hydrostatic equilibrium (\ref{g4}) avoids
dealing explicitly with an infinite constant since it disappears in the pressure
gradient. Indeed, a constant term in the equation of state $P(\rho)$ has no effect in nonrelativistic mechanics.\footnote{This is no more true in general relativity \cite{epjp}.}

For the logotropic equation of state (\ref{g41}), the energy functional (\ref{g6}) associated with the Euler-Poisson equations reads
\begin{equation}
E_{\rm tot}=\frac{1}{2}\int \rho {\bf u}^2\, d{\bf r}-A\int \ln\rho\, d{\bf r}+\frac{1}{2}\int\rho\Phi\, d{\bf r}.
\label{g47}
\end{equation}
It can also be obtained from the energy functional (\ref{g9})  of a polytrope  in the limit $\gamma\rightarrow 0$ and $K\rightarrow \infty$ with
$A=K\gamma$ fixed. It can be interpreted as a generalized free energy in physical space $E_{\rm tot}=\Theta_c+W-AS$, where $\Theta_c+W$ is the kinetic $+$ potential energy, $A$ plays the role of a generalized temperature called the logotropic temperature, and $S_L=\int\ln\rho\, d{\bf r}$ is a generalized entropy called the log-entropy \cite{cslogo}.  We also have (up to a possibly infinite additive constant) 
\begin{equation}
E_{\rm tot}=\frac{1}{2}\int \rho {\bf u}^2\, d{\bf r}-\int P\, d{\bf r}+\frac{1}{2}\int\rho\Phi\, d{\bf r}.
\label{g48}
\end{equation}
The logotropic equilibrium density profile can be obtained by extremizing the energy at fixed mass. Writing $\delta E_{\rm tot}-\mu\delta M=0$, we get
\begin{equation}
\rho=\frac{A}{\Phi-\mu}.
\label{g49}
\end{equation}
This distribution can also be deduced from the polytropic (Tsallis) distribution (\ref{g11}) in the limit $\gamma\rightarrow 0$ and $K\rightarrow \infty$ with $A=K\gamma$ fixed, after removing or absorbing an infinite constant in $\mu$.

For the logotropic equation of state
(\ref{g41}),  the fundamental differential equation of hydrostatic
equilibrium (\ref{g5}) becomes
\begin{eqnarray}
A\Delta\left(\frac{1}{\rho}\right )=S_d G\rho.
\label{g50}
\end{eqnarray}
Assuming that the system is spherically symmetric and making the change of variables
\begin{equation}
\label{g51}
\theta=\frac{\rho_0}{\rho},\qquad \xi=\left (\frac{S_d
G\rho_0^2}{A}\right )^{1/2}r=r/r_0,
\end{equation}
where $\rho_0$ is the central density and $r_0=({A}/{S_d
G\rho_0^2})^{1/2}$ is the logotropic core radius, we find that Eq.
(\ref{g50})
reduces to the Lane-Emden equation of index $n=-1$:
\begin{equation}
\label{g52}
\frac{1}{\xi^{d-1}}\frac{d}{d\xi}\left (\xi^{d-1}
\frac{d\theta}{d\xi}\right )=\frac{1}{\theta},
\end{equation}
with the boundary conditions $\theta=1$ and $\theta'=0$ at $\xi=0$.\footnote{The Lane-Emden equation of index $n=-1$ cannot be obtained from the equation of state of a polytrope of index $\gamma=0$ (i.e. $n=-1$) since it has a vanishing pressure gradient which cannot balance the gravitational attraction. Therefore, the  condition of hydrostatic equilibrium (\ref{g4}) has no solution in that case (there is no equilibrium state) and the  Lane-Emden equation of index $n=-1$ is ill-defined (the scaled radius $r_0$ defined by Eq. (\ref{g13}) vanishes for $n=-1$). In order to justify the Lane-Emden equation of index $n=-1$ one has to consider the limit $\gamma\rightarrow 0$
and $K\rightarrow \infty$ with $A=K\gamma$ fixed, leading to the logotropic
equation of state (\ref{g41}).  In this sense, the logotropic equation of state completes the class of standard polytropes and gives a physical meaning to the Lane-Emden equation of index $n=-1$ which is excluded by the usual polytropic model.} This equation has been studied in detail in  \cite{cslogo,epjp}. The function $\theta(\xi)$ increases and tends to $+\infty$ at infinity.  The density $\rho\propto 1/\theta$ decreases and tends to $0$ at infinity while the pressure $P\propto -\ln\theta+C$ decreases, becomes negative, and tends to $-\infty$ at infinity.

In dimensions $d>1$, there exists an exact
analytical solution 
\begin{equation}
\label{g53}
\theta_s=\frac{\xi}{\sqrt{d-1}},
\end{equation}
corresponding to the density profile
\begin{equation}
\label{g54}
\rho_s=\left\lbrack \frac{(d-1)A}{S_d G}\right \rbrack^{1/2}\frac{1}{r}
\end{equation}
called the singular logotropic sphere. This singular density profile is integrable for $r\rightarrow 0$ but not for $r\rightarrow +\infty$. 

The regular logotropic density profiles (with a finite central density $\rho_0$) must be
computed numerically. The normalized density profile $(\rho/\rho_0)(r/r_0)$ is
universal as a consequence of the homology invariance of
the solutions of the Lane-Emden equation. It is plotted in Fig. 2 of \cite{cslogo} and in Fig. 18 of \cite{epjp}. The density
profile of a regular logotropic sphere has a core
($\rho\rightarrow {\rm cst}$ when $r\rightarrow 0$) and decreases at large
distances as $\rho\sim
r^{-1}$. More precisely, for $r\rightarrow +\infty$ we have
\begin{eqnarray}
\label{g55}
\rho\sim \left \lbrack \frac{(d-1)A}{S_d G}\right \rbrack^{1/2}\frac{1}{r},
\end{eqnarray}
like for the singular logotropic sphere. This profile has an infinite mass
because the density does not decrease sufficiently rapidly with the distance.

In $d=1$ dimension, the solution of the Lane-Emden equation (\ref{g52}) is given by Eq. (\ref{g39}) and the logotropic density profile is $\rho(r)=\rho_0/\theta(\xi)$ with $\xi=(2G\rho_0^2/A)^{1/2}r$. It decreases asymptotically as $\rho\sim (A/G)^{1/2}/(2r\sqrt{\ln r})$ for $r\rightarrow +\infty$.  It is plotted in Fig. 1 of \cite{cslogo}.  The integrated density behaves as $M(r)\propto \sqrt{\ln r}$ for $r\rightarrow +\infty$ so that the mass of the configuration is infinite.

In $d<1$, the logotrope is unbounded but its mass is finite.

{\it Remark:} For $d>1$, the density profile of self-gravitating logotropes decreases at large distances as $r^{-1}$ with damped oscillations superimposed as in the case of isothermal spheres and incomplete polytropes.  This asymptotic behavior is studied in detail in  \cite{cslogo} (see in particular Eq. (65) of \cite{cslogo}).
When plotted in terms of  the Milne variables the solution of the Emden equation forms a spiral in the $(u,v)$ plane.  A similar behavior is obtained in terms of thermodynamic variables.  Indeed, the  caloric curve of box-confined logotropes has the form of a spiral \cite{cslogo}.

\subsection{Dynamical stability of gaseous polytropes}
\label{sec_stabg}

In this section, we summarize the stability results obtained for barotropic stars (see \cite{aaISO,aaPOLY,sc2002,grand,anomalous,csTsallis,aastab,cslogo,wdsD,assise,nyquistHMF,cc,GR1,kinVR} and references therein for a detailed discussion of these stability results).

A barotropic star is (nonlinearly) dynamically stable with respect to the Euler-Poisson equations (\ref{g1})-(\ref{g3}) if, and only if, it is a minimum of energy $E_{\rm tot}$ at fixed mass $M$.  This is similar to a condition of generalized thermodynamical stability in the canonical ensemble (one constraint problem). For small perturbations (linear stability), we require that $\delta^2E_{\rm tot}>0$ for all perturbations that conserve mass. This is equivalent to linearizing the Euler-Poisson equations about an equilibrium state and requiring that the squared pulsation is positive ($\omega^2>0$) for such perturbations (spectral stability) \cite{aastab,wdsD}.

We can therefore use a thermodynamical analogy to investigate the dynamical stability of a steady state of the Euler-Poisson equations \cite{aastab}. In particular, we can use the Poincar\'e turning point argument \cite{poincare,katztopo,ijmpb} to determine the stability of the system from the topology of the generalized caloric curve (see a summary of the Poincar\'e theory of linear series of equilibria  in Appendix C of \cite{acf}).

In $d=3$ dimensions,\footnote{See \cite{wdsD} for results valid in other dimensions of space. In $d$ dimensions, it can be shown that gaseous polytropes are stable if, and only if, $1/n>1/n_3=(d-2)/d$. When $d>2$, gaseous polytropes are stable if, and only if, $0\le n<n_3$. When $d\le 2$, gaseous polytropes are stable if, and only if, $n\ge 0$ or $n<n_3=d/(d-2)$.} it can be shown that complete gaseous polytropes are stable for $0\le n<3$ and unstable for $3<n\le 5$ (they are marginally stable for $n=3$).\footnote{These results can be established by a direct calculation (see, e.g., Appendix B of \cite{wdsD}) or from the topology of their generalized caloric curve without making any calculation \cite{aastab}.} In that case, they undergo gravitational collapse. For box-confined gaseous polytropes, it can be shown that the series of equilibria (generalized caloric curve) is stable before the first turning point of generalized temperature and that it becomes unstable afterwards \cite{aaPOLY}. In particular, the instability of complete polytropes in the range $3\le n\le 5$ can be related to the existence of a region of negative specific heats in the generalized caloric curve. Indeed, we know that negative specific heats are forbidden (unstable) in the canonical ensemble \cite{aastab}. Therefore, box-confined polytropes with index $n>3$ become unstable after the first turning point of generalized temperature (corresponding to a minimum temperature) and undergo a collapse \cite{aaPOLY,anomalous} like isothermal systems \cite{aaISO,sc2002}.

{\it Remark:} The stability of logotropic stars is a bit subtle because these configurations have an infinite mass and must be artificially confined within a box or surrounded by an envelope. This is the same problem as for isothermal stars and unbounded polytropic stars. These gaseous configurations are dynamically stable with respect to the Euler-Poisson equations if the central density is sufficiently small and they become unstable above a critical central density corresponding to the first turning point of generalized temperature in the series of equilibria. We refer to \cite{aaISO,aaPOLY,sc2002,grand,anomalous,csTsallis,aastab,cslogo} for a detailed study of box-confined isothermal, polytropic and logotropic gaseous configurations in various dimensions of space.

\subsection{Effect of a central black hole}
\label{sec_cbh}

We can take into account the effect of a massive central object (e.g.  a compact sub-halo or a central black hole) in a polytropic sphere by adding the external potential $\Phi_{\rm BH}=-GM_{\rm BH}/r$ induced by this object (we restrict ourselves here to the dimension $d=3$) to the self-generated gravitational potential $\Phi$. This amounts to making the transformation $\Phi\rightarrow \Phi+\Phi_{\rm BH}$ in the foregoing equations, except in Eq. (\ref{g6}) where the potential energy associated with the external potential reads $W_{\rm BH}=\int\rho\Phi_{\rm BH}\, d{\bf r}$ without the factor $1/2$. The case of a polytrope of index $n=1$ surrounding a central body can be solved analytically as shown in Appendix I of \cite{becbh}. The case of an isothermal envelope ($n\rightarrow +\infty$) surrounding a central body is studied in \cite{corp1,corp2}. Considering a polytrope of arbitrary index $n>0$ and neglecting the self-gravity of the system (which is valid if we are sufficiently close to the central object), we find from Eq. (\ref{g11a}) that\footnote{For an index $n<0$, using Eq. (\ref{g11b}),  the density displays a singularity at a finite radial distance $r_c=GM_{\rm BH}/\epsilon_m$.} 
\begin{eqnarray}
\label{g56}
\rho=C\left (\epsilon_m+\frac{GM}{r}\right )^n\sim r^{-n},
\end{eqnarray}
\begin{eqnarray}
\label{g56b}
T=KC^{1/n}\left (\epsilon_m+\frac{GM}{r}\right )\sim r^{-1}.
\end{eqnarray}
The polytrope displays a density cusp scaling as $r^{-n}$. In particular for $n=3/2$, corresponding to the Fermi distribution (see Sec. \ref{sec_spol}), we get $\rho\propto r^{-3/2}$.  On the other hand, for $n=1$ we get a density cusp $\rho\propto r^{-1}$ (see Appendix I of \cite{becbh}) like the NFW cusp or the singular logotropic profile.

{\it Remark:} Generalizing these arguments in $d$ dimensions, we find that $\rho\propto r^{-(d-2)n}$ when $d>2$ and $\rho\propto (-\ln r)^{n}$ when $d=2$.

\section{Self-gravitating systems described by a DF: stellar systems}
\label{sec_gal}

\subsection{Equilibrium states of spherically symmetric and isotropic stellar systems}
\label{sec_sta}

We consider a collisionless self-gravitating system (stellar system) described by the Vlasov-Poisson equations
\begin{eqnarray}
\frac{\partial f}{\partial t}+{\bf v}\cdot \frac{\partial f}{\partial {\bf r}}-\nabla\Phi\cdot \frac{\partial f}{\partial {\bf v}}=0,
\label{g57}
\end{eqnarray}
\begin{eqnarray}
\Delta\Phi=S_d G\int f\, d{\bf v}.
\label{g58}
\end{eqnarray} 
It can be shown that spherically symmetric and isotropic steady states of the Vlasov-Poisson equations are of the form $f=f(\epsilon)$, where $\epsilon=v^2/2+\Phi({\bf r})$ is the individual energy of the constituents (stars) by unit of mass. This is a particular case of the Jeans theorem \cite{jeansT,bt}. Introducing the density
\begin{eqnarray}
\rho=\int f\, d{\bf v}
\label{g59}
\end{eqnarray} 
and the pressure
\begin{eqnarray}
P=\frac{1}{d}\int f v^2\, d{\bf v},
\label{g60}
\end{eqnarray} 
and assuming that $f=f(\epsilon)$, we find that $\rho=\rho(\Phi)$ and $P=P(\Phi)$. Eliminating formally the gravitational potential between these two relations we obtain $P=P(\rho)$. Therefore, the equation of state associated with a spherically symmetric and isotropic stellar system is the same as for a barotropic star.  Furthermore, it is easy to show that the condition $f=f(\epsilon)$ together with the definitions (\ref{g59}) and (\ref{g60}) of $\rho$ and $P$ implies the condition of hydrostatic equilibrium from Eq. (\ref{g4}). Indeed
\begin{eqnarray}
\nabla P=\frac{1}{d}\int \frac{\partial f}{\partial {\bf r}} v^2\, d{\bf v}=\frac{1}{d}\nabla \Phi \int f'(\epsilon) v^2\, d{\bf v}\nonumber\\
=\frac{1}{d}\nabla \Phi \int  \frac{\partial f}{\partial {\bf v}} \cdot {\bf v}\, d{\bf v}=-\nabla\Phi\int f\, d{\bf v}=-\rho\nabla\Phi,
\label{g60he}
\end{eqnarray} 
where we have used an integration by parts to obtain the last equalities. One can then obtain the Eddington \cite{edd1916,bt} formula (see Appendix F of \cite{nfp} for its expression in $d$ dimensions) which allows us to determine the DF $f(\epsilon)$ for a prescribed barotropic equation of state $P(\rho)$ or density-potential relation $\rho(\Phi)$.

The Vlasov-Poisson equations conserve the total energy 
\begin{equation}
E=\frac{1}{2}\int f {\bf v}^2\, d{\bf r}d{\bf v}+\frac{1}{2}\int\rho\Phi\, d{\bf r}
\label{g61}
\end{equation}
and an infinite number of  Casimir functionals
\begin{equation}
I_h=\int h(f)\, d{\bf r}d{\bf v},
\label{g62}
\end{equation}
where $h(f)$ is any regular function of $f$. This includes the total mass $M=\int f\, d{\bf r}d{\bf v}$ and the generalized entropy\footnote{For commodity, we call it a generalized ``entropy'' in a loose sense. This functional could also be called  an effective entropy, a pseudo entropy, a generalized $H$-functional etc. We stress that this functional is not directly connected to thermodynamics in the usual sense since we are considering here a dynamical problem.} 
\begin{equation}
S=-\int C(f) \, d{\bf r}d{\bf v},
\label{g63}
\end{equation}
where $C(f)$ is any convex function (i.e. $C''>0$).

Using general arguments based on these conservation laws, one can show that an extremum of $S$ at fixed mass and energy is a steady state of the Vlasov equation. Writing the first variations as $\delta S-\beta\delta E-\alpha\delta M=0$, where $\beta$ and $\alpha$ are Lagrange multipliers, we get
\begin{equation}
C'(f)=-\beta\epsilon-\alpha.
\label{g64}
\end{equation}
Since $C$ is convex, this relation can be inverted to yield
\begin{equation}
 f=F(\beta\epsilon+\alpha)\quad {\rm  with}\quad  F(x)=(C')^{-1}(-x). 
 \label{g64b}
\end{equation}
Here $\beta=1/T$ is similar to an inverse temperature\footnote{Note that the
uniform temperature $T=1/\beta$ is different from the kinetic
temperature $T({\bf r})$ defined in footnote 13, except for the Boltzmann
distribution.} and $-\alpha/\beta$ is similar to a chemical potential.
Differentiating Eq. (\ref{g64}) with respect to $\epsilon$ we obtain 
\begin{equation}
f'(\epsilon)=-\frac{\beta}{C''(f)}.
\label{g64ca}
\end{equation}
Since $\beta>0$ in the usual case we conclude that $f'(\epsilon)<0$ so that the DF is a monotonically decreasing function of the energy. Therefore, an extremum of $S$ at fixed mass and energy determines a spherically symmetric and isotropic steady state of the Vlasov-Poisson equations of the form 
\begin{equation}
f=f(\epsilon),\qquad f'(\epsilon)<0.
\label{g65}
\end{equation}
Conversely, any spherically symmetric and isotropic steady state of the Vlasov-Poisson equations which is a monotonically decreasing function of the energy extremizes a particular generalized entropy $S$ of the form of Eq. (\ref{g63}) at fixed mass $M$ and energy $E$. If we write the DF as  $f=F(\beta\epsilon+\alpha)$ the generalized entropy is given by
\begin{equation}
C(f)=-\int F^{-1}(f)\, df.
 \label{g64c}
\end{equation}
The dynamical stability of these solutions with respect to the Vlasov-Poisson equations is discussed in Secs. \ref{sec_ssp}-\ref{sec_nafl} below.\footnote{For stellar systems, the total energy is $E=E_{\rm kin}+W$. Using the equilibrium virial theorem from Eqs. (\ref{vir1}) and (\ref{vir2}), we get $E=(4-d)W/2$ ($d\neq 2$) and $E=GM^2/4+W$ ($d=2$). From these results, one can show that stellar systems are always unstable in $d\ge 4$ \cite{wdsD}.}

{\it Remark:} According to Eq. (\ref{g64b}) we have
\begin{equation}
 f({\bf r},{\bf v})=F\left \lbrack\beta\frac{v^2}{2}+\beta\Phi({\bf r})+\alpha\right \rbrack.
 \label{pagny1}
\end{equation}
Substituting Eq. (\ref{pagny1}) into Eqs. (\ref{g59}) and (\ref{g60}), and making the change of variables ${\bf x}=\sqrt{\beta/2}\, {\bf v}$, we obtain 
\begin{eqnarray}
\rho=\left (\frac{2}{\beta}\right )^{d/2}\int F(x^2+\beta\Phi+\alpha)\, d{\bf x},
\label{pagny2}
\end{eqnarray} 
\begin{eqnarray}
P=\frac{1}{d}\left (\frac{2}{\beta}\right )^{1+d/2}\int F(x^2+\beta\Phi+\alpha) x^2\, d{\bf x}.
\label{pagny3}
\end{eqnarray} 
Eliminating formally $\beta\Phi({\bf r})+\alpha$ between these equations, we
find that the equation of state is of the form
\begin{eqnarray}
P(\rho,T)=T^{1+d/2} {\cal P}\left (\frac{\rho}{T^{d/2}}\right ),
\label{pagny4}
\end{eqnarray} 
where the dependence on the temperature $T$ has been made explicit (see also
Sec. 12.3 in \cite{GR1}). Using Eq. (\ref{pagny4}), we find that the internal
energy from Eq. (\ref{g6}) can be written as
\begin{eqnarray}
U=T\int\rho \int^{\frac{\rho}{T^{d/2}}}\frac{{\cal P}(\nu)}{\nu^2}\, d\nu
d{\bf r}.
\label{pagny5}
\end{eqnarray}

\subsection{Stellar polytropes}
\label{sec_spol}

We consider the Tsallis entropy\footnote{In the present context, it is an
``effective'' entropy because it is related to dynamics, 
not thermodynamics (see footnote 26).}
\begin{equation}
S=-\frac{1}{q-1}\int (f^q-f)\, d{\bf r}d{\bf v}.
\label{g66}
\end{equation}
Its extremization at fixed mass and energy leads to the polytropic DF
\begin{equation}
f=\left \lbrack \frac{1}{q}-\frac{q-1}{q}(\beta\epsilon+\alpha)\right\rbrack_+^{1/(q-1)}.
\label{g67}
\end{equation}
The index $n$ of the polytrope (see below) is related to the Tsallis parameter $q$ by
the relation
\begin{equation}
\frac{1}{q-1}=n-\frac{d}{2}.
\label{g68}
\end{equation}
The isothermal (Maxwell-Boltzmann) distribution $f=e^{-\beta\epsilon-\alpha}$ is
recovered in the limit $q\rightarrow 1$ (i.e. $n\rightarrow +\infty$) where the
Tsallis  entropy reduces to the Boltzmann  entropy $S=-\int f\ln f\, d{\bf
r}d{\bf v}$.  In the following, we shall consider $q>0$ so that
$C(f)=(f^q-f)/(q-1)$ is convex. We also assume $\beta>0$. We have to distinguish
two cases:

(i) When $q>1$ (i.e. $n\ge d/2$), we can rewrite the polytropic DF (\ref{g67}) as
\begin{equation}
f=A_*(\epsilon_{m}-\epsilon)_+^{n-d/2} 
\label{g69}
\end{equation}
with
\begin{equation}
A_*=\left \lbrack \frac{(q-1)\beta}{q}\right\rbrack^{1/(q-1)},\qquad
\epsilon_{m}=\frac{1-(q-1)\alpha}{(q-1)\beta}.
\label{g70}
\end{equation}
This DF has a compact support since $f$ is defined only for
$\epsilon\le\epsilon_m$ (it vanishes at $\epsilon=\epsilon_m$). For $\epsilon\ge
\epsilon_m$, we set $f=0$.\footnote{For $q\rightarrow +\infty$ (i.e.
$n=n_{3/2}=d/2$), $f$ is the Heaviside function. It corresponds to the Fermi
distribution (i.e. the Fermi-Dirac distribution at $T=0$) that arises, for
example, in the theory of white 
dwarf stars \cite{chandrabook}, neutron stars \cite{ov}, and fermionic DM halos
\cite{mg16F,modeldmF}. For $n=n_5=(d+2)/(d-2)$ when $d\le (4+\sqrt{32})/2$, the DF $f(\epsilon)$ is usually called the Plummer distribution function although it was actually introduced by
Eddington (see the introduction). For $q=2$ (i.e. $n=n_{5/2}=(d+2)/2$), $S$ is minus the enstrophy $M_2=\int f^2\, d{\bf r}d{\bf v}$ and $f=A_*(\epsilon_m-\epsilon)_+$ is linear in $\epsilon$. The maximum entropy state is equivalent to the minimum enstrophy state.} The density and the pressure are always
well-defined. Therefore, case (i) corresponds to
\begin{equation}
q>1,\qquad n\ge d/2. 
\label{g71}
\end{equation}
Substituting the DF from  Eq. (\ref{g69}) into Eqs. (\ref{g59}) and (\ref{g60}), we find that
\begin{equation}
\rho=A_* S_d (\epsilon_m-\Phi)^n
2^{d/2-1}\frac{\Gamma(d/2)\Gamma(1-d/2+n)}{\Gamma(1+n)},
\label{g72}
\end{equation}
\begin{equation}
P=\frac{A_*S_d}{n+1}(\epsilon_m-\Phi)^{n+1}
2^{d/2-1}\frac{\Gamma(d/2)\Gamma(1-d/2+n)}{\Gamma(1+n)}.
\label{g73}
\end{equation}
Eliminating the potential $\Phi$ between these two expressions, we obtain the polytropic equation of state (\ref{g8}) with the polytropic constant
\begin{equation}
K=\frac{1}{n+1}\left\lbrack A_*S_d
2^{d/2-1}\frac{\Gamma(d/2)\Gamma(1-d/2+n)}{\Gamma(1+n)}\right\rbrack^{-1/n}.
\label{g74}
\end{equation}
We note that $K$ depends on $\beta$ through Eqs. (\ref{g70}) and (\ref{g74}) so that it can really be interpreted as a generalized (polytropic) temperature. More precisely, $K\propto T^{1-\frac{d}{2}(\gamma-1)}$, implying $P=K\rho^{\gamma}\propto T^{1+d/2}(\rho/T^{d/2})^{\gamma}$, in agreement with Eq. (\ref{pagny4}).

(ii) When $0<q<1$ (i.e. $n<d/2-1$), we can rewrite the polytropic DF (\ref{g67}) as
\begin{equation}
f=A_*(\epsilon_{m}+\epsilon)^{n-d/2} 
\label{g75}
\end{equation}
with
\begin{equation}
A_*=\left \lbrack -\frac{(q-1)\beta}{q}\right\rbrack^{1/(q-1)},\qquad
\epsilon_{m}=-\frac{1-(q-1)\alpha}{(q-1)\beta}.
\label{g76}
\end{equation}
This DF is defined for all energies. It behaves as $f\sim \epsilon^{-(d/2-n)}$ for $\epsilon\rightarrow +\infty$, hence as $f\sim v^{-(d-2n)}$ for $v\rightarrow +\infty$ (at fixed $r$). The density is defined for $n<0$ (i.e. $(d-2)/d<q<1$) and the pressure is defined for $n<-1$ (i.e. $d/(d+2)<q<1$). Therefore, case (ii) corresponds to
\begin{equation}
\frac{d}{d+2}<q<1,\qquad n<-1. 
\label{g77}
\end{equation}
Substituting the DF from Eq. (\ref{g75})  into Eqs. (\ref{g59}) and (\ref{g60}), we find that
\begin{equation}
\rho=A_* S_d (\epsilon_m+\Phi)^n
2^{d/2-1}\frac{\Gamma(d/2)\Gamma(-n)}{\Gamma(d/2-n)},
\label{g78}
\end{equation}
\begin{equation}
P=-\frac{A_*S_d}{n+1}(\epsilon_m+\Phi)^{n+1}
2^{d/2-1}\frac{\Gamma(d/2)\Gamma(-n)}{\Gamma(d/2-n)}.
\label{g79}
\end{equation}
Eliminating the potential $\Phi$ between these two expressions, we obtain the polytropic equation of state (\ref{g8}) with the polytropic constant
\begin{equation}
K=-\frac{1}{n+1}\left\lbrack A_*S_d
2^{d/2-1}\frac{\Gamma(d/2)\Gamma(-n)}{\Gamma(d/2-n)}\right\rbrack^{-1/n}.
\label{g80}
\end{equation}
We note that $K$ depends on $\beta$ through Eqs. (\ref{g76}) and (\ref{g80})  so that it can really be interpreted as a generalized (polytropic) temperature. More precisely, $K\propto T^{1-\frac{d}{2}(\gamma-1)}$, implying $P=K\rho^{\gamma}\propto T^{1+d/2}(\rho/T^{d/2})^{\gamma}$, in agreement with Eq. (\ref{pagny4}).

In conclusion, the barotropic star associated with a stellar polytrope is the polytropic star.\footnote{In particular, we can check that Eqs. (\ref{g72}) and (\ref{g78}) are equivalent to Eq. (\ref{g11}) if we use Eqs. (\ref{g74}) and (\ref{g80}) and set $\epsilon_m=Kn+\mu$ in case (i) and $\epsilon_m=-Kn-\mu$ in case (ii), leading to Eqs. (\ref{g11a}) and (\ref{g11b}).} We can therefore use the results established in Sec. \ref{sec_baro} to study the structure of stellar polytropes. We note that the DF $f(\epsilon)$ and the free energy $F[f]=E[f]-TS[f]$ (see below) of stellar polytropes are similar to the density $\rho(\Phi)$ and the energy $E_{\rm tot}[\rho]$ of  polytropic stars. This is the only case (including isothermal systems with $n\rightarrow +\infty$) where there is such a similarity.

\subsection{Stellar logotropes}
\label{sec_slog}

Let us specifically consider the index $n=-1$ (i.e. $q=d/(d+2)$). The corresponding effective entropy reads
\begin{equation}
S=\frac{d+2}{d}\int \left \lbrack f^{d/(d+2)}-f\right \rbrack\, d{\bf r}d{\bf v}.
\label{g81}
\end{equation}
Its extremization at fixed mass and energy leads to the logotropic DF 
\begin{equation}
f=\frac{A_*}{(\epsilon_{m}+\epsilon)^{d/2+1}} 
\label{g82}
\end{equation}
with
\begin{equation}
A_*=\left (\frac{d}{2\beta}\right )^{(d+2)/2},\qquad
\epsilon_{m}=\frac{d+2+2\alpha}{2\beta}.
\label{g83}
\end{equation}
This DF is defined for all energies. It behaves as $f\sim \epsilon^{-(d/2+1)}$ for $\epsilon\rightarrow +\infty$, hence as $f\sim v^{-(d+2)}$ for $v\rightarrow +\infty$ (at fixed $r$). The density is well-defined but the pressure (hence the density of kinetic energy) diverges logarithmically. Substituting the DF from Eq.  (\ref{g82}) into Eq. (\ref{g59}) we find that
\begin{equation}
\rho=\frac{A_* S_d}{\epsilon_m+\Phi}
2^{d/2-1}\frac{\Gamma(d/2)\Gamma(1)}{\Gamma(d/2+1)}.
\label{g84}
\end{equation}
We can also directly obtain this result by taking $n=-1$ in Eq. (\ref{g78}).
Comparing Eq. (\ref{g84}) with Eq. (\ref{g49}) we can make the
identifications
\begin{equation}
A= A_*S_d 2^{d/2-1}\frac{\Gamma(d/2)\Gamma(1)}{\Gamma(d/2+1)},\qquad
\epsilon_{m}=-\mu.
\label{g85}
\end{equation}
To evaluate the pressure, we consider the limit $n\rightarrow -1^{-}$ in Eq. (\ref{g79}). Writing $n=-1-\nu$, we get
\begin{equation}
P=\frac{A_*S_d}{\nu}(\epsilon_m+\Phi)^{-\nu}
2^{d/2-1}\frac{\Gamma(d/2)\Gamma(1)}{\Gamma(d/2+1)}.
\label{g86}
\end{equation}
The pressure diverges when $\nu=0$. Making the expansion
\begin{eqnarray}
\frac{1}{\nu}(\epsilon_m+\Phi)^{-\nu}=\frac{1}{\nu}e^{-\nu\ln(\epsilon_m+\Phi)}\nonumber\\
=\frac{1}{\nu}(1-\nu\ln(\epsilon_m+\Phi)+...)=\frac{1}{\nu}-\ln(\epsilon_m+\Phi)+...\quad
\label{g87}
\end{eqnarray}
for $\nu\rightarrow 0^+$ and discarding ``by hand''  the infinite constant term,\footnote{As before, this is justified because only the gradient of the pressure matters in nonrelativistic mechanics.  This procedure may also be related to a form of renormalization (see the Remark at the end of this section).} there remains
\begin{equation}
P=-A_*S_d
2^{d/2-1}\frac{\Gamma(d/2)\Gamma(1)}{\Gamma(d/2+1)}\ln(\epsilon_m+\Phi)+...
\label{g88}
\end{equation}
Eliminating the potential $\Phi$ between Eqs. (\ref{g84}) and (\ref{g88}), we obtain 
\begin{equation}
P=A_*S_d 2^{d/2-1}\frac{\Gamma(d/2)\Gamma(1)}{\Gamma(d/2+1)}\ln\rho+...
\label{g89}
\end{equation}
This returns the logotropic equation of state (\ref{g41}) with the logotropic
constant $A$ given by Eq. (\ref{g85}).
Using Eq. (\ref{g80}) and recalling the calculation after Eq. (\ref{g44}), we can check that the logotropic equation of state is consistent with the polytropic equation of state with $\gamma\rightarrow 0$, $K\rightarrow +\infty$ and 
\begin{equation}
K\gamma\rightarrow A_*S_d
2^{d/2-1}\frac{\Gamma(d/2)\Gamma(1)}{\Gamma(d/2+1)}\equiv A.
\label{g90}
\end{equation}
We note that $A$ depends on $\beta$ through Eqs. (\ref{g83}) and (\ref{g85}) 
so that it can really be interpreted as a generalized (logotropic) temperature.
This gives more strength to our thermodynamical interpretation of $A$ in Sec.
\ref{sec_logog}. More precisely, $A\propto T^{1+d/2}$, implying
$P=A\ln\rho\propto T^{1+d/2}\ln(\rho/T^{d/2})$, in agreement with Eq.
(\ref{pagny4}).

In conclusion, the barotropic star associated with a stellar logotrope is the logotropic star. We can therefore use the results established in Sec. \ref{sec_logog} to study the structure of stellar logotropes. We note that the DF $f(\epsilon)$ and the free energy $F[f]$ of stellar logotropes are different from the density $\rho(\Phi)$ and the energy $E_{\rm tot}[\rho]$ of  logotropic stars, even though logotropes are related to polytropes of index $n=-1$.

{\it Remark:} Although the pressure defined by Eq. (\ref{g60}) is positive, our procedure yields at the end a pressure that may be negative.\footnote{The pressure of logotropic spheres is effectively negative at sufficiently large distances (see Sec. \ref{sec_logog}).} This is because we have subtracted an infinite constant. This is reminiscent of the procedure of renormalization which leads to the equation of state of the vacuum $P=-\epsilon$ with a negative pressure starting from integrals for $\epsilon$ and $P$ which are positive but diverging (see, e.g., \cite{zeldovichA,martin,ouf}). This is another argument that the logotropic equation of state may be related to dark energy (DE) (see \cite{epjp,lettre,jcap,preouf1,action,logosf,preouf2,ouf,gbi} and Secs. \ref{sec_logodm} and \ref{sec_value}).

\subsection{Dynamical stability of stellar polytropes}
\label{sec_ssp}

In this section, we summarize the stability results obtained for collisionless stellar systems (see
\cite{aaISO,aaPOLY,sc2002,grand,anomalous,csTsallis,aastab,cslogo,wdsD,assise,nyquistHMF,cc,GR1,kinVR} and references therein for a detailed discussion of these stability results). 

A stellar system is (nonlinearly) dynamically stable with respect to the Vlasov-Poisson equations (\ref{g57}) and (\ref{g58}) if, and only if, it is a minimum of energy $E[f]$ with respect to symplectic perturbations (i.e. perturbations that conserve all the Casimirs) \cite{ko,bartholomew,dfb,ih,kandrup91}. This energy principle (infinite-constraint problem) is the most refined condition of dynamical stability \cite{cc}.  It is similar to the Kelvin–Arnold energy principle \cite{kelvin,arnold} for two-dimensional (2D) inviscid incompressible hydrodynamical flows governed by the 2D Euler–Poisson equations \cite{varp}. For small perturbations (linear stability), we require that $\delta^2E>0$ for all perturbations that conserve all the Casimirs at first order. This is equivalent to linearizing the Vlasov-Poisson equations about an equilibrium state and requiring that the squared pulsation is positive ($\omega^2>0$) for such perturbations (spectral stability) \cite{antolin,cc}.

It can be shown that all the DFs of the form (\ref{g65}) are minima of energy  with respect to symplectic perturbations  so they are dynamically stable with respect to the Vlasov-Poisson equations \cite{doremus71,doremus73,gillon76,sflp,ks,kandrup91,lmr}. In particular, all the stellar polytropes are dynamically stable. This result is valid in Newtonian gravity but it is no more true in general relativity (see \cite{ipserGR} and the discussion in Sec. 5.2 of \cite{GR1}).

\subsection{Thermodynamical analogy}
\label{sec_than}

As a corollary of the energy principle, a spherically symmetric and isotropic stellar system is dynamically stable if it is a minimum of energy $E$ at fixed generalized entropy $S$ (for a specified convex function $C$) and mass $M$.  Equivalently,  a spherically symmetric and isotropic stellar system is dynamically stable if it is a maximum of a generalized entropy $S$ (for a specified convex function $C$) at fixed mass $M$ and energy $E$ \cite{cc}.  This is similar to a condition of generalized thermodynamical stability in the microcanonical ensemble (two-constraint problem) \cite{aastab}. This microcanonical stability criterion is less refined than the energy principle since it takes into account only two constraints instead of an infinity. As a result, this two-constraint problem just provides a sufficient condition of dynamical stability \cite{ih,ipser}.

We can define a generalized free energy by the Legendre transform $F[f]=E[f]-TS[f]$ with $T=1/\beta$. Explicitly,
\begin{equation}
F=\frac{1}{2}\int f {\bf v}^2\, d{\bf r}d{\bf v}+\frac{1}{2}\int\rho\Phi\, d{\bf r}+T\int C(f) \, d{\bf r}d{\bf v}.
\label{g91}
\end{equation}
As a corollary of the microcanonical criterion, a spherically symmetric and isotropic stellar system is dynamically stable if it is a minimum of free energy $F$ at fixed mass $M$.\footnote{If a DF minimizes $F$ at fixed mass, then it automatically maximizes $S$ at fixed mass and energy \cite{cc}.} This is similar to a condition of generalized thermodynamical stability in the canonical ensemble (one-constraint problem) \cite{aastab}.  This canonical stability criterion is less refined than the microcanonical criterion (and thus less refined than the energy principle) since it takes into account only one constraint. As a result, this one-constraint problem just provides a sufficient condition of dynamical stability. It corresponds to the energy-Casimir method \cite{bernstein,rosenbluthstab,fowlerstab,gardner,rowlands,sanitt} (see the reviews in \cite{holm,morrison}). 
The energy-Casimir method was  used by Arnold \cite{arnoldEC} in the context of 2D incompressible flows \cite{varp}.

According to the foregoing results,  we can therefore use a thermodynamical analogy to investigate the dynamical stability of a spherically symmetric and isotropic steady state of the Vlasov-Poisson equations \cite{aastab}. In particular, we can use the Poincar\'e turning point argument \cite{poincare,katztopo,ijmpb} to determine the stability of the system from the topology of the generalized caloric curve (see a summary of the Poincar\'e theory of linear series of equilibria  in Appendix C of \cite{acf}). The fact that the canonical criterion is less refined than the microcanonical criterion is  similar to the situation of ensemble inequivalence in the statistical mechanics of systems with long-range interactions \cite{paddy,found,ijmpb,campabook}.

\subsection{Nonlinear Antonov first law}
\label{sec_nafl}

There is an interesting application of these results. It can be shown \cite{aastab} that the minimization of free energy $F[f]$ at fixed mass $M[f]$ for a stellar system is equivalent to the minimization of energy $E_{\rm tot}[\rho]$ at fixed mass $M[\rho]$ for the corresponding barotropic gas determined by Eqs. (\ref{g59}) and (\ref{g60}).\footnote{We can show in full generality  that the generalized free energy $F[\rho]$ of the stellar system expressed in terms of the density $\rho$ coincides with the energy $E_{\rm tot}[\rho]$ of the barotropic gas (see Appendix B of \cite{aastab}).} Therefore, a spherically symmetric and isotropic stellar system with $f'(\epsilon)<0$ is dynamically stable with respect to the Vlasov-Poisson equations (\ref{g57}) and (\ref{g58}) if the corresponding barotropic gas is dynamically stable with respect to the Euler-Poisson equations (\ref{g1})-(\ref{g3}) \cite{aastab}. This provides an original derivation \cite{aastab} of the Antonov \cite{antolaw} first law.\footnote{Furthermore, this justification is valid for the nonlinear stability problem. Therefore, it leads to the nonlinear Antonov first law \cite{aastab}.} We stress that this is just a sufficient condition of stability for a collisionless stellar system. More refined stability conditions can be obtained as discussed above.\footnote{We may wonder about the interest of these sufficient conditions of dynamical stability since we know that all spherically symmetric and isotropic stellar systems with $f'(\epsilon)<0$ are dynamically stable \cite{doremus71,doremus73,gillon76,sflp,ks,kandrup91,lmr}. The point is that these sufficient conditions of dynamical stability remain valid in general relativity and for other systems with long-range interactions with a nongravitational potential, for which there is no general result about their dynamical stability such as the one obtained by  \cite{doremus71,doremus73,gillon76,sflp,ks,kandrup91,lmr} in Newtonian gravity.}

In $d=3$ dimensions,\footnote{See \cite{wdsD} for results valid in other dimensions of space.} we know (see Sec. \ref{sec_stabg}) that complete polytropic stars are stable for $3/2\le n<3$ (they are marginally stable for $n=3$). We conclude from the canonical criterion and the Antonov first law that complete stellar polytropes are stable for $3/2\le n<3$. In addition, it can be shown from the microcanonical criterion that complete stellar polytropes are stable for $3/2\le n<5$ (they are marginally stable for $n=5$).\footnote{This result can be established by a direct calculation \cite{antolaw} or from the topology of their generalized caloric curve without making any calculation \cite{aastab}.} Therefore, we do not need to use the most refined criterion (energy principle) to prove the dynamical stability of all the complete stellar polytropes. The microcanonical criterion is sufficient in that case.

For box-confined gaseous polytropes, we know (see Sec. \ref{sec_stabg}) that the series of equilibria (generalized caloric curve) is stable before the first turning point of generalized temperature. We conclude from the canonical criterion and the Antonov first law that box-confined  stellar polytropes are stable before the first turning point of generalized temperature. In addition, it can be shown from the microcanonical criterion that box-confined  stellar polytropes are stable before the first turning point of energy \cite{anomalous,aastab}. We note that the region of negative specific heat between the first turning point of temperature and the first turning point of energy can be shown to be dynamically stable with the microcanonical criterion while the canonical criterion does not allow us to conclude since it only yields a sufficient condition of dynamical stability. This is similar to a situation of ensemble inequivalence in statistical mechanics for systems with long-range interactions \cite{paddy,found,ijmpb,campabook}. In particular, the fact that complete stellar polytropes with index   $3\le n\le 5$ are stable with respect to the Vlasov-Poisson equations (through the microcanonical ensemble) while  complete gaseous polytropes with index   $3\le n\le 5$ are unstable with respect to the Euler-Poisson equations (through the canonical ensemble) can be related to a situation of ensemble inequivalence \cite{aastab}. 

Similar results concerning the stability of gaseous and stellar polytropes (Tsallis distributions) have been obtained in parallel to our works \cite{aaPOLY,grand,anomalous,aastab} by Taruya and Sakagami \cite{ts1,ts2,ts3}  but they connected the stability of these systems to a true notion of thermodynamics in Tsallis generalized sense instead of a notion of effective thermodynamics used to investigate their dynamical stability with respect to the Euler-Poisson  and Vlasov-Poisson equations as in \cite{aaPOLY,grand,anomalous,aastab}. It is possible that the generalized thermodynamical interpretation is relevant in other contexts as discussed in \cite{csTsallis,assise,nfp,kaniadakis,kingen,frank,gen,ggpp,tsprl,ccgm} and in Sec. IV of \cite{anomalous}.\footnote{It should be emphasized that Sec. II of \cite{anomalous} is concerned with the dynamical stability of the Euler-Poisson and Vlasov-Poisson equations while Sec. IV of \cite{anomalous} is concerned with the generalized thermodynamics of nonlinear Fokker-Planck equations (Smoluchowski-Poisson equations).} In the collisional case, the microcanonical criterion provides a necessary and sufficient condition of thermodynamical stability. Therefore, box-confined polytropes with index $n>5$ become unstable after the first turning point of energy (corresponding to a minimum energy) and undergo a form of gravothermal catastrophe \cite{anomalous} like globular clusters and isothermal stellar systems \cite{antonov,lbw,sc2002}.

{\it Remark:} The stability of stellar logotropes is a bit subtle because these configurations have an infinite mass and must be artificially confined within a box or surrounded by an envelope. This is the same problem as for isothermal galaxies and unbounded stellar polytropes.  According to the microcanonical criterion, these systems are dynamically stable with respect to the Vlasov-Poisson equations below a critical central density corresponding to the first turning point of energy in the series of equilibria. They may also be always stable (for arbitrary central density) if the result of \cite{doremus71,doremus73,gillon76,sflp,ks,kandrup91,lmr} applies to box-confined galaxies. We refer to \cite{aaISO,aaPOLY,sc2002,grand,anomalous,csTsallis,aastab,cslogo}
 for a detailed study of box-confined isothermal, polytropic and logotropic stellar configurations in various dimensions of space.

\section{Derivation of the polytropic DF from the Lynden-Bell theory}
\label{sec_ew}

In this section, we first summarize the statistical theory of Lynden-Bell \cite{lb} on the violent relaxation of collisionless stellar systems (see our papers \cite{csr,grand,superstat,assise,kinVR} for a detailed exposition of this theory with the notations used in the present paper and some complements). We consider the general (multilevel) case. Then, we show how one can derive the polytropic (Tsallis) DF as a particular case of the Lynden-Bell theory. In this sense, we ``justify'' the Tsallis entropy from the Lynden-Bell entropy. We stress, however, their fundamental differences.   Finally, following Ewart {\it et al.} \cite{ewart1,ewart2}, we show that the Lynden-Bell theory selects a ``universal'' polytropic distribution of index  $n=-1$ that turns out to correspond to stellar logotropes.

 \subsection{General case}
\label{sec_gc}

We consider a collisionless stellar system described by the Vlasov-Poisson equations (\ref{g57}) and (\ref{g58}). We decompose the initial DF $f_{0}({\bf r},{\bf v})$ into a continuum of phase levels $\lbrace\eta\rbrace$ (in practice, the continuum is replaced by a set of ${\cal N}$ discrete levels $\eta_i$). After a complicated mixing process,  the system reaches a quasistationary state (QSS) on the coarse-grained scale. This corresponds to a phase of violent collisionless relaxation. According to the Lynden-Bell theory, the local distribution of phase levels $\rho({\bf r},{\bf v},\eta)$ at statistical equilibrium is obtained by maximizing a mixing entropy 
\begin{equation}
S[\rho]=-\int \rho\ln\rho\, d\eta d{\bf r}d{\bf v},
\label{mix}
\end{equation}
while conserving the energy [see Eq. (\ref{g61})] and all the Casimirs [see Eq. (\ref{g62})].\footnote{The Lynden-Bell statistical theory relies on an assumption of ergodicity (efficient mixing), which is not always fulfilled in practice. This is the complicated problem of incomplete relaxation \cite{lb,incomplete}.} The conservation of the Casimirs  is equivalent to the conservation of the phase space hypervolumes $\gamma(\eta)$ occupied  by the levels $\eta$ [see Eq. (\ref{a1})]. Indeed, the Casimirs can be expressed as $I_h=\int h(\eta)\gamma(\eta)\, d\eta$. This imposes the constraint
\begin{equation}
\gamma(\eta)=\int \rho({\bf r},{\bf v},\eta)\, d{\bf r}d{\bf v}.
\label{g106}
\end{equation}
The extremization problem leads to the Gibbs state (see \cite{lb,csr,grand,superstat,assise,kinVR}  for details) 
\begin{equation}
\rho({\bf r},{\bf v},\eta)=\frac{1}{Z(\epsilon)}g(\eta)e^{-\beta \eta\epsilon},
\label{g92}
\end{equation}
where $\beta$ and $g(\eta)$ are the Lagrange multipliers associated with the conservation of energy $E$ and phase space hypervolumes $\gamma(\eta)$.\footnote{We could redefine $g(\eta)$ by writing $g(\eta)=\chi(\eta)e^{-\alpha\eta}$ so as to display a term $\beta\epsilon+\alpha$ in the exponential. Such a decomposition makes a distinction between the Lagrange multipliers $\alpha$ and $\beta$ associated with the ``robust'' constraints $M$ and $E$ and the Lagrange multipliers $\chi(\eta)$ associated with the ``fragile'' constraints $M_{n>1}^{\rm f.g.}=\int \overline{f^n}\, d{\bf r}d{\bf v}$ (see \cite{csr,grand,superstat,assise,kinVR} for details). Here, we leave the term  $e^{-\alpha\eta}$ implicit in $g(\eta)$. If necessary, we can always replace $g(\eta)$ by $\chi(\eta)$ and $\beta\epsilon$ by
$\beta\epsilon+\alpha$.} The ``partition function'' $Z$ is obtained from the normalization condition $\int\rho\, d\eta=1$. Actually, as explained in \cite{superstat}, we have to treat the level $\eta=0$ (vacuum) specifically. Therefore, the partition function must be decomposed into the form  $Z=g(0)+\int_{0^+}g(\eta)e^{-\beta\eta\epsilon}\, d\eta$. We can take $g(0)=1$ without restriction of generality. Therefore, we shall write the partition function as
\begin{equation}
Z(\epsilon)=1+\int_{0^+}^{+\infty} g(\eta)e^{-\beta\eta\epsilon}\, d\eta.
\label{g93}
\end{equation}
The coarse-grained DF is given by
\begin{equation}
\overline{f}({\bf r},{\bf v})=\int\rho({\bf r},{\bf v},\eta)\eta\, d\eta.
\label{g94}
\end{equation}
Substituting the Gibbs state from Eq. (\ref{g92}) into Eq. (\ref{g94}) we obtain
\begin{equation}
\overline{f}({\bf r},{\bf v})=\frac{1}{Z(\epsilon)}\int_0^{+\infty} g(\eta)\eta e^{-\beta \eta\epsilon}\, d\eta.
\label{g95}
\end{equation}
This relation determines the coarse-grained DF at statistical equilibrium in the framework of Lynden-Bell's theory. It can be interpreted as a generalized form of superstatistics \cite{superstat}. Eq. (\ref{g95}) can be rewritten as 
\begin{equation}
\overline{f}({\bf r},{\bf v})=-\frac{\partial\ln Z}{\partial (\beta \epsilon)}.
\label{g96}
\end{equation}
We note that the Lynden-Bell coarse-grained DF depends only on the energy:
\begin{equation}
\overline{f}({\bf r},{\bf v})=F(\beta \epsilon)=\overline{f}(\epsilon)\quad {\rm with}\quad F(x)=-(\ln Z)'(x).
\label{g96b}
\end{equation}
Therefore, the maximum entropy principle selects a spherically symmetric and isotropic steady state of the Vlasov-Poisson equations. Furthermore, we show in Appendix \ref{sec_fdt} that  \cite{grand,kingen,superstat,assise,kinVR}
\begin{equation}
\overline{f}'(\epsilon)=-\beta f_2,
\label{g96c}
\end{equation}
where
\begin{equation}
f_2=\overline{f^2}-\overline{f}^2>0
\label{g96cprime}
\end{equation}
is the local centered variance of the distribution. Therefore, in the usual case where $\beta>0$, the coarse-grained DF predicted by Lynden-Bell's theory is a monotonically decreasing function of the energy: $\overline{f}=\overline{f}(\epsilon)$ with $\overline{f}'(\epsilon)<0$.

In principle, the function $g(\eta)$ is determined {\it a posteriori} by the initial condition. This is a Lagrange multiplier that has to be related to the constraints of the dynamics $E$ and $\gamma(\eta)$. However, in certain cases, it can be specified {\it a priori}. A given function  $g(\eta)$  may correspond  to a specific type of initial conditions belonging to the same ``class of equivalence'' or it may result from the balance between forcing and dissipation (see \cite{gen,superstat,kinVR} for a detailed discussion). If we assume that $g(\eta)$ is given by the $\chi$-squared distribution \cite{superstat} 
\begin{equation}
g(\eta)=\frac{1}{b\Gamma(c)}\left (\frac{\eta}{b}\right )^{c-1}e^{-\eta/b},
\label{g97}
\end{equation}
with $c>0$ and $b>0$ we find that 
\begin{equation}
Z=1+\frac{1}{(1+\beta b\epsilon)^c}.
\label{g98}
\end{equation}
Then, using  Eq. (\ref{g96}), we obtain 
\begin{equation}
\overline{f}=\frac{bc}{(1+\beta b\epsilon)\left\lbrack 1+(1+\beta b\epsilon)^c\right\rbrack}.
\label{g99}
\end{equation}
This may be seen as a generalized polytropic distribution \cite{superstat}. If we just  take
\begin{equation}
g(\eta)=\frac{1}{b\Gamma(c)}\left (\frac{\eta}{b}\right )^{c-1},
\label{bill1}
\end{equation}
we obtain 
\begin{equation}
\overline{f}=\frac{bc}{\beta b\epsilon\left\lbrack 1+(\beta b\epsilon)^c\right\rbrack}.
\label{g99b}
\end{equation}
The $\chi$-squared distribution from Eq. (\ref{g97}) was considered in \cite{superstat} together with other  examples of functions $g(\eta)$.

\subsection{Fermi-Dirac-like distribution}
\label{sec_fdl}

In the general case, the coarse-grained Lynden-Bell distribution can be written as
\begin{equation}
\overline{f}({\bf r},{\bf v})=\frac{\int_0^{+\infty} g(\eta)\eta e^{-\beta \eta\epsilon}\, d\eta}{1+\int_{0^+}^{+\infty} g(\eta) 
e^{-\beta \eta\epsilon}\, d\eta}.
\label{fdl1}
\end{equation}
It is similar to a superposition of Fermi-Dirac-like distributions. In the nondegenerate limit (see Sec. \ref{sec_ndg}), the denominator can be approximated by unity and we obtain a superposition of Maxwell-Boltzmann-like distributions. 

More specifically, in the single level (+ vacuum) case $f\in \lbrace 0,\eta_0\rbrace$, Eq. (\ref{fdl1})  reduces to the Fermi-Dirac-like distribution
\begin{equation}
\overline{f}({\bf r},{\bf v})=\frac{\eta_0}{1+e^{\eta_0(\beta\epsilon+\alpha)}}.
\label{fdl2}
\end{equation}
In the nondegenerate limit, we obtain the Maxwell-Boltzmann-like distribution
\begin{equation}
\overline{f}({\bf r},{\bf v})=\eta_0 e^{-\eta_0(\beta\epsilon+\alpha)}.
\label{fdl3}
\end{equation}

\subsection{Nondegenerate limit}
\label{sec_ndg}

In the nondegenerate limit, we can make the approximation
\begin{equation}
Z\simeq 1.
\label{g101}
\end{equation}
Therefore, the Gibbs state from Eq. (\ref{g92}) reduces to
\begin{equation}
\rho({\bf r},{\bf v},\eta)=g(\eta)e^{-\beta \eta\epsilon}.
\label{g102}
\end{equation}
The corresponding  coarse-grained DF reads
\begin{equation}
\overline{f}({\bf r},{\bf v})=\int_{0}^{+\infty} g(\eta)\eta e^{-\beta \eta\epsilon}\, d\eta.
\label{g103}
\end{equation}
Under this form, it has the standard form of a  superstatistics  \cite{superstat}.   If we define
\begin{equation}
{G}(x)=\int_{0^+}^{+\infty} g(\eta) e^{-\eta x}\, d\eta,
\label{sac1}
\end{equation}
we can rewrite Eq. (\ref{g103}) as
\begin{equation}
\overline{f}({\bf r},{\bf v})=-G'(\beta\epsilon).
\label{sac2}
\end{equation}
Therefore,
\begin{equation}
\overline{f}({\bf r},{\bf v})=F(\beta \epsilon)=\overline{f}(\epsilon)\quad {\rm with}\quad F(x)=-G'(x).
\label{sca5}
\end{equation}
We also have
\begin{equation}
\overline{f}'(\epsilon)=-\beta\overline{f^2},
\label{sac4}
\end{equation}
where
\begin{equation}
\overline{f^2}=\int_{0}^{+\infty} g(\eta)\eta^2 e^{-\beta \eta\epsilon}\, d\eta>0
\label{sac4prime}
\end{equation}
is the second moment of the distribution. As before, Eq. (\ref{sca5}) determines a spherically symmetric and isotropic DF $\overline{f}=\overline{f}(\epsilon)$ with $\overline{f}'(\epsilon)<0$, which is a decreasing function of the energy. For the $\chi$-squared distribution from Eq. (\ref{g97}) we obtain 
\begin{equation}
\overline{f}=\frac{bc}{(1+\beta b\epsilon)^{c+1}}.
\label{g104}
\end{equation}
Interestingly, this is a polytropic DF of index 
\begin{equation}
q=\frac{c}{c+1},\qquad n=\frac{d}{2}-c-1.
\label{g105}
\end{equation}
It corresponds to case (ii) of Sec. \ref{sec_spol} provided that $c>d/2$.  In conclusion, for a function $g(\eta)$ given by the $\chi$-squared distribution from  Eq. (\ref{g97}), the Lynden-Bell theory can justify the Tsallis distributions (stellar polytropes). If we just use  Eq. (\ref{bill1}) we obtain
\begin{equation}
\overline{f}=\frac{bc}{(\beta b\epsilon)^{c+1}}.
\label{bill2}
\end{equation}

{\it Remark:} We can recover these results from 
the general equations of Sec. \ref{sec_gc} as follows. In the nondegenerate
limit $Z\simeq 1$, the second term in Eq.  (\ref{g93}) is small so,  in Eq.
(\ref{g96}), we can expand $\ln Z$ to first order in this small parameter and
recover Eq. (\ref{g103}).
Equivalently, since $Z=1+G$ with $G\ll 1$, we see that Eq. (\ref{g96}) reduces
to Eq. (\ref{sac2}). In addition, Eq. (\ref{g104}) can be recovered from Eq.
(\ref{g99}) for $\beta b\epsilon \gg 1$ [see Eq. (\ref{g98})]. This shows that
the nondegenerate
limit is always valid for high energies $\epsilon\rightarrow
+\infty$. In particular, for the $\chi$-squared distribution, the DFs
from Eqs. (\ref{g99b}) and (\ref{g104}) have the same asymptotic behavior
$\overline{f}\propto \epsilon^{-(c+1)}$.

\subsection{Cauchy distribution}

When $c\rightarrow 0$, the DF from Eq. (\ref{g104}), which is valid in the nondegenerate limit,  reduces to
\begin{equation}
\overline{f}=\frac{bc}{1+\beta b\epsilon}.
\label{g100}
\end{equation}
It corresponds to a polytropic DF of index $n=d/2-1$.  It decreases with the velocity as $\overline{f}\sim v^{-2}$. The density is defined for $d<2$ while the pressure diverges in any dimension. The effective entropy associated with this DF is
\begin{equation}
S=\int \ln f\, d{\bf r}d{\bf v}.
\label{g100b}
\end{equation}
It may be related to the Tsallis entropy of parameter $q=0$ if we make the expansion $f^q=e^{q\ln f}\simeq 1+q\ln f$ in Eq. (\ref{g66}) and remove the diverging constant term.\footnote{This is justified by the fact that the entropy is defined up to an additive  constant. Indeed, a constant term has no effect on the variational principle associated with the extremization of the entropy.} This generalized entropy is called the log-entropy \cite{cslogo}. The  DF (\ref{g100}) is the Cauchy (or Lorentz) distribution. It occurs in the statistical mechanics of bosons in the limit of large occupation numbers where it is sometimes called the Rayleigh-Jeans distribution \cite{kinQ}.

{\it Remark:} The Cauchy distribution (\ref{g100}) and the corresponding log-entropy (\ref{g100b}) can also be obtained from the Lynden-Bell entropy [Eq. (\ref{mix})] with the $\chi$-squared distribution from Eq. (\ref{g97}) for any value of $c$ if we remove the $+1$ term (vacuum) in the partition function from Eq. (\ref{g93}) (see Sec. 3.4 of \cite{superstat}).

\subsection{Derivation of the Tsallis entropy $S[\overline{f}]$ from the Lynden-Bell entropy $S[\rho]$}

The Lynden-Bell entropy  (\ref{mix}) is a functional of $\rho({\bf r},{\bf v},\eta)$. We have seen that its extremization at fixed energy and Casimirs leads to a coarse-grained DF $\overline{f}({\bf r},{\bf v})=\overline{f}(\epsilon)$, given by Eq. (\ref{g95}), which depends only on the individual energy $\epsilon$. Such a DF describes a spherically symmetric and isotropic stellar system \cite{bt}. Furthermore, we have shown that $\overline{f}'(\epsilon)<0$, so that $\overline{f}(\epsilon)$ is a monotonically decreasing function of the energy. Therefore, according to the results of Sec. \ref{sec_sta}, the coarse-grained DF predicted by the theory of Lynden-Bell extremizes a generalized entropy of the form 
\begin{equation}
S[\overline{f}]=-\int C(\overline{f})\, d{\bf r}d{\bf v}
\label{gens}
\end{equation}
at fixed mass and energy. This ``entropy'' is a functional of the coarse-grained DF $\overline{f}$. According to Eqs. (\ref{g64c}) and (\ref{g96b}) the function $C$ is given by\footnote{For the DF from Eq. (\ref{g99}) with $c=1$ we find
\begin{equation}
C(\overline{f})=\frac{3\overline{f}}{2b}-\frac{1}{2b}\sqrt{1+\frac{4b}{\overline{f}}}-2\tanh^{-1}\sqrt{1+\frac{4b}{\overline{f}}}.
\label{cun}
\end{equation}
For the Fermi-Dirac-like DF from Eq.  (\ref{fdl2}) we obtain the Fermi-Dirac-like entropy $C(\overline{f})=\overline{f}\ln \overline{f}+(\eta_0-\overline{f})\ln(\eta_0-\overline{f})$, which reduces to the Boltzmann-like  entropy $C(\overline{f})=\overline{f}\ln \overline{f}$ in the nondegenerate (dilute) limit.} 
\begin{equation}
C(\overline{f})=-\int \left\lbrack -(\ln Z)'\right\rbrack^{-1}(\overline{f})\, d\overline{f}.
\label{gensb}
\end{equation}
In the nondegenerate limit, it reduces to
\begin{equation}
C(\overline{f})=-\int  (-G')^{-1}(\overline{f})\,d\overline{f}.
\label{sac3}
\end{equation}
The generalized entropy (\ref{gens}) with Eq. (\ref{gensb}) can also be directly obtained from the  Lynden-Bell entropy (\ref{mix}) by following the procedure developed in Appendix C of \cite{kinVR}. From Eqs. (\ref{g64ca}) and (\ref{g96c}) we obtain the important identity
\begin{equation}
f_2=\frac{1}{C''(\overline{f})}.
\label{imp}
\end{equation}
This identity is rigorously valid at equilibrium.\footnote{When $g(\eta)$ is treated canonically, i.e. when it is prescribed {\it a priori}, we can show  \cite{kinVR} that $\rho({\bf r},{\bf v},\eta)$ is a maximum of $S[\rho]$ at fixed mass and energy if, and only if, $\overline{f}({\bf r},{\bf v})$ is a maximum of $S[\overline{f}]$ at fixed mass and energy. Furthermore, in that case, the identity from Eq. (\ref{imp}) is also valid out-of-equilibrium.}

For the  polytropic DF $f$ given by Eq. (\ref{g67})  the associated generalized entropy $S[f]$ is the Tsallis entropy (\ref{g66}).  Since we have shown that the Lynden-Bell  theory can justify, in certain cases, a coarse-grained polytropic DF $\overline{f}$ [see Eq. (\ref{g104})] we conclude that, in the sense of Appendix C of \cite{kinVR}, the  Tsallis entropy
\begin{equation}
S[\overline{f}]=-\frac{1}{q-1}\int (\overline{f}^q-\overline{f})\, d{\bf r}d{\bf v}
\label{tsa}
\end{equation}
obtained from Eq. (\ref{g66}) by replacing $f$ by $\overline{f}$  can be justified from the Lynden-Bell entropy $S[\rho]$. We stress, however, that the Lynden-Bell entropy $S[\rho]$ and the Tsallis ``entropy'' $S[\overline{f}]$ operate in different spaces of functions $\rho({\bf r},{\bf v},\eta)$ and $\overline{f}({\bf r},{\bf v})$ so they fundamentally have a different status.  We refer to \cite{assise} for an interpretation of the different types of functionals appearing in the statistical mechanics of systems with long-range interactions.

{\it Remark:} Let us give a quick derivation of Eq. (\ref{tsa}) from the Lynden-Bell entropy (\ref{mix}) with the $\chi$-squared distribution from Eq. (\ref{g97}). A more rigorous and more general derivation (valid out-of-equilibrium) is given in Appendix C of \cite{kinVR}. If we assume that $g(\eta)$ is prescribed, we can introduce the relative Lynden-Bell entropy (see Appendix C of \cite{kinVR} for a justification)
\begin{equation}
S[\rho]=-\int \rho\ln \left \lbrack \frac{\rho}{g(\eta)}\right \rbrack\, d\eta d{\bf r}d{\bf v}.
\label{mb1}
\end{equation}
Substituting Eq. (\ref{g92}) into Eq. (\ref{mb1}), we obtain an entropy of the form of Eq. (\ref{gens}) with
\begin{equation}
C(\overline{f})=-\beta \overline{f}\epsilon-\ln Z.
\label{mb2}
\end{equation}
In the nondegenerate limit, we can make the approximation $Z\simeq 1+G$ with $G\ll 1$, so that
\begin{equation}
C(\overline{f})=-\beta \overline{f}\epsilon-G(\beta\epsilon).
\label{mb3}
\end{equation}
Taking the derivative of Eq. (\ref{mb3}) with respect to $\overline{f}$ and using Eq. (\ref{sac2}), we obtain 
\begin{equation}
C'(\overline{f})=-\beta\epsilon-\beta \overline{f}\frac{1}{\overline{f}'(\epsilon)}-\beta G'(\beta\epsilon)\frac{1}{\overline{f}'(\epsilon)}=-\beta \epsilon,
\label{mb4}
\end{equation}
which coincides with Eq. (\ref{g64}) with $\alpha=0$ (see footnote 42). This equation is general (in the nondegenerate limit) and leads to Eq. (\ref{sac3}). For the $\chi$-squared distribution from Eq. (\ref{g97}), inverting Eq. (\ref{g104}) and substituting the result into Eq. (\ref{mb4}), we get
\begin{equation}
C'(\overline{f})=-\frac{1}{b}\left \lbrack \left (\frac{bc}{\overline{f}}\right )^{1/(c+1)}-1\right\rbrack.
\label{mb5}
\end{equation}
After integration, we obtain
\begin{equation}
C(\overline{f})=-(c+1)\left \lbrack \left (\frac{\overline{f}}{bc}\right )^{c/(c+1)}-\frac{\overline{f}}{b(c+1)}\right\rbrack,
\label{mb6}
\end{equation}
which is equivalent to Eq. (\ref{tsa}) with Eq. (\ref{g105}).

\subsection{Computation of $g(\eta)$}
\label{sec_fgalle}

In principle, the function $g(\eta)$ is determined by the phase space hypervolumes $\gamma(\eta)$ occupied  by the levels $\eta$ through Eqs. (\ref{g106}) and (\ref{g92}). For spatially homogeneous systems, Eq. (\ref{g106}) reduces to
\begin{equation}
\gamma(\eta)=\int \rho({\bf v},\eta)\, d{\bf v},
\label{g107}
\end{equation}
where we have absorbed the domain volume in $\gamma(\eta)$. In the nondegenerate limit, using [see Eq. (\ref{g102})]
\begin{equation}
\rho({\bf v},\eta)=g(\eta)e^{-\beta \eta v^2/2},
\label{g108}
\end{equation}
we obtain
\begin{equation}
\gamma(\eta)=\int g(\eta)e^{-\beta \eta v^2/2}\, d{\bf v}=g(\eta)\left (\frac{2\pi}{\beta\eta}\right )^{d/2}.
\label{g109}
\end{equation}
Therefore, the function $g(\eta)$ is determined by the initial condition through the formula
\begin{equation}
g(\eta)=\left (\frac{\beta\eta}{2\pi}\right )^{d/2}\gamma(\eta).
\label{g110}
\end{equation}
The coarse-grained DF is then given by [see Eq. (\ref{g103})]
\begin{equation}
\overline{f}({\bf v})=\left (\frac{\beta}{2\pi}\right )^{d/2}\int_0^{+\infty} \gamma(\eta) \eta^{1+d/2} e^{-\beta \eta v^2/2}\, d\eta.
\label{g111}
\end{equation}
If $\gamma(\eta)\sim \eta^k$, we find from Eq. (\ref{g110}) that $g(\eta)\sim \eta^{d/2+k}$. This  corresponds to Eq. (\ref{g97}) or Eq. (\ref{bill1}) with  an exponent
\begin{equation}
c=\frac{d}{2}+k+1.
\label{ck}
\end{equation}
According to Eq. (\ref{g105}), this selects a polytropic coarse-grained distribution [see Eq. (\ref{g104})] of index
\begin{equation}
n=-k-2.
\label{nk}
\end{equation}
In order to be in case (ii) of Sec. \ref{sec_spol}, the condition $n<-1$ (or $c>d/2$)  imposes $k>-1$.

{\it Remark:} We can also obtain a formula valid for inhomogeneous systems. In the nondegenerate limit, substituting Eq. (\ref{g102}) into Eq. (\ref{g106}), we obtain
 \begin{equation}
\gamma(\eta)=\int g(\eta)e^{-\beta \eta \lbrack v^2/2+\Phi({\bf r})\rbrack} \, d{\bf r}d{\bf v}.
\label{g112}
\end{equation}
Integrating over the velocity, we get
\begin{equation}
\gamma(\eta)=g(\eta)\left (\frac{2\pi}{\beta\eta}\right )^{d/2}\int e^{-\beta \eta\Phi({\bf r})}\, d{\bf r}.
\label{g113}
\end{equation}
Therefore, the function $g(\eta)$ is determined by the initial condition through the formula
\begin{equation}
g(\eta)=\left (\frac{\beta\eta}{2\pi}\right )^{d/2}\frac{\gamma(\eta)}{\int e^{-\beta \eta\Phi({\bf r})}\, d{\bf r}}.
\label{g114}
\end{equation}
The coarse-grained DF (\ref{g103}) is then given by
\begin{equation}
\overline{f}({\bf r},{\bf v})=\left (\frac{\beta}{2\pi}\right )^{d/2}\int_0^{+\infty} \frac{\gamma(\eta)}{\int e^{-\beta \eta\Phi({\bf r})}\, d{\bf r}} \eta^{1+d/2} e^{-\beta \eta\epsilon}\, d\eta.
\label{g115}
\end{equation}
We stress that the equilibrium potential $\Phi({\bf r})$ depends itself on $\gamma(\eta)$. It is determined  by solving the differential equation obtained by substituting Eq. (\ref{g115}) into the Poisson equation (\ref{g58}). This leads, in the inhomogeneous case, to a complicated system of coupled equations that has to be solved self-consistently.

{\it Remark:} The nondegenerate equilibrium state of a collisionless stellar
system is similar to the collisional equilibrium state of a multi-species
stellar system (see Sec. 2 of \cite{superstat}). In that case, $\eta$ plays the
role of the mass $m$ of a star and $\gamma(\eta)$ corresponds to the total
number $N(m)$ of stars of mass $m$. A mass spectrum
$M(m)=N(m)m\propto m^{k+1}$ leads to a polytropic DF of index $n=-k-2$ [see Eq.
(\ref{nk})]. In particular, a mass spectrum $M(m)\sim m$
obtained when the number of stars of each species is typically the same
($N(m)\sim 1$) corresponds to $k=0$ 
and leads to  a polytropic DF of index $n=-2$.

\subsection{Universal DF from the Lynden-Bell theory: logotropes}
\label{sec_udf}

In very interesting recent papers, Ewart {\it et al.} \cite{ewart1,ewart2} argued that the Lynden-Bell theory, when applied in the multilevel case, predicts a universal coarse-grained DF with a power-law tail $\overline{f}\sim \epsilon^{-(d/2+1)}$.\footnote{By contrast, in the single level case, the Lynden-Bell DF decreases exponentially rapidly (see Sec. \ref{sec_fdl}).} In brief (see their papers for a more detailed discussion), their approach amounts to computing $\gamma(\eta)$ defined by Eq. (\ref{a1}) for a Gaussian fine-grained distribution $f({\bf v})$ (see Appendix \ref{sec_jumax} for a possible justification of this distribution) and arguing that the resulting expression of $\gamma(\eta)$ for $\eta\rightarrow 0$ is representative of more general situations.  From the expression of $\gamma(\eta)$ one can obtain $g(\eta)$ from Eq. (\ref{g110}) and determine the equilibrium coarse-grained DF from Eq. (\ref{g111}). These results assume that the nondegenerate limit of the Lynden-Bell theory is applicable. They also assume that the system is spatially homogeneous. However, Ewart {\it et al.} \cite{ewart1,ewart2} argue that their results remain approximately valid for inhomogeneous distributions. Although we  are not convinced about the strict ``universality'' of their distribution (in principle other forms of functions $g(\eta)$ and other DFs $\overline{f}(\epsilon)$ could be obtained from the theory of Lynden-Bell as discussed in \cite{superstat}), it is clear that the specific case treated in \cite{ewart1,ewart2} is worth considering. The calculation of $\gamma(\eta)$ for a Gaussian (Maxwell) distribution is detailed in Appendix \ref{sec_sadm} leading to Eq. (\ref{a17}).\footnote{In Appendix \ref{sec_sadt} we carry out a similar calculation for polytropic (Tsallis) distributions.} Substituting this result into Eq. (\ref{g110}) we obtain
\begin{equation}
g(\eta)=\left (\frac{\beta\eta}{2\pi}\right )^{d/2}\frac{S_d}{\beta\eta}\left\lbrace -\frac{2}{\beta}\ln\left\lbrack \left (\frac{2\pi}{\beta}\right )^{d/2}\frac{1}{\rho}\eta\right\rbrack\right\rbrace^{d/2-1}.
\label{g116}
\end{equation}
Up to logarithmic corrections, this gives 
\begin{equation}
\gamma(\eta)\sim \frac{1}{\eta}\quad \Rightarrow \quad g(\eta)\sim \eta^{d/2-1}.
\label{g117}
\end{equation}
Comparing Eq. (\ref{g117}) with Eq. (\ref{g97}) or Eq. (\ref{bill1}) we see that this function selects $c=d/2$. According to the results of Sec. \ref{sec_ndg}, it leads to an equilibrium coarse-grained DF which is (at least asymptotically) a  polytropic DF  with an index  $n=-1$ i.e. $q=d/(d+2)$ [see Eq. (\ref{g105})].\footnote{These results correspond to Eqs. (\ref{ck}) and (\ref{nk}) with $k=-1$ [see Eq. (\ref{a17mag})].} The coarse-grained DF $\overline{f}(\epsilon)$ develops an $\epsilon^{-(d/2+1)}$ tail in $d$ dimensions ($\epsilon^{-5/2}$ tail in 3D), while the energy distribution $N(\epsilon)$ develops an $\epsilon^{-2}$ tail (in the homogeneous case) irrespective of the dimensionality of space (see Appendix  \ref{sec_dsg}). According to Sec. \ref{sec_slog}, this ``universal'' DF turns out to be the one of stellar logotropes!

{\it Remark:} This result suggests that the ``universal'' DF 
of collisionless systems with long-range interactions is the logotropic DF in
the same manner that the  universal DF of collisionlal systems with long-range
interactions is the isothermal (Boltzmann) DF.\footnote{As mentioned above the
logotropic DF in the collisionless regime is probably less ``universal'' than
the isothermal DF in the collisional regime.} They both correspond to a form of
statistical equilibrium state (Lynden-Bell versus Boltzmann). This analogy
strengthens our interpretation of the logotropic constant $A$ as a logotropic
``temperature''. In virtue of the Remark at the end of Sec. \ref{sec_fgalle}, we
also note that the logotropic DF can be obtained from a multi-species
collisional stellar system with a flat mass spectrum
$M(m)=N(m)m\sim 1$, i.e. $N(m)\sim m^{-1}$, corresponding to $k=-1$ and leading
to
$n=-1$.

\section{Universal DF from the SDD equation: logotropes}
\label{sec_banik}

\subsection{SDD equation}
\label{sec_homsdd}

The evolution of an isolated system with long-range interactions (e.g. a self-gravitating system, a Coulombian plasma, the HMF model...) is governed, when $N\rightarrow +\infty$, by the Vlasov equation 
\begin{eqnarray}
\frac{\partial f}{\partial t}+{\bf v}\cdot \frac{\partial f}{\partial {\bf r}}-\nabla\Phi\cdot \frac{\partial f}{\partial {\bf v}}=0,
\label{g57sdd}
\end{eqnarray}
coupled to a mean field potential of the form
\begin{equation}
\Phi({\bf r},t)=\int u({\bf r}-{\bf r}')f({\bf r}',{\bf v}',t)\, d{\bf r}'d{\bf v}',
\label{ba0}
\end{equation}
where $u({\bf r}-{\bf r}')$ is a binary potential of interaction. The Vlasov
equation is valid in the collisionless regime of the dynamics where the
evolution of the system is only due to mean field effects. A collisionless
system with long-range interactions can reach, on the coarse-grained scale, a
steady state as a result of a process of violent relaxation (see Sec.
\ref{sec_ew}). The Vlasov equation admits an infinite number of stable
stationary solutions \cite{bt} and the general prediction of the QSS actually
reached by the system for a given initial condition remains an open problem.
Indeed, the Lynden-Bell prediction does not always work because of the problem
of 
incomplete relaxation \cite{lb,incomplete} (see footnote 41).

In practice,  a system is never isolated from the surrounding. Let us therefore consider a collisionless system with long-range interactions stochastically  forced by an external medium. If the noise is weak we can make a quasilinear analysis of the perturbed Vlasov equation.  In that case, it can be shown that the evolution of the DF is governed by  a quasilinear diffusion equation with a diffusion coefficient $D[f]$ that depends on the DF itself through the ``dielectric'' function, which takes into account collective effects. This is the so-called SDD equation derived in \cite{epjp1,sdduniverse}.  It applies to any type of systems with long-range interactions submitted to an external stochastic forcing such as plasmas, stellar systems, the Hamiltonian Mean Field (HMF) model...

For a spatially homogeneous system, the SDD equation is given by \cite{epjp1,sdduniverse}
\begin{equation}
\frac{\partial f}{\partial t}=\frac{\partial}{\partial v_i}\left (D_{ij}\frac{\partial f}{\partial v_j}\right ),
\label{ba1}
\end{equation}
with the diffusion tensor
\begin{equation}
D_{ij}=\frac{1}{2}\int d{\bf k}\, k_i k_j\frac{{\hat C}({\bf k},{\bf k}\cdot {\bf v})}{|\epsilon({\bf k},{\bf k}\cdot {\bf v})|^2}
\label{ba2}
\end{equation}
involving the dielectric function
\begin{equation}
\epsilon({\bf k},\omega)=1-(2\pi)^d{\hat u}(k)\int \frac{{\bf k}\cdot \frac{\partial f}{\partial {\bf v}}}{{\bf k}\cdot {\bf v}-\omega}\, d{\bf v}.
\label{ba3}
\end{equation}
Here ${\hat u}(k)$ is the Fourier transform of the binary potential of interaction and ${\hat C}({\bf k},\omega)$ is the Fourier transform of the correlation function $C({\bf r}-{\bf r}',t-t')$ of the external stochastic force (see \cite{epjp1,sdduniverse} for details). Since the diffusion coefficient depends on the DF itself,  Eq. (\ref{ba1}) with Eqs. (\ref{ba2}) and (\ref{ba3}) is a very complicated and very nonlinear integral equation in general. It can, however, be simplified in certain limits.

\subsection{Homogeneous systems}
\label{sec_hom}

Recently, Banik {\it et al.} \cite{banik1} used the SDD equation (\ref{ba1})-(\ref{ba3}) introduced by Chavanis in \cite{epjp1,sdduniverse} to make interesting predictions about the high velocity tail of the DF. Their argumentation proceeds as follows. For an isotropic external forcing with a small correlation time (white noise-like) peaked around a given wavenumber, they showed that an intermediate  regime of velocities exists, $v_{\rm min}\ll v\ll v_{\rm max}$, in which  the diffusion coefficient scales as  $D(v)\sim v^4$ in any dimension of space.

Actually, this $v^4$ scaling law can be understood very simply by computing the diffusion coefficient from Eq. (\ref{ba2}) with a Dirac distribution  $f({\bf v})=\rho \delta({\bf v})$. In that case, after an integration by parts, the dielectric function from Eq. (\ref{ba3}) reduces to \cite{nyquistgrav}
\begin{equation}
\epsilon({\bf k},\omega)=1-(2\pi)^d{\hat u}(k)\rho \frac{k^2}{\omega^2}.
\label{ba4}
\end{equation}
Substituting this expression into Eq. (\ref{ba2}) and assuming that the external perturbation is a white noise for which the Fourier transform of the correlation function $C({\bf k},\omega)$ is independent of $\omega$, we get $D_{ij}=D\delta_{ij}$ with
\begin{equation}
D(v)=\frac{1}{6}\int k^2  {\hat C}(k) \frac{v^4}{\left\lbrack v^2-(2\pi)^d{\hat u}(k)\rho\right\rbrack^2}\, d{\bf k}.
\label{ba5}
\end{equation}
This diffusion coefficient behaves as $v^4$ for $v_{\rm min}\ll v\ll v_{\rm max}$. This corresponds to the behavior obtained in \cite{banik1} for a more general DF than the Dirac distribution in a specific range of velocities (see their paper for details). Under these assumptions the SDD equation (\ref{ba1}) reduces to
\begin{equation}
\frac{\partial f}{\partial t}=\frac{1}{v^{d-1}}\frac{\partial}{\partial v}\left (v^{d-1} D(v)\frac{\partial f}{\partial v}\right )
\label{ba6}
\end{equation}
with $D(v)\sim v^4$ in the regime of interest.

Then, Banik {\it et al.} \cite{banik1} looked for a self-similar solution of Eq. (\ref{ba6}) with $D\sim v^4$ and found that the quasi-steady-state $f$ develops a $v^{-(d+2)}$ tail in $d$ dimensions ($v^{-5}$ tail in 3D), while the energy distribution $N(\epsilon)$  (with $\epsilon=v^2/2$ in the homogeneous case) develops an $\epsilon^{-2}$ tail irrespective of the dimensionality of space (see Appendix  \ref{sec_dsg}).  One can also obtain this result by looking for a stationary solution of the SDD equation. To be general, let us assume that $D\sim v^{\alpha}$. Taking $\partial f/\partial t=0$ in Eq. (\ref{ba6}), we find that
\begin{equation}
v^{d-1+\alpha}\frac{\partial f}{\partial v}=C,
\label{ba7}
\end{equation}
where $C$ is a constant. After integration we obtain
\begin{equation}
f\propto v^{-(d-2+\alpha)},\qquad f\propto \epsilon^{-(d-2+\alpha)/2},
\label{dait}
\end{equation} 
yielding $N(\epsilon)\propto \epsilon^{-\alpha/2}$ (see Appendix  \ref{sec_dsg}). The same results are obtained by looking for a self-similar solution of Eq. (\ref{ba6})  with $D\sim v^{\alpha}$  (see Appendix \ref{sec_sol}). For $\alpha=4$, we recover the  $f\sim v^{-(d+2)}$ and $N\sim\epsilon^{-2}$ tails \cite{banik1}.

Collisionless systems with long-range interactions often exhibit nonthermal (non-Boltzmannian) power-law tails in their DFs. These power-law tails are sometimes interpreted in terms of polytropic distributions also called Tsallis distributions or kappa distributions \cite{leubner1,leubner2}.  Banik {\it et al.}  \cite{banik1} applied their results to the ion population of the solar wind which features a $v^{-5}$ tail in their velocity distribution, whose origin has been a long-standing puzzle. According to their theory, this power-law tail turns out to be a natural outcome of the collisionless relaxation of electrostatic plasmas driven by a stochastic force with  a small correlation time. They emphasized that self-consistency (ignored in former test-particle treatments) is crucial for the emergence of the universal $v^{-5}$ tail.  We shall discuss this power-law tail in more detail in the following section where we consider the case of spatially inhomogeneous systems.

{\it Remark:} For a colored (correlated) noise, the correlation function generically reads
\begin{equation}
{\hat C}({\bf k},t-t')=\frac{{\hat C}({\bf k})}{2\tau_{c,k}}e^{-|t-t'|/\tau_{c,k}},
\label{color1}
\end{equation}
where $\tau_{c,k}$ is the finite correlation time for the mode ${\bf k}$. Its
Fourier transform 
in time is 
\begin{equation}
{\hat C}({\bf k},\omega)=\frac{{\hat C}({\bf k})}{1+(\omega \tau_{c,k})^2}.
\label{color2}
\end{equation}
The diffusion tensor of the SDD equation [see Eq. (\ref{ba2})] then takes the form \cite{sdduniverse}
\begin{equation}
D_{ij}=\frac{1}{2}\int d{\bf k}\, k_i k_j\frac{{\hat C}({\bf k})}{\left\lbrack 1+({\bf k}\cdot {\bf v} \tau_{c,k})^2\right\rbrack |\epsilon({\bf k},{\bf k}\cdot {\bf v})|^2}.
\label{color3}
\end{equation}
A finite correlation
time can lead to a more complicated dependence of the diffusion coefficient with
${\bf v}$ and possibly affect the previous results relying on a white noise. The
case of a colored noise has been investigated in \cite{banik1}.

\subsection{Inhomogeneous systems}

In a subsequent paper, Banik {\it et al.} \cite{banik2} applied their results to DM halos. Collisionless self-gravitating systems such as CDM halos are known to display a remarkable universal density profile despite the intricate nonlinear physics of hierarchical structure formation. Early cosmological $N$-body simulations showed that the  density profiles of CDM halos can be fitted by the Navarro-Frenk-White (NFW) profile \cite{nfw}
\begin{equation}
\rho=\frac{\rho_sr_s}{r(1+r/r_s)^2},
\label{ba8}
\end{equation}
irrespective of the halo mass, concentration, power-law index of the initial power spectrum, and cosmology. This profile displays a central cusp scaling like $r^{-1}$ and a density tail decaying as $r^{-3}$. The origin of this attractor state has been a persistent mystery, particularly because the physics of collective collisionless relaxation is not well understood.

In general, DM halos are not isolated but they experience an external perturber potential sourced by other galaxies or halos or by substructures within the system. In addition, it is necessary to take into account the spatial inhomogeneity of the self-gravitating system. Therefore,  in order to understand the origin of the $r^{-1}$ cusp, Banik {\it et al.} \cite{banik2} used the inhomogeneous SDD equation with angle-action variables, which was  formulated in \cite{sdduniverse}.\footnote{The possibility to use the SDD equation in the context of DM halos was mentioned in the conclusion of \cite{sdduniverse}, although it was conjectured that the effect of stochastic forcing would be to transform cusps into cores as originally argued by Weinberg \cite{weinberg}. Banik {\it et al.} \cite{banik2}, on the contrary, report a situation where the stochastic forcing creates an $r^{-1}$ cusp. It is possible that different regimes are relevant in DM halos depending on the forcing and response function, and lead to a wealth of results.} This  quasilinear diffusion equation describes the secular evolution of the DF of a halo due to the nonlinear coupling of the linear fluctuations induced by random perturbations in the gravitational potential. The density fluctuations are collectively dressed by self-gravity, a phenomenon that is described by the response matrix.  Assuming that the DM halo is spherically symmetric and isotropic, Banik {\it et al.} \cite{banik2}  reduced the original SDD equation \cite{sdduniverse} to a one-dimensional diffusion equation for the DF $f=f(\epsilon,t)$.

Then, by looking for a steady-state solution of the inhomogeneous SDD equation, they found that the DF behaves as $f\sim \epsilon^{-5/2}$ for high energies. This implies that the density profile behaves as $\rho\sim r^{-1}$. Therefore, the $r^{-1}$  NFW cusp emerges as a quasi-steady state attractor solution of the SDD equation when the halo is forced by small-scale white noise-like fluctuations modeled as an external perturbation. This result assumes that there is a constant flux of matter into the halo, i.e., that the halo is accreting at a constant rate.  Banik {\it et al.} \cite{banik2} also stressed that collective effects are crucial for the emergence of the $r^{-1}$ NFW cusp.

Banik {\it et al.} \cite{banik2} also considered the zero-flux steady state
solution of the SDD  equation which naturally selects a DF that is flat in
energy space before dropping to zero  (Heaviside function). This DF can be
interpreted as a polytropic distribution of index $n=3/2$ (see Sec.
\ref{sec_spol}).  It corresponds to the Fermi distribution that occurs in the
theory of white dwarf stars \cite{chandrabook}, neutron stars \cite{ov}, and 
fermionic DM halos \cite{mg16F,modeldmF}. The density is a cored profile (like
the Einasto or Burkert profile) with a compact support, i.e., the density
vanishes at a certain distance $R$ identified with the radius of the system. In
the presence of  a central black hole or a massive compact subhalo, the DM halo
develops an $r^{-3/2}$ cusp near the origin (see Sec. \ref{sec_cbh}).  This cusp
is an attractor of the SDD equation with appropriate boundary conditions.  Such
a profile also emerges naturally in the secondary infall model of Bertschinger
\cite{bert}, which predicts the formation of an $r^{-3/2}$ cusp from the
self-similar accretion of a collisionless fluid onto a central black hole.
Therefore, depending on the inner boundary conditions, one can have either a
central core or an $r^{-3/2}$ cusp \cite{banik2}.

In conclusion,  the quasilinear approach (SDD equation) selects a subclass of steady solutions the Vlasov equations among an infinity of possible solutions. It can be an  NFW  $r^{-1}$ cusp (constant flux solution), an Einasto-like central core (zero flux solution in the absence of a central object), or an  $r^{-3/2}$ prompt cusp (zero flux solution in the presence of a central object).

Concerning the $f(\epsilon)\sim \epsilon^{-5/2}$ power-law tail we can make the connection with the results of Secs. \ref{sec_gal} and \ref{sec_hom} as follows. First we note that, according to the Jeans theorem \cite{jeansT,bt}, the steady DF of a spherically symmetric and isotropic self-gravitating system depends only on the energy $\epsilon=v^2/2+\Phi({\bf r})$, i.e., $f=f(\epsilon)$. Assuming that the universal velocity tail $f\sim v^{-(d+2)}$ obtained from the SDD equation in the homogeneous case (see Sec. \ref{sec_hom}) remains valid in the inhomogeneous case, and using the Jeans theorem, we get $f(\epsilon)\sim \epsilon^{-(d+2)/2}$, which returns the result of Banik {\it et al.} \cite{banik2} for $d=3$.

We now remark that the power-law tail $f(\epsilon)\sim \epsilon^{-(d+2)/2}$ corresponds to the asymptotic behavior of the logotropic DF from Eq. (\ref{g82}).  Therefore, the quasilinear theory tends to select the logotropic DF (or a DF close to it) among all the steady states of the Vlasov equation.\footnote{It is an interesting open problem to know if the SDD equation rigorously selects the logotropic DF or a distinct DF which has the same asymptotic behavior. Banik {\it et al.} \cite{banik1} report that their numerically obtained DF is in rather good agreement with the kappa distribution of index $\kappa=3/2$ which is precisely a polytropic (or Tsallis) DF of index $n=-1$ (logotrope). Kappa distributions have been used in studies of the solar wind.}  We can thus apply the results of Sec. \ref{sec_logog} leading to the $\rho\sim r^{-1}$ decay law of the density profile in any dimension $d$ (with a logarithmic correction in $d=1$).\footnote{More generally, a tail $f(\epsilon)\sim \epsilon^{-(d/2-1+\alpha/2)}$ [see Eq. (\ref{dait})] corresponds to the  asymptotic behavior of a polytropic DF  of index $n=-(\alpha-2)/2$ [see Eq. (\ref{g75})] when $\alpha>4$. In that case, the density profile decreases as $\rho\sim r^{-2(\alpha-2)/\alpha}$ (see Sec. \ref{sec_abi}).} As shown in our previous works (see \cite{epjp,lettre,jcap,preouf1,action,logosf,preouf2,ouf,gbi}  and Appendix 1 of \cite{gbi}), this power law decay can describe either the  $r^{-1}$ NFW cusp of a CDM halo or the  $r^{-1}$ envelope of a logotropic DM halo. Actually, the $r^{-1}$ NFW cusp is not consistent with the observations of DM halos which favor a central core (like in the Burkert profile \cite{observations}) rather than a cusp (like in the NFW profile \cite{nfw}). This is the so-called core-cusp problem \cite{moore}. Therefore, the interpretation of  the $\rho\sim r^{-1}$ decay law as a logotropic envelope may be more relevant for astrophysical applications than its interpretation as a NFW cusp.\footnote{In this connection see the Remark at the end of Sec. \ref{sec_dpro}.} Following our previous works \cite{epjp,lettre,jcap,preouf1,action,logosf,preouf2,ouf,gbi}, we show in the following section that the logotropic distribution gives very interesting results when applied to DM halos. Therefore, the kinetic theory based on the SDD equation may help justifying the logotropic distribution postulated in our previous papers and give a physical meaning to the logotropic constant (or logotropic temperature) $A$. Indeed, in the kinetic theory, this constant is related to the correlation function of the noise that sources the secular evolution of the system. Since this noise is of cosmological origin it may be related in a manner or the other to the cosmological constant $\Lambda$. It is also possible that the noise accounts for vacuum fluctuations. These considerations may justify the relation between the logotropic constant (or logotropic temperature) $A$ and the cosmological constant $\Lambda$ [see Eq. (\ref{logomodel8}) below] that we found in \cite{epjp,lettre,jcap,preouf1,action,logosf,preouf2,ouf,gbi}.

\section{Logotropic DM halos}
\label{sec_logodm}

In the previous sections, we have seen that arguments from statistical mechanics and kinetic theory suggest that collisionless systems with long-range interaction have a universal DF, which is the logotropic distribution. In this section, we specifically apply these results to DM halos. We describe in detail the structure of logotropic DM halos and relate their universal surface density to the cosmological constant.  We use a nonrelativistic approach that is appropriate to DM halos.

 \subsection{Logotropic equation of state}

In a series of papers \cite{epjp,lettre,jcap,preouf1,action,logosf,preouf2,ouf,gbi}  we
have
developed a UDME model based on the
logotropic equation of state  
\begin{eqnarray}
\label{logomodel1}
P=A\ln \left (\frac{\rho}{\rho_P}\right ),
\end{eqnarray}
where $A$ is a new fundamental constant of physics replacing the cosmological constant (see below) and
$\rho_P=c^5/G^2\hbar=5.16\times 10^{99}\, {\rm
g\, m^{-3}}$ is the Planck density. As discussed in Secs. \ref{sec_logog} and \ref{sec_slog}, the constant $A$ can be interpreted as a sort of ``logotropic
temperature'' in a generalized thermodynamical framework
\cite{cslogo,epjp,lettre,jcap}.

By applying the logotropic
equation of state (\ref{logomodel1}) to the Universe as a whole (``large''
scales) in a cosmological framework\footnote{The logotropic model can account
for the present acceleration of the Universe and is indistinguishable from the
standard $\Lambda$CDM model up to the present time. The two models will differ
in about $27$ Gyrs when the logotropic model  becomes phantom (the energy
density increases logarithmically with the scale factor) and presents a super de
Sitter behavior ($a\sim e^{t^2}$)  leading to a little rip (the energy density and the scale factor become infinite in infinite time) while the $\Lambda$CDM model has a de Sitter
behavior ($a\sim e^t$) corresponding to a constant energy density
$\rho_{\Lambda}c^2$. We could also consider the possibility that $A$ 
depends on time. We do not develop the logotropic
cosmology further in the present paper and refer to our previous papers
\cite{epjp,lettre,jcap,preouf1,action,logosf,preouf2,ouf,gbi} for details.}  we
found
that the universal constant $A$ is related to the present value of the DE density $\rho_{\Lambda}\equiv \rho_{\rm de,0}=5.96\times
10^{-24}\, {\rm
g\, m^{-3}}$ by
\begin{eqnarray}
\label{logomodel4}
{A}/{c^2}=\frac{\rho_\Lambda}{\ln \left (\frac{\rho_P}{\rho_{\Lambda}}\right
)}=2.10\times 10^{-26}\, {\rm g}\, {\rm m}^{-3}.
\end{eqnarray}
Writing
\begin{eqnarray}
\label{logomodel4j}
\rho_{\Lambda}=\frac{\Lambda c^2}{8\pi G},
\end{eqnarray}
where $\Lambda=1.11\times
10^{-52}\, {\rm m^{-2}}$ is the usual cosmological constant,\footnote{Although we express the logotropic constant $A$ in terms of the widely used cosmological constant $\Lambda$, we stress that the logotropic model is different from the $\Lambda$CDM model. In particular, in the logotropic model, the DE density is not constant. Therefore, its value coincides (by definition) with the value of the constant DE density of the $\Lambda$CDM model only at the present 
time ($\rho_{\rm de,0}=\rho_{\Lambda}$).} we can rewrite the foregoing equation as\footnote{We have absorbed the  logarithmic term $\ln(\rho_P/\rho_\Lambda)\simeq 283$ in the prefactor. It could be written explicit if necessary.}
\begin{eqnarray}
\label{logomodel8}
A=1.40\times 10^{-4}\, \frac{c^4\Lambda}{G}.
\end{eqnarray}
As a result, the logotropic equation of state (\ref{logomodel1}) takes the form
\begin{eqnarray}
\label{logomodel1rw}
P=B\rho_{\Lambda}c^2\ln \left (\frac{\rho}{\rho_P}\right ),
\end{eqnarray}
where $B=1/\ln(\rho_P/\rho_\Lambda)=3.53\times 10^{-3}$. We note that the logotropic equation of state (\ref{logomodel1rw}) involves both the Planck density $\rho_P$ (which serves as a normalization of the density) and the cosmological density $\rho_{\Lambda}$ (which serves as a normalization of the pressure). Therefore, it involves all the fundamental constants of physics. The Planck density, which is very high, is usually invoked to describe  the very early Universe and the cosmological density, which is very low, is used to explain the present acceleration of the Universe. They arise as upper and lower bounds in a cosmological model based on a quadratic equation of state \cite{aip,universe,vacuumon}. The dimensionless constant $B$ is approximately equal (up to a conversion from natural to common logarithm) to the inverse of the famous number $120$ (the exponent in the ratio $\rho_P/\rho_\Lambda\sim 10^{120}$) that appears in connection to the cosmological constant problem and in the theory of large numbers \cite{ouf}. The ratio $10^{120}$ between the Planck density and the cosmological density may be interpreted as the ``largest large number'' in Nature \cite{ouf}. The $\Lambda$CDM model is recovered in the limit $B\rightarrow 0$, which can be interpreted as a semiclassical limit $\hbar\rightarrow 0$ or $\rho_P\rightarrow +\infty$ (see \cite{epjp,lettre,jcap,preouf1,action,logosf,preouf2,ouf,gbi} for details).

\subsection{Density profile: The structure  of logotropic DM halos}
\label{sec_dpro}

When applied to DM halos (``small'' scales), the logotropic model leads to DM halos with a density profile that is flat in the center (thereby solving the core-cusp problem of the CDM model \cite{moore}) and that decreases at large distances as (see Sec. \ref{sec_logog})
\begin{eqnarray}
\label{logomodel2}
\rho\sim \left (\frac{A}{2\pi G}\right )^{1/2}\frac{1}{r},
\end{eqnarray}
like the singular logotropic solution. The regular logotropic density profile is plotted in Fig. 18  of \cite{epjp}. We believe that the $r^{-1}$ law is consistent with the observations. Indeed, this $r^{-1}$ decay can be seen in
Fig. 6 (right) of \cite{oh} and in Fig. 3 (plate U11583) of  \cite{rm}. The fact that the
slope of the regular density profile of DM halos close to the core  radius $r_h$ is
approximately equal to $-1$ has
also been pointed out by Burkert \cite{burkert2} (see in particular the upper
right panel of his Fig. 1).

A logotropic DM halo has an infinite mass.\footnote{This infinite mass problem does not rule out the logotropic
model. Actually, we
have the same problem with isothermal self-gravitating systems. The isothermal density profile
decreases at large distances as $\rho\sim 1/(2\pi G \beta m r^2)$, like the
singular isothermal sphere \cite{chandrabook}. It has an infinite mass. Despite
this problem, the isothermal density profile has often been used to model DM
halos because it provides a good fit of their central parts (up to a few halo
radii) and it can be
justified by Lynden-Bell's statistical theory of violent collisionless  relaxation in the single level case \cite{lb,clm1,clm2}.
In reality, the density of DM halos
decreases more rapidly than $r^{-2}$ at large distances, typically as $r^{-3}$, like for the
Burkert \cite{observations} and NFW \cite{nfw} profiles, or even more rapidly. This can be explained in
terms of incomplete relaxation, stochastic forcing, and tidal effects (see, e.g., Appendix B of \cite{modeldmB}). In
\cite{cslogo,epjp,lettre,jcap} (see also Secs. \ref{sec_logog} and \ref{sec_slog}) we have suggested that the logotropic model could be justified
by a notion of generalized thermodynamics. In this context, the constant $A$ in
the logotropic distribution plays the role of a generalized temperature which is
the counterpart of the temperature $T$ in the isothermal distribution. This generalized thermodynamical 
interpretation strengthens the analogy between the isothermal and the logotropic distributions.} Therefore, the logotropic density profile (\ref{logomodel2}) and the logotropic equation of state (\ref{logomodel1}) cannot be valid at infinitely large distances (corresponding to low
densities). At large distances, the density of real DM halos decreases more rapidly than $r^{-1}$, typically as $r^{-2}$ consistently with the asymptotic behavior of the isothermal sphere \cite{chandrabook} or as $r^{-3}$  consistently with the asymptotic behaviors of the Burkert \cite{observations} and NFW \cite{nfw}
profiles. It can even fall off more rapidly in order to guarantee a finite mass. This fast decrease of the density may be interpreted as a result of incomplete violent relaxation \cite{preouf2,logosf}, especially if we justify the logotropic model from the Lynden-Bell theory (see Sec. \ref{sec_ew}). In addition, if the DM halos are submitted to tidal effects from surrounding galaxies, their outer region is described by the King distribution $f=A(e^{-\beta \epsilon}-e^{-\beta\epsilon_m})$ which is asymptotically  equivalent to a polytrope of index $n=5/2$ for which the density drops to zero at a finite (tidal) radius (close to the escape velocity, the relation $f\sim A\beta e^{-\beta\epsilon_m}(\epsilon_m-\epsilon)$ is linear).  Therefore, the logotropic ``core'' is expected to be surrounded by an extended
``envelope''  resulting from a process of violent collisionless relaxation \cite{lb,wignerPH,logosf} or from tidal effects \cite{lb,clm1,clm2}, in which the density decreases more rapidly that $r^{-1}$. More precisely, a realistic logotropic DM halo would possess a core ($\rho\sim r^0$) $+$ an intermediate logotropic profile ($\rho\sim r^{-1}$) $+$ an
extended isothermal ($\rho\sim r^{-2}$), or NFW ($\rho\sim
r^{-3}$), or even steeper, envelope.\footnote{An even more general model would take into account the presence of a quantum core. This can be done by developing a complex scalar field (SF) model with a logarithmic potential associated with a logotropic equation of state \cite{logosf}. In this manner, logotropes can be justified by a SF model. Note that a purely quantum core (without logarithmic self-interaction), like a soliton in the fuzzy dark matter (FDM) model or like a fermion ball in the warm dark matter (WDM) model,  cannot account for
the observed constant surface density of DM halos (see Sec. 6.2 of \cite{frontiers}). A logarithmic self-interaction potential leading to a logotropic envelope seems to be necessary.} We believe that such a general structure is in agreement with the observations (see, e.g.,
\cite{oh,rm,burkert2}). In the following, we shall consider the logotropic profile up to a
few halo radii $r_h$ so we do not have to worry about  the envelope. The envelope is considered in Sec. 
\ref{sec_interpol}.

{\it Remark:} We note that the logotropic profile
$\rho\sim r^{-1}$ may be
wrongly interpreted in certain observations as a NFW cusp  $r^{-1}$ if the
logotropic core is not sufficiently well-resolved.  Indeed, in that case, we see
only the $r^{-1}$ tail of the logotropic distribution, not the core ($\rho\sim
r^0$). This may
lead to the illusion that certain DM halos are cuspy in agreement
with the NFW prediction while they are not \cite{blok}.

\subsection{Halo radius and halo mass}
\label{sec_hma}

The halo radius $r_h$ is defined as the distance at which the
central density  $\rho_0$  is divided by $4$. For logotropic DM halos, using Eq. (\ref{g51}), it is
given by
\begin{eqnarray}
\label{lel4}
r_h=\left (\frac{A}{4\pi
G\rho_0^2}\right )^{1/2}\xi_h,
\end{eqnarray} 
where
$\xi_h$ is determined by the equation
\begin{eqnarray}
\label{lel5}
\theta(\xi_h)=4.
\end{eqnarray} 
The normalized density profile $(\rho/\rho_0)(r/r_h)$ of logotropic DM halos is
plotted in Fig. 19 of \cite{epjp}. The  halo mass
$M_h$,
which is the
mass $M_h=\int_0^{r_h} \rho(r') 4\pi {r'}^2\, dr'$ contained within the sphere
of radius $r_h$, is given by \cite{epjp} 
\begin{eqnarray}
\label{lel6}
M_h=4\pi\frac{\theta'(\xi_h)}{{\xi_h}}\rho_0 r_h^3.
\end{eqnarray} 
Solving the Lane-Emden equation of index $n=-1$ [see Eq. (\ref{g52})] numerically, we find \cite{epjp} 
\begin{eqnarray}
\xi_h=5.85,\qquad \theta'_h=0.693.
\label{lel7}
\end{eqnarray}
This yields
\begin{eqnarray}
r_h=5.85\, \left (\frac{A}{4\pi G}\right )^{1/2}\frac{1}{\rho_0}
\label{lel8}
\end{eqnarray}
and
\begin{eqnarray}
M_h=1.49\, \rho_0 r_h^3.
\label{lel9}
\end{eqnarray}

\subsection{Universal surface density}

Eliminating the central density between Eqs. (\ref{lel8}) and (\ref{lel9}), we
obtain the logotropic halo mass-radius relation
\begin{eqnarray}
M_h=8.71 \, \left (\frac{A}{4\pi G}\right
)^{1/2} r_h^2.
\label{lel11}
\end{eqnarray}
Since $M_h\propto r_h^2$ we see that the surface density $\Sigma_0$ is 
constant.\footnote{This is a consequence of the fact that the density of a
logotropic DM halo decreases as $\rho\propto r^{-1}$ at large distances (i.e. $\rho\propto \Sigma_0/r$).} This is a very
important property of logotropic DM
halos \cite{epjp}. The logotropic model implies that the
DM halos (``small'' scales) have a universal surface density given by [see Eq. (\ref{lel8})]
\begin{eqnarray}
\label{logomodel3}
\Sigma_0\equiv \rho_0 r_h=5.85\, \left (\frac{A}{4\pi G}\right )^{1/2}.
\end{eqnarray}
Therefore, all the logotropic DM halos have the same surface density, whatever
their size, provided that
$A$ is interpreted as a universal constant of Nature \cite{epjp}.

Using the expression of $A$ from Eq. (\ref{logomodel8}), we can
express the
universal surface density of DM halos [see Eq. (\ref{logomodel3})] in terms of the  cosmological
constant $\Lambda$ of the $\Lambda$CDM model by
\begin{eqnarray}
\label{logomodel5}
\Sigma_0=0.01955\, \frac{c^2\sqrt{\Lambda}}{G}=133\,
M_{\odot}/{\rm pc}^2.
\end{eqnarray}
This value turns out to be in very good agreement with the observational value $\Sigma_0^{\rm
obs}=141_{-52}^{+83}\,
M_{\odot}/{\rm pc}^2$ \cite{kormendy,spano,donato}.  On the other hand, Eq.
(\ref{lel9}) may
 be rewritten as 
\begin{eqnarray}
M_h=1.49\, \Sigma_0 r_h^2=1.49\, \frac{\Sigma_0^3}{\rho_0^2}.
\label{lel13}
\end{eqnarray}
The ratio $M_h/(\Sigma_0 r_h^2)=1.49$ in Eq. (\ref{lel13}) is  in
good
agreement with the ratio $M_h/(\Sigma_0 r_h^2)=1.60$ obtained from the
observational Burkert profile \cite{observations} (see Appendix D.4 of \cite{modeldmB})
\begin{eqnarray}
\rho=\frac{\rho_0}{(1+r/r_h)\lbrack 1+(r/r_h)^2\rbrack}.
\label{bprof}
\end{eqnarray}
This is an
additional argument in favor of the logotropic model.

{\it Remark:} We stress that there is no adjustable
parameter in our theory, which leads to a definite prediction of $\Sigma_0$  in
terms of the fundamental constants of physics $G$, $c$,  $\hbar$ and $\Lambda$
(or $A$). In particular, the important prefactor $0.01955$ in Eq.
(\ref{logomodel5}) is determined from the theory. Therefore, the agreement with
the observational value $\Sigma_0^{\rm
obs}$ is remarkable. We present below other agreements which can be seen as a consequence of the fundamental result (\ref{logomodel5}).

\subsection{Universal gravitational acceleration}

We can define an average DM halo surface density by the relation
\begin{eqnarray}
\label{lel14}
\langle\Sigma\rangle=\frac{M_h}{\pi r_h^2}.
\end{eqnarray}
For logotropic DM halos, using Eq. (\ref{lel13}), we find
\begin{equation}
\label{lel15}
\langle\Sigma\rangle=\frac{M_h}{\pi
r_h^2}=\frac{1.49}{\pi}\Sigma_0=0.474\,
\Sigma_0=63.1\,
M_{\odot}/{\rm pc}^2.
\end{equation}
This theoretical value is in good agreement with the
value
$\langle\Sigma\rangle_{\rm
obs}=0.51\, \Sigma_0^{\rm
obs}=72_{-27}^{+42}, M_{\odot}/{\rm pc}^2$ obtained from the observations
\cite{gentile}.

The logotropic model
therefore implies a universal gravitational acceleration 
\begin{equation}
\label{logomodel6}
g=\frac{GM_h}{r_h^2}=\pi G\langle\Sigma\rangle=1.49\, G\Sigma_0=8.72\, G\left (\frac{A}{4\pi G}\right )^{1/2}.
\end{equation}
This is the gravitational acceleration at the halo radius: $g=g(r_h)$. Using Eq. (\ref{logomodel5}), the
universal gravitational
acceleration  can be expressed in terms of the cosmological constant as
\begin{eqnarray}
\label{logomodel7}
g=0.0291\, c^2\sqrt{\Lambda}=2.76\times 10^{-11}\, {\rm m\, s^{-2}}.
\end{eqnarray}
This theoretical value is in good agreement with the observational value $g_{\rm obs}=\pi G
\langle\Sigma\rangle_{\rm obs}=3.2_{-1.2}^{+1.8}\times 10^{-11}\, {\rm
m/s^2}$ of the gravitational acceleration \cite{gentile}.

\subsection{Universal NFW cusp}

Using the foregoing relations, the
asymptotic behavior of the logotropic density from Eq. (\ref{logomodel2})
can be rewritten as
\begin{eqnarray}
\label{logomodel9b}
\rho\sim 0.242\, \frac{\Sigma_0}{r}\sim 0.162\, \frac{g}{Gr}\sim 
0.00472\, \frac{c^2\sqrt{\Lambda}}{Gr}. 
\end{eqnarray}
As noted in \cite{logosf} this $r^{-1}$ behavior is similar to the
density cusp in the NFW model $\rho=\rho_sr_s/[r(1+r/r_s)^2]\sim \rho_s r_s/r$ for $r\rightarrow 0$ \cite{nfw}.
We find that 
\begin{eqnarray}
\label{logomodel10}
\rho_s r_s= 0.242\, \Sigma_0= 0.162\, \frac{g}{G}= 
0.00472\, \frac{c^2\sqrt{\Lambda}}{G}\nonumber\\
=1.42\, \left (\frac{A}{4\pi G}\right )^{1/2}=32.0\, M_{\odot}/{\rm pc}^2.
\end{eqnarray}
Therefore, the logotropic equation of state may describe or mimic  the $r^{-1}$ cusp of the NFW profile (see Sec. \ref{sec_banik}  and the Remark at the end of Sec. \ref{sec_dpro}).

\subsection{Tully-Fisher relation}

The circular velocity at the halo radius is
\begin{eqnarray}
\label{logomodel10b}
v_h^2=\frac{GM_h}{r_h}.
\end{eqnarray}
Combining the foregoing relations, we find that
\begin{equation}
\label{logomodel11}
\frac{v_h^4}{M_h}=Gg=\pi\langle\Sigma\rangle G^2=1.49\, \Sigma_0
G^2=8.72\, G^2\left (\frac{A}{4\pi G}\right )^{1/2}.
\end{equation}
Using Eq. (\ref{logomodel7}) we obtain
\begin{equation}
\label{logomodel7g}
\frac{v_h^4}{M_h}=0.0291\, G c^2\sqrt{\Lambda}=3.66\times 10^{-3}\,
{\rm km}^{4}{\rm s}^{-4}M_{\odot}^{-1}.
\end{equation}
Therefore, the ratio $v_h^4/M_h$ is universal (the same for all the DM halos). This relation is connected to the Tully-Fisher relation $v_h\propto M_b^{1/4}$ \cite{tf,tfmcgaugh} which involves the baryon mass $M_b$ instead of the DM halo mass
$M_h$. Introducing the baryon fraction $f_b=M_b/M_h\sim 0.17$, we obtain
$M_{\rm
b}/v_h^4=f_b/(1.49\, \Sigma_0 G^2)=46.4\,
M_{\odot}{\rm km}^{-4}{\rm s}^4$, which is close to the
observed value 
$(M_{\rm b}/v_h^4)^{\rm obs}=47\pm 6 \, M_{\odot}{\rm km}^{-4}{\rm s}^4$
\cite{mcgaugh}.

\subsection{Rotation curve}

The rotation curve of a DM halo is given by
\begin{eqnarray}
\label{logomodel12}
v^2(r)=\frac{GM(r)}{r}.
\end{eqnarray}
For a logotropic DM halo, we have for $r\rightarrow +\infty$: 
\begin{eqnarray}
\label{logomodel13}
M(r)\sim 2\pi \left (\frac{A}{2\pi G}\right )^{1/2}r^2
\sim 1.52\, \Sigma_0 r^2\nonumber\\
\sim 1.02\, \frac{gr^2}{G}\sim 
0.0296\, \frac{c^2\sqrt{\Lambda}r^2}{G},
\end{eqnarray}
implying
\begin{eqnarray}
\label{logomodel14}
v^2(r)\sim 2\pi G \left (\frac{A}{2\pi G}\right )^{1/2}r
\sim 1.52\, G\Sigma_0 r\nonumber\\
\sim 1.02\, gr\sim 
0.0296\, c^2\sqrt{\Lambda}r.
\end{eqnarray}
In the logotropic model, the circular velocity increases as $r^{1/2}$. On the other hand, asymptotically,  the gravitational acceleration (the gravitational force
$F(r)=m g(r)$ by unit of mass) produced by the
logotropic distribution tends to a constant
\begin{eqnarray}
\label{logomodel15}
g(r)=\frac{d\Phi}{dr}=\frac{GM(r)}{r^2}\rightarrow 1.02\,
g\equiv g_{\infty}.
\end{eqnarray}
We find
$M(r)/M_{\odot}\sim 202\, (r/{\rm pc})^2$, $\log(\frac{v}{\rm km\,
s^{-1}})=1.47+\frac{1}{2}\log(\frac{r}{\rm kpc})$ and $g_{\infty}=2.82\times
10^{-11}\, {\rm m \, s^{-2}}$ to be compared with the observational expressions
$M(r)^{\rm obs}/M_{\odot}=200_{-120}^{+200}\, (r/{\rm pc})^2$,
$\log(\frac{v_{\rm obs}}{\rm km\, 
s^{-1}})=1.47_{-0.19}^{+0.15}+0.5\log(\frac{r}{\rm kpc})$ and
$g_{\infty}^{\rm obs}=3_{-2}^{+3}\times
10^{-11}\, {\rm m \, s^{-2}}$ \cite{walker}. The agreement (without
adjustable parameter) is remarkable.\footnote{In our
previous paper \cite{gbi} we found a discrepancy between theory and observations
by a factor $2$ (theory gave half the observational values). Actually, in our
former paper \cite{cslogo} on which \cite{gbi} relied, we had made a mistake
(typo) in the asymptotic expression of the logotropic density. The reported
expression was half the correct expression from Eq. (\ref{logomodel2}). When
this mistake is corrected, as in the present paper, we obtain a perfect
agreement with the observational results.}

\subsection{Expression of the results in terms of the asymptotic gravitational acceleration}
\label{sec_aga}

Starting from the logotropic equation of state (\ref{logomodel1}), we have found that the gravitational acceleration $g(r)$ tends for $r\rightarrow +\infty$  to a universal constant $g_{\infty}$. It is useful to rewrite the foregoing relations in terms of the gravitational acceleration at infinity  $g_{\infty}$ instead of the gravitational constant at the halo radius $g$. We find
\begin{eqnarray}
\Phi(r)\sim g_{\infty} r,
\label{aga1}
\end{eqnarray}
\begin{eqnarray}
v^2(r)=\frac{GM(r)}{r}=r\frac{d\Phi}{dr}\sim g_{\infty}r,
\label{aga2}
\end{eqnarray}
\begin{eqnarray}
\rho(r)=\frac{\Delta\Phi}{4\pi G}=\frac{M'(r)}{4\pi r^2}\sim \frac{g_{\infty}}{2\pi Gr},
\label{aga3}
\end{eqnarray}
\begin{eqnarray}
\rho_sr_s=\frac{g_{\infty}}{2\pi G},
\label{aga4}
\end{eqnarray}
\begin{eqnarray}
\label{aga5}
\langle \Sigma\rangle(r)=\frac{M(r)}{\pi r^2}=\frac{g(r)}{\pi G}\rightarrow \langle\Sigma\rangle_{\infty}=\frac{g_{\infty}}{\pi G}.
\end{eqnarray}
We also have the relations
\begin{eqnarray}
\label{aga6}
g_{\infty}=2\pi G\left (\frac{A}{2\pi G}\right )^{1/2}
=1.52\, G\Sigma_0\nonumber\\
=1.02\, g=
0.0296\, c^2\sqrt{\Lambda}.
\end{eqnarray}
We stress that the logotropic model determines $g_{\infty}$ in terms of  the cosmological constant $\Lambda$ without free or {\it ad hoc} parameter.

\subsection{MOND theory}
\label{sec_modtheo}

The above results can be expressed in terms of the present value
of the Hubble
parameter $H_0=(8\pi G\rho_0/3)^{1/2}=2.195\times 10^{-18}\, {\rm s}^{-1}$
instead of the cosmological constant $\Lambda$ by
using the relation $\rho_{\Lambda}=\Omega_{\Lambda,0}\rho_0$ (where $\rho_0=8.62\times 10^{-24}\, {\rm g\, m^{-3}}$ is the present density of the Universe and $\Omega_{\Lambda,0}=0.6911$ is the present fraction of DE) giving 
\begin{eqnarray}
\label{logomodel15b}
\Lambda c^2=3\Omega_{\Lambda,0}H_0^2=2.07\, H_0^2.
\end{eqnarray}
For example, the universal surface density of DM halos can be written as
$\Sigma_0=0.02815 H_0 c/G$ and the universal gravitational acceleration can be
written as $g=0.0419\, H_0 c$.

The MOND (modification of Newtonian dynamics) theory \cite{mond} has also been advocated to explain the flat rotation curves of the galaxies, the universal value of their surface density, and the Tully-Fisher relation (see Appendix \ref{sec_another}).  This theory involves a transition acceleration $a_0\simeq 1.2\times 10^{-10}\, {\rm m/s^2}$, treated as a fundamental constant of physics, at which the equation of dynamics changes. If we make the  identification  $a_0=g/f_b$ with $f_b\sim 0.17$ (see above), our  relation $g=0.0419\, H_0 c$ {\it derived} from the logotropic model, and giving $a_0=H_0 c/4.06$, may explain why the fundamental
constant $a_0$ that appears in
the MOND theory is of order
$a_0\simeq H_0 c/4=1.65\times
10^{-10}\, {\rm m/s^2}$ (or $a_0\simeq 0.171\, c^2\sqrt{\Lambda}$) in good agreement with the observational
value $a_0^{\rm obs}=(1.3\pm 0.3)\times
10^{-10}\, {\rm m/s^2}$ (see the Remark in Sec. 3.3. of \cite{preouf1} for a more
detailed discussion). We emphasize, however, that our logotropic model is completely different
from the MOND theory.

\subsection{Universalities}

The identities from Eqs. (\ref{lel11})-(\ref{logomodel15}) derived from the logotropic model express the universal surface density of DM halos, their universal gravitational acceleration, the universal NFW cusp, the Tully-Fisher relation, and the MOND acceleration $a_0$ in terms of
the fundamental constants of physics $G$, $c$,  $\hbar$ and $\Lambda$ (or $A$). We
stress that the prefactors are determined by our model so there is no adjustable
parameter in our theory. In this sense, it is
fully predictive. We note that the 
identities from Eqs. (\ref{lel11})-(\ref{logomodel15}), which can be checked by
a
direct numerical application, are interesting in themselves 
even in the case where the logotropic model would
turn out to be wrong.

\subsection{Simple interpolation formula for the logotropic density profile}
\label{sec_interpol}

We can propose a simple analytical interpolation formula for the logotropic density profile under the form
\begin{eqnarray}
\rho\simeq \rho_0\qquad (r\le r_*),
\label{inter1}
\end{eqnarray}
\begin{eqnarray}
\rho\simeq \left (\frac{A}{2\pi G}\right )^{1/2}\frac{1}{r}\qquad (r\ge r_*),
\label{inter2}
\end{eqnarray}
where $r_*=(A/2\pi G\rho_0^2)^{1/2}$ is the radial distance at which the two
pieces of the profile connect 
each other. According
to this approximate formula the surface density is
\begin{equation}
\Sigma_0^{\rm approx}=\rho_0 r_h^{\rm approx}\simeq 4\sqrt{2}\left (\frac{A}{4\pi G}\right )^{1/2}\simeq 129\, M_{\odot}/{\rm pc}^2.
\label{inter3}
\end{equation}
The prefactor $4\sqrt{2}\simeq 5.66$ is close to the exact value $5.85$ from Eq.
(\ref{logomodel3}).

At short distances, the logotropic profile displays a central core like the Burkert profile [see Eq. (\ref{bprof})]. In Eq. (\ref{inter1}) the core is simply approximated  by a constant density $\rho_0$.  This is in contradistinction with the NFW profile [see Eq. (\ref{ba8})], which presents a central $r^{-1}$ cusp. Note that the core of DM halos could also be due to quantum effects as discussed in \cite{logosf} with a logotropic SF model.\footnote{The core could also be due to the presence of baryons.} This leads to a more complicated model involving a quantum core (soliton) that we do not consider here.

At intermediate distances, the logotropic density profile decreases as $r^{-1}$. This is modeled in Eq. (\ref{inter2}) by the asymptotic density profile from Eq. (\ref{logomodel2}).

At large distances, the density profile must decrease more rapidly than $r^{-1}$, for instance as $r^{-2}$ like for the isothermal sphere \cite{chandrabook}, or as $r^{-3}$ like for the Burkert \cite{observations} or NFW \cite{nfw} profiles, or even more rapidly in order to have a finite mass.\footnote{For example, an envelope described by the  King \cite{king} distribution is equivalent to a polytrope of index $n=5/2$ whose density drops to zero at a finite distance identified with the radius of the DM halo \cite{clm1,clm2}. Such a confinement may take into account incomplete violent relaxation or tidal interactions with surrounding galaxies (see, e.g., Appendix B of \cite{modeldmB}).}

Since the logotropic density profile decreases at intermediate distances as $r^{-1}$, precisely like the NFW profile  from Eq. (\ref{ba8}) at short distances, its connection to the NFW profile is straightforward. Therefore, a convenient interpolation formula for the density profile of a DM halo is
\begin{eqnarray}
\rho=\rho_0\qquad (r\le r_*),
\label{inter4}
\end{eqnarray}
\begin{eqnarray}
\rho=\left (\frac{A}{2\pi G}\right )^{1/2}\frac{1}{r(1+r/r_s)^2}\qquad (r\ge r_*).
\label{inter5}
\end{eqnarray}
This is a nice compromise between the Burkert profile, the logotropic profile, and the NFW profile at sufficiently large distance (excluding the cusp).\footnote{More generally, the exponent $2$ in Eq. (\ref{inter5}) could be replaced by some other exponent $\alpha$ (e.g., $\alpha=1$ for the isothermal sphere with $r_s=(k_B T/2\pi Gm)(2\pi G/A)^{1/2}$ in order to recover the asymptotic behavior $\rho\sim k_B T/(2\pi Gmr^2)$ [see Eq. (\ref{kam2})]).} This kind of regularization of the NFW profile to avoid the cusp has been proposed before but the novelty of our approach is to make the connection with logotropes and predict the universal value of $A$ [see Eq. (\ref{logomodel4})].

\begin{figure}[!h]
\begin{center}
\includegraphics[clip,scale=0.3]{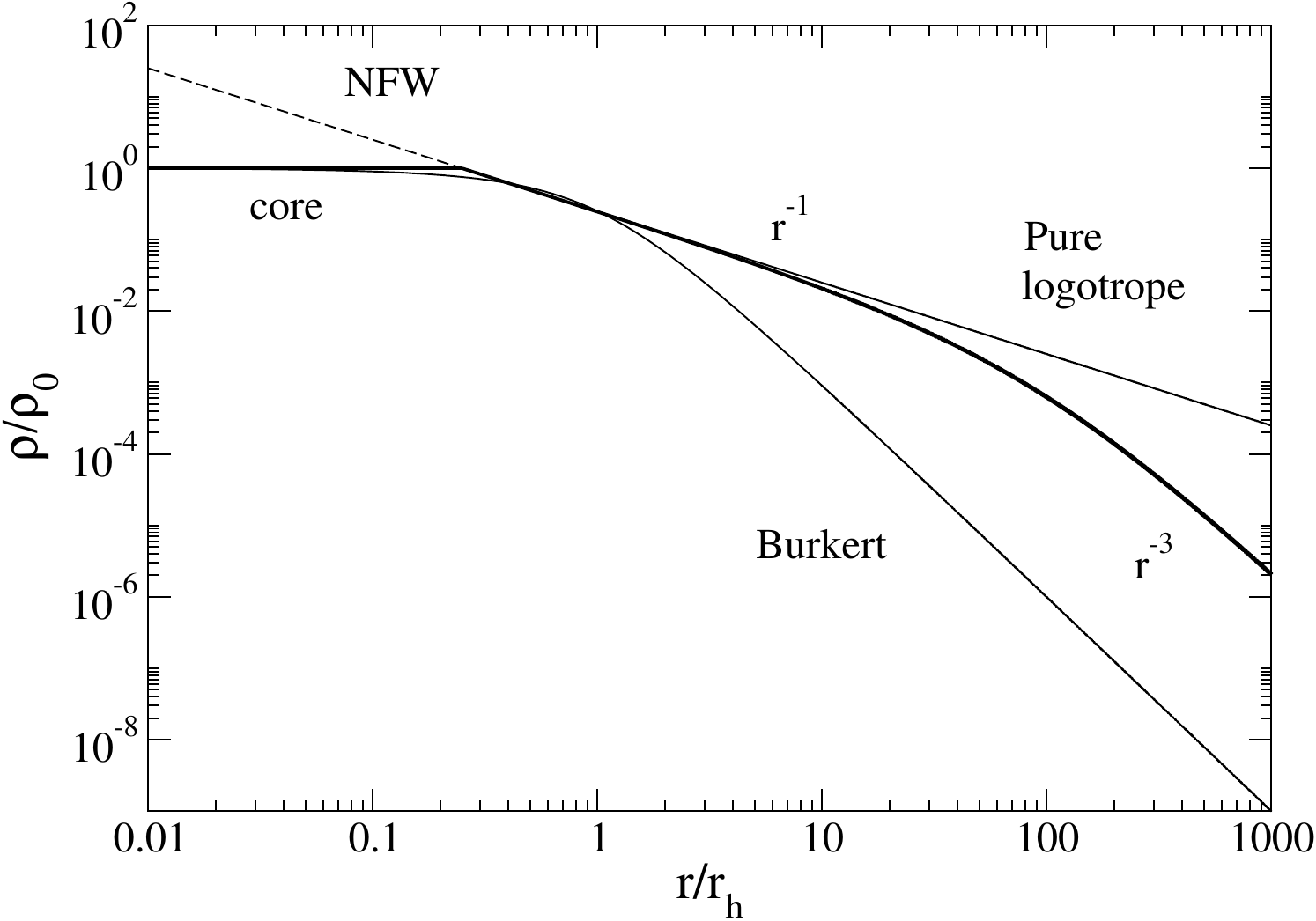}
\caption{Normalized density profile of a DM halo according
to Eqs. (\ref{inter4}) and (\ref{inter5}). It is given by $f(x)=1$ if $x\le 1/4$
and $f(x)=(1+\lambda)^2/[4x(1+\lambda x)^2]$  if $x\ge 1/4$ (approximately),
where $f(x)=\rho(r)/\rho_0$, $x=r/r_h$ and $\lambda=r_h/r_s$. It is compared to
the cored Burkert profile $f_{B}(x)=1/[(1+x)(1+x^2)]$ and to the cuspy NFW
profile $f_{\rm NFW}(x)=(1+\lambda)^2/[4x(1+\lambda x)^2]$ for all $x$. The pure
logotropic profile is approximately given by $f_{L}(x)=1$ if $x\le 1/4$ and
$f_{L}(x)=1/(4x)$ if $x\ge 1/4$. We have taken $\lambda=0.01$ for illustration.}
\label{app}
\end{center}
\end{figure}

\begin{figure}[!h]
\begin{center}
\includegraphics[clip,scale=0.3]{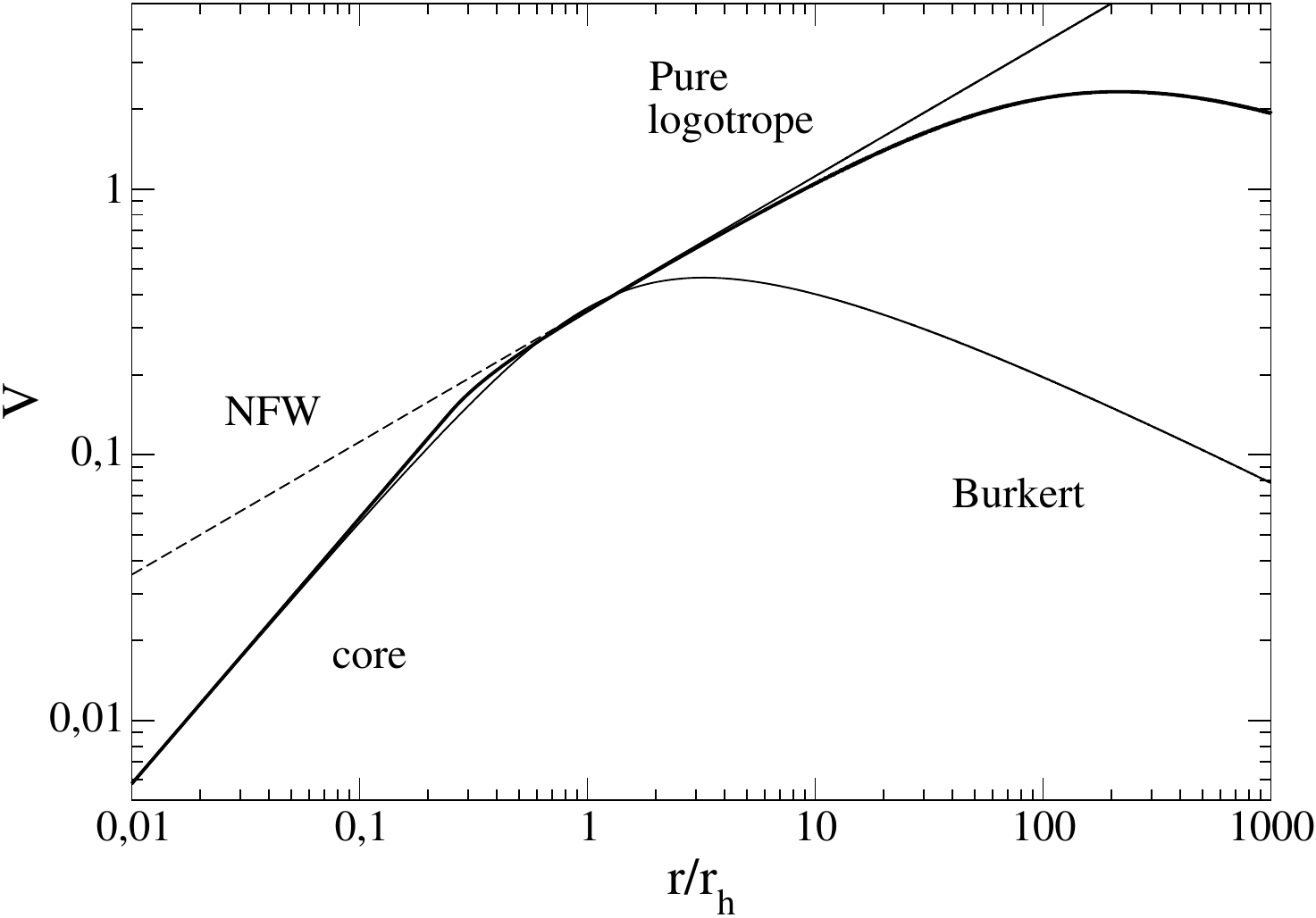}
\caption{Normalized
circular velocity profile of a DM halo according to Eq. (\ref{logomodel12}) with
Eqs. (\ref{inter4}) and (\ref{inter5}). It is given by
$V^2(x)=\frac{1}{x}\int_0^x f(y)y^2\, dy$, where $V(x)=v(r)/\sqrt{4\pi G\rho_0
r_h^2}$. It is compared to the cored Burkert profile
$V_{B}^2(x)=\frac{1}{4x}\lbrack -2\tan^{-1}(x)+2\ln(1+x)+\ln(1+x^2)\rbrack$ and
to the cuspy NFW profile $V_{\rm NFW}^2(x)=\frac{(1+\lambda)^2}{4\lambda^2
x}\lbrack -1+\frac{1}{1+\lambda x}+\ln(1+\lambda x)\rbrack$. The pure logotropic
profile is approximately given by $V_{L}^2(x)=x^2/3$ if $x\le 1/4$ and
$V_{L}^2(x)=\frac{1}{8x}(x^2-\frac{1}{48})$ if $x\ge 1/4$. We have taken
$\lambda=0.01$ for illustration.}
\label{vitesse}
\end{center}
\end{figure}

In Figs. \ref{app} and  \ref{vitesse} we have represented the normalized density profile and the rotation curve of the DM model defined by Eqs. (\ref{inter4}) and (\ref{inter5}). We have also added the Burkert profile and the pure logotropic profile. We see that the Burkert profile is close to the logotropic profile up to a few halo radii $r_h$. In particular,  the Burkert profile has an effective $r^{-1}$ region \cite{burkert2}. Therefore, the logotropic model may justify the empirical Burkert profile in this region where the observations are made. In Figs. \ref{app} and  \ref{vitesse} we have taken a small value of $\lambda=r_h/r_s$ in order to clearly illustrate the three regions where the density scales as $r^{0}$ (core), $r^{-1}$ (logotrope), and $r^{-3}$ (Burkert/NFW). For this unrealistically small value of $\lambda$ the density profile at large distances is higher than the Burkert profile although they both decay as $r^{-3}$. For larger values of $\lambda$ the two profiles approach each other and coincide when $\lambda=1$.

{\it Remark:} As in Sec. \ref{sec_aga} it may be useful to express the results in terms of $g_{\infty}$ instead of $A$. Using Eq. (\ref{aga6}) we find that Eqs. (\ref{inter4}) and (\ref{inter5}) are replaced by
\begin{eqnarray}
\rho=\rho_0\qquad (r\le r_*),
\label{inter6}
\end{eqnarray}\begin{eqnarray}
\rho=\frac{g_{\infty}}{2\pi Gr(1+r/r_s)^2}\qquad (r\ge r_*),
\label{inter7}
\end{eqnarray}
with $r_*=g_{\infty}/(2\pi G\rho_0)$. According to this approximate formula the surface density is
\begin{eqnarray}
\Sigma_0^{\rm approx}=\rho_0 r_h^{\rm approx}\simeq \frac{2g_{\infty}}{\pi G}.
\label{inter8}
\end{eqnarray}
The prefactor $0.637$ is close to the exact value $0.658$ from Eq. (\ref{aga6}).

\section{The value of the cosmological constant in terms of the mass of the electron}
\label{sec_value}

In this section, we summarize and complete the results of our previous investigations which provide an accurate (or possibly exact) value of the cosmological constant in terms of the mass of the electron \cite{preouf1,preouf2,ouf,gbi}. We point out the connection with the logotropic model.

\subsection{The Universe}
\label{sec_uni}

In 1917, Einstein \cite{einsteincosmo}  introduced a cosmological constant
$\Lambda$ in his equations of general relativity in order to have a static
Universe. After the discovery of the expansion of the Universe, he considered
the cosmological constant as his ``biggest blunder'' \cite{raif} and
banished it (see a short history of modern cosmology in the introduction of \cite{cosmopoly1}). However, the importance of the
cosmological constant was revived with the discovery of the present accelerating
expansion of the Universe \cite{novae1,novae2,novae3,novae4}. The cosmological
constant could be the
source of the DE responsible for this acceleration. We may therefore
regard the cosmological constant $\Lambda$ as a fundamental constant of physics.

By using general arguments based on physical considerations and dimensional analysis, we
can introduce cosmological scales.\footnote{The following results
can be derived from the Friedmann equations \cite{oufsuite} in the framework of the $\Lambda$CDM model by using the fact
that the
present
density of the Universe is of the order of the cosmological density on
account of the cosmic coincidence.} The cosmological density\footnote{In this section, we define cosmological scales without any prefactor. As a result, the cosmological density from Eq. (\ref{uni1}) differs from the exact one from Eq. (\ref{logomodel4j}) by a factor $8\pi$.}
\begin{equation}
\rho_{\Lambda}=\frac{\Lambda c^2}{G}=1.50\times 10^{-22}\, {\rm g\, m^{-3}}
\label{uni1}
\end{equation}
is of the order of
the present density of the Universe, the cosmological time 
\begin{equation}
t_{\Lambda}=
\frac{1}{\sqrt{G\rho_{\Lambda}}}=\frac{1}{c\sqrt{\Lambda}}=3.16\times 10^{17}\, {\rm s}
\label{uni2}
\end{equation}
is
of the order of
the age of the Universe, the cosmological length  
\begin{equation}
R_{\Lambda}= c
t_{\Lambda}=\frac{1}{\sqrt{\Lambda}}=9.49\times 10^{25}\, {\rm m}
\label{uni3}
\end{equation}
is of the
order of the present size of the visible Universe (the distance travelled by a photon on
a timescale $t_{\Lambda}$), and the cosmological mass
\begin{equation}
M_{\Lambda}=\rho_{\Lambda}R_{\Lambda}^3=\frac{c^2}{G\sqrt{\Lambda}}=1.28\times 10^{56}\, {\rm g}
\label{uni4}
\end{equation}
is of the order of the present mass of the Universe. In astronomical units,
$t_{\Lambda}=10.0\, {\rm Gyrs}$,
$R_{\Lambda}=3.07\, {\rm
Gpc}$, and 
$M_{\Lambda}=6.42\times 10^{22}\, M_{\odot}$.  We note that the
cosmological
scales satisfy the relation
$R_{\Lambda}= GM_{\Lambda}/c^2$, which is similar to the Schwarzschild
radius of an object of mass $M_{\Lambda}$. This
suggests that the Universe may be a huge black hole inside which we live (this
idea is further
developed in \cite{oufsuite}).  Using the foregoing relations, we find that the surface density of the present Universe is 
\begin{equation}
\Sigma_\Lambda=\frac{M_{\Lambda}}{R_{\Lambda}^2}=\frac{c^2\sqrt{\Lambda}}{G}=6800\, M_{\odot}/{\rm pc}^2.
\label{uni5}
\end{equation}
We note that the typical scale of the universal surface density $\Sigma_0$ of DM halos predicted by the logotropic model [see Eq. (\ref{logomodel5})] is given by the surface density of the Universe $\Sigma_\Lambda$. More precisely,
\begin{equation}
\Sigma_0=0.01955\, \Sigma_\Lambda.
\label{uni6}
\end{equation}

\subsection{The electron}
\label{sec_ele}

The classical radius $r_e$ of the electron is defined  through the relation
\begin{eqnarray}
\label{el1}
m_e c^2=\frac{e^2}{r_e},
\end{eqnarray}
where $m_e$ is its mass and $-e$ is its charge. This equation expresses the equality (in order of magnitude) between the
rest-mass energy of the
electron and its electrostatic energy, assuming that the electron is a little sphere. This is a convenient manner to define the
classical ``radius'' of the electron. This relation first appeared in the Abraham-Lorentz
\cite{abraham,lorentz} model of the extended electron and later in the Born-Infeld
\cite{born1933,borninfeld} theory of nonlinear electrodynamics. In these theories, the mass of the electron has an electromagnetic origin (see Appendix F of \cite{massmaxrel} for a short
review of these old theories).  Recalling the value of the charge of the electron
$e=4.80\times 10^{-13}\, {\rm
g^{1/2}\, m^{3/2}\, s^{-1}}$ and its mass $m_e=9.11\times 10^{-28}\, {\rm
g}=0.511\, {\rm MeV/c^2}$,
we obtain
\begin{eqnarray}
\label{el2}
r_e=\frac{e^2}{m_e c^2}=2.82\times 10^{-15}\, {\rm m}.
\end{eqnarray}
The Compton wavelength of the electron is
$\lambda_e=\hbar/(m_e c)=3.86\times 10^{-13}\, {\rm m}$. It is related to the
classical radius of the electron by
\begin{eqnarray}
\label{el3}
\lambda_e=\frac{r_e}{\alpha}\simeq 137\, r_e,
\end{eqnarray}
where
\begin{eqnarray}
\label{el4}
\alpha=\frac{e^2}{\hbar c}\simeq
\frac{1}{137}\simeq
7.30\times 10^{-3}
\end{eqnarray}
is Sommerfeld's fine-structure constant \cite{sommerfeld1}. Using the foregoing relations, the typical surface density of the electron is
\begin{eqnarray}
\label{el5}
\Sigma_e=\frac{m_e}{r_e^2}=\frac{m_e^3c^2}{\alpha^2\hbar^2}=\frac{\alpha\hbar}{r_e^3 c}=54.9\,
M_{\odot}/{\rm pc^2}.
\end{eqnarray}

\subsection{Qualitative Eddington relation}

In previous works \cite{preouf1,preouf2,ouf} we observed that the surface density $\Sigma_0=133\,
M_{\odot}/{\rm pc}^2$ of DM halos (and
the surface density of the
Universe $\Sigma_\Lambda=6800\,
M_{\odot}/{\rm pc}^2$) is of
the same order of magnitude as the surface density of the electron $\Sigma_e=54.9\, M_{\odot}/{\rm pc^2}$. This coincidence is amazing. Comparatively, the volume density of the electron and the volume density of the Universe differ by $40$ orders of magnitude \cite{ouf}. Writing $\Sigma_e\sim \Sigma_\Lambda$ and using Eqs. (\ref{uni5}) and (\ref{el5}) leads to a relation between the cosmological constant and the mass of the electron (and the other fundamental constants of physics) \cite{preouf1,preouf2,ouf}: 
\begin{equation}
m_e\sim \left (\frac{\Lambda
\hbar^4}{G^2}\right )^{1/6},\qquad \Lambda\sim \frac{G^2m_e^6}{\hbar^4}.
\label{qer1}
\end{equation}
This is the so-called qualitative Eddington relation, which was originally obtained in a different manner \cite{eddington1931lambda} (see \cite{ouf} for a short review).\footnote{In his paper, Eddington \cite{eddington1931lambda} obtained $\Lambda=(m_e/\alpha\hbar)^4(2Gm_p/\pi)^2=9.92\times 10^{-51}\, {\rm m}^{-2}$. It differs from the observational value $\Lambda^{\rm obs}=1.11\times 10^{-52}\, {\rm m^{-2}}$ by a factor $100$. We just give here the scaling of this relation $\Lambda\sim G^2m_e^6/\hbar^4\sim 2.06\times 10^{-65}\, {\rm m}^{-2}$ treating $\mu=m_p/m_e\simeq 1836$ and $\alpha=e^2/\hbar c\simeq 1/137\simeq 7.30\times 10^{-3}$ as quantities of order unity \cite{ouf}. The qualitative Eddington relation differs from  the observational value by a factor $10^{13}$. We note that the speed of light $c$ does not appear in the Eddington relation but it could be introduced in the definition of the cosmological constant (it is just a scaling factor). In this sense, the Eddington relation involves all the fundamental constants of physics.} Using the relation $H_0^2\sim\Lambda c^2$ [see Eq. (\ref{logomodel15b})] which is based on the cosmic coincidence we get
\begin{eqnarray}
m_e\sim \left (\frac{H_0\hbar^2}{Gc}\right )^{1/3},\qquad H_0\sim \frac{m_e^3 Gc}{\hbar^2}.
\label{wein}
\end{eqnarray}
This returns the empirical Weinberg relation (again in a qualitative sense) \cite{weinbergbook}. The
mysterious Eddington and Weinberg relations 
provide an intriguing connection between microphysics and macrophysics (i.e.
between atomic physics and cosmology) which is
further discussed in \cite{preouf1,preouf2,ouf,gbi,oufsuite}. Curiously, the Weinberg 
relation (\ref{wein}) involves the {\it present} value of the Hubble constant suggesting that this 
relation is valid only {\it now}, i.e., that our epoch is particular. This fact may be explained in terms of the anthropic principle \cite{dicke}. Then, by identifying the Eddington and the Weinberg relations (considered as two physically distinct relations), we can explain the observed cosmic coincidence $H_0^2\sim\Lambda c^2$ (see Appendix D of \cite{ouf} for details).

\subsection{Zeldovich relation}

Lema\^itre \cite{lemaitre1934} was the first to understand
that the effect of the cosmological constant $\Lambda$ is
equivalent to that of a fluid with a constant density
$\rho=\rho_\Lambda=\Lambda c^2/8\pi G$ described by an
equation of state $P=-\rho c^2$. He interpreted $\rho_{\Lambda}c^2$ as
the vacuum energy density. However, he did not connect his interpretation with
the zero-point energy, nor relate it to quantum mechanics.\footnote{He also
mentioned incorrectly that the density of vacuum is negative and that its
pressure is positive.} The origin of the vacuum energy was
later discussed by Zeldovich \cite{zeldovich,zeldovichA} and Sakharov \cite{sakharov}
in relation to quantum field theory. When one tries to compute the vacuum energy density $\rho_{\Lambda}c^2$
from first principles, one encounters a severe problem of divergence at small scales (UV divergence) \cite{martin}. It is therefore necessary to introduce, one way or the other, a regularization at some minimum length $l_{\rm min}$ (see below).

By considering the gravitational interaction energy
between virtual pairs of the quantum electrodynamic vacuum, Zeldovich 
\cite{zeldovich,zeldovichA} obtained a qualitative formula for the vacuum 
energy density
\begin{equation}
\rho_{\Lambda}c^2\sim \frac{Gm^2}{\lambda_C}\times \frac{1}{\lambda_C^3},
\label{zel1}
\end{equation}
where $\lambda_C=\hbar/(mc)$ is the Compton wavelength of the particle of mass $m$ involved in the pair. This expression assumes that the vacuum contains virtual pairs of particles with an effective density $n\sim
1/\lambda_C^3$ and that these pairs have a gravitational energy of interaction
$Gm^2/\lambda_C$. Zeldovich therefore wrote the vacuum energy density under the form\footnote{Zeldovich proposed another formula for the vacuum energy density, which is discussed in \cite{ouf,gbi}.}
\begin{equation}
\label{zel2}
\rho_\Lambda c^2\sim \frac{G m^6c^4}{\hbar^4}\sim
\frac{G\hbar^2}{c^2\lambda_C^6}.
\end{equation}
We note that this expression of the
vacuum density explicitly involves the gravitational constant $G$. Therefore,
it may be related to a theory of quantum gravity. In
his original papers, Zeldovich \cite{zeldovich,zeldovichA}  used Eq.
(\ref{zel2}) with the proton
mass $m_p=1.67\times 10^{-24}\, {\rm g}$ and obtained a discrepancy of $7$ orders of magnitude with the empirical
cosmological density $\rho_{\Lambda}^{\rm obs}=5.96\times 10^{-24}\, {\rm g\, m^{-3}}$. If we use the Planck mass $M_P=(\hbar c/G)^{1/2}=2.18\times 10^{-5}\, {\rm g}$ and
the Planck length $l_P=(G\hbar/c^3)^{1/2}=1.62\times
10^{-35}\, {\rm m}$ in Eq. (\ref{zel2}), we find that $\rho_{\Lambda}\sim \rho_P$,
yielding a discrepancy of $120$ orders of magnitude with the empirical
value of the cosmological constant. This is the usual  cosmological constant problem \cite{weinbergcosmo,paddycosmo}. If we use the electron mass $m_e$ 
we recover the qualitative Eddington relation (\ref{qer1}) which provides a better prediction of the cosmological constant.

To obtain a more accurate expression of the cosmological constant we shall write the Zeldovich relation under the form
\begin{equation}
\label{zel3}
\Lambda=\frac{G^2\hbar^2}{c^6 l_{\rm min}^6},
\end{equation}
where $l_{\rm min}$  represents a relevant small scale cut-off (minimum length) to be determined.

\subsection{Refined Eddington relation}

If we take for $l_{\rm min}$ the classical radius of the electron from Eq. (\ref{el2}):
\begin{equation}
l_{\rm min}=r_e=\frac{\alpha\hbar}{m_e c},
\label{rer1alive}
\end{equation}
we obtain \cite{preouf1,preouf2,ouf,gbi}
\begin{equation}
\Lambda=\frac{G^2\hbar^2}{r_e^6c^6}=\frac{G^2m_e^6}{\alpha^6\hbar^4}=1.36\times
10^{-52}\, {\rm m^{-2}}
\label{rer1}
\end{equation}
or, using Eq. (\ref{logomodel4j}),
\begin{equation}
\rho_\Lambda=\frac{G\hbar^2}{8\pi r_e^6c^4}=\frac{Gm_e^6c^2}{8\pi\alpha^6\hbar^4}=7.29\times
10^{-24}\, {\rm g m^{-3}}.
\label{rer1bis}
\end{equation}
Eq. (\ref{rer1}) can be viewed as a refined or accurate form of the Eddington
relation. It gives a remarkable agreement with the empirical value of the
cosmological constant $\Lambda^{\rm obs}=1.11\times 10^{-52}\, {\rm m^{-2}}$ (or
$\rho_{\Lambda}^{\rm obs}=5.96\times 10^{-24}\, {\rm g\, m^{-3}}$). It would be
desirable to know if this formula is {\it exact} or just a good approximate
relation (possibly the leading term in an expansion of $\Lambda$ in powers of
$\alpha$). If it is exact, it provides the value of the cosmological constant in
terms of the electron mass which is known very precisely. We note that,
according to Eq. (\ref{rer1}), the cosmological constant can be written as
$\Lambda=l_P^4/r_e^6$ instead of $\Lambda=1/l_P^2$. They differ by a factor
$(l_P/r_e)^6\sim 10^{-120}$ \cite{ouf}, which solves the cosmological constant
problem. Eq. (\ref{rer1}) 
is certainly one of the most beautiful and intriguing formulas of physics as it
connects microphysics and macrophysics\footnote{It relates the
cosmological
constant $\Lambda$, or equivalently the size $R$ of the visible
universe ($\Lambda\sim 1/R^2$), to the mass $m_e$ of the electron.} and remains
largely mysterious and enigmatic.

Conversely, we can express the mass and the radius of the electron in terms of the cosmological constant as
\begin{equation}
m_e= \alpha \left (\frac{\Lambda
\hbar^4}{G^2}\right )^{1/6}= \alpha
(m_{\Lambda}M_P^2)^{1/3},
\label{rer2}
\end{equation}
\begin{equation}
r_e= \left (\frac{G^2\hbar^2}{\Lambda c^6}\right
)^{1/6}=
(R_{\Lambda}l_P^2)^{1/3}.
\label{rer3}
\end{equation}
To get the second equalities, we have introduced the cosmon mass $m_{\Lambda}=\hbar\sqrt{\Lambda}/c=3.71\times 10^{-66}\, {\rm g}=2.08\times 10^{-33},{\rm eV/c^2}$, which is the Compton mass associated with the radius of the Universe $R_{\Lambda}$ \cite{ouf}. As discussed in \cite{ouf}, the cosmon is the particle with the smallest mass in the Universe (it represents the quantum of mass and entropy). It could be the particle of DE.\footnote{In terms of the cosmon mass $m_{\Lambda}=\hbar\sqrt{\Lambda}/c$, the fundamental constant $A\sim c^4\Lambda/G$ of the logotropic model [see Eq. (\ref{logomodel8})] can be rewritten as $A\sim \frac{c^4}{G}(m_{\Lambda}c/\hbar)^2$. Introducing the ``vacuum temperature'' $T_{\Lambda}\sim m_{\Lambda}c^2/k_B\sim \hbar c\sqrt{\Lambda}/k_B=2.41\times 10^{-29}\, {\rm K}$ (the smallest temperature in the Universe), which is equal to the rest mass of the cosmon, and which can be related to the Gibbons-Hawking \cite{gibbonshawking} temperature $k_B T_H\sim \hbar H_0$ with $H_0\sim c\sqrt{\Lambda}$ and to the Unruh \cite{unruh} temperature $k_B T_U\sim g \hbar/c$ with $g\sim G\Sigma\sim c^2\sqrt{\Lambda}$ (see \cite{ouf} for details), we get $A\sim \frac{c^2}{G\hbar^2}(k_B T_{\Lambda})^2$. This is another manner to see that $A$ plays the role of a generalized temperature.}

Using the accurate Eddington relation (\ref{rer1}) and Eqs. (\ref{logomodel5}), (\ref{uni5}) and (\ref{el5}) we find that
\begin{eqnarray}
\label{rer4}
\Sigma_e=
\alpha\frac{c^2\sqrt{\Lambda}}{G}= \alpha\Sigma_\Lambda= 0.373\,
\Sigma_0.
\end{eqnarray}
This relation justifies the close equality between the surface density of DM
halos and the surface density of the electron. The universality of the surface
density $\Sigma$ of these objects and the Eddington relation from Eq.
(\ref{rer1})  may be related to 
the holographic principle stating that the entropy is proportional to the
surface not to the volume (see \cite{preouf1,ouf} for details). Using Eqs.
(\ref{logomodel5}), (\ref{el5}) and (\ref{rer4}) we can also express the
universal surface density of DM halos in terms of the mass or classical radius
of the electron as 
\begin{equation}
\Sigma_0=0.01955\, \frac{c^2m_e^3}{\alpha^3\hbar^2}=0.01955\, \frac{\hbar}{r_e^3 c}=147\, M_{\odot}/{\rm pc}^2.
\label{rer5}
\end{equation}
If Eq. (\ref{rer1}) is exact, this expression provides the exact value of the surface density of DM halos in the logotropic model. Similarly, the exact gravitational acceleration is
\begin{equation}
g=0.0291\, \frac{Gc^2m_e^3}{\alpha^3\hbar^2}=0.0291\, \frac{G\hbar}{r_e^3 c}=3.05\times 10^{-11}\, {\rm m\, s^{-2}}.
\label{rer5gil}
\end{equation}
These results are in very good agreement with the observational values $\Sigma_0^{\rm
obs}=141_{-52}^{+83}\,
M_{\odot}/{\rm pc}^2$ and 
$g_{\rm obs}=3.2_{-1.2}^{+1.8}\times 10^{-11}\, {\rm
m/s^2}$.

We now have to justify why we should take the classical electron radius $r_e$ as the minimum length $l_{\rm min}$ in the Zeldovich relation (\ref{zel3}).

\subsection{Karolyhazy relation}

Some authors have tried to determine the limit on the measurability of spacetime distances in quantum gravity or the minimum uncertainty of spacetime geodesics. This limit arises when taking into account the quantum properties of devices used for measurement. According to Karolyhazy \cite{karolyhazy}, the minimum uncertainty in the measure of the length of an object of size $l$ due to quantum fluctuations is given by\footnote{Other approaches leading to different results are discussed in \cite{ouf}.}
\begin{equation}
\delta l\sim (l l_P^2)^{1/3}.
\label{ka1}
\end{equation}
This is the
condition that the device used to make the measurement does not turn into a
black hole. This expression was later related to the holographic principle
and to the theory of quantum information. The minimum total uncertainty in the
measurement of a length equal to the
size of the Universe  ($l\sim R_{\Lambda}$), which is a
consequence of
combining the principles of quantum mechanics and general relativity, is given
by
\begin{equation}
l_{\rm min}=(R_{\Lambda}l_P^2)^{1/3}.
\label{ka2}
\end{equation}
It corresponds to Eq. (\ref{rer3}), which results from the  refined Eddington relation (\ref{rer1}) \cite{ouf}. This shows that the minimum length is equal to the classical radius of the electron: $l_{\rm min}=r_e=2.82\times 10^{-15}\, {\rm m}$.  This minimum length is much larger than
the Planck length $l_P=1.62\times
10^{-35}\, {\rm m}$ showing
that quantum
gravity is not a Planck scale phenomenon contrary to common belief. Conversely, assuming that the minimum length $l_{\rm min}$ is equal to the classical electron radius $r_e$, Eq. (\ref{ka2}) leads to the refined Eddington relation (\ref{rer1}). As a matter of fact, using Eq. (\ref{uni3}), the Karolyhazy relation (\ref{ka2}) turns out to be equivalent to the Zeldovich relation (\ref{zel3}).

\subsection{Nonlinear electrodynamics}

\begin{figure}[!h]
\begin{center}
\includegraphics[clip,scale=0.3]{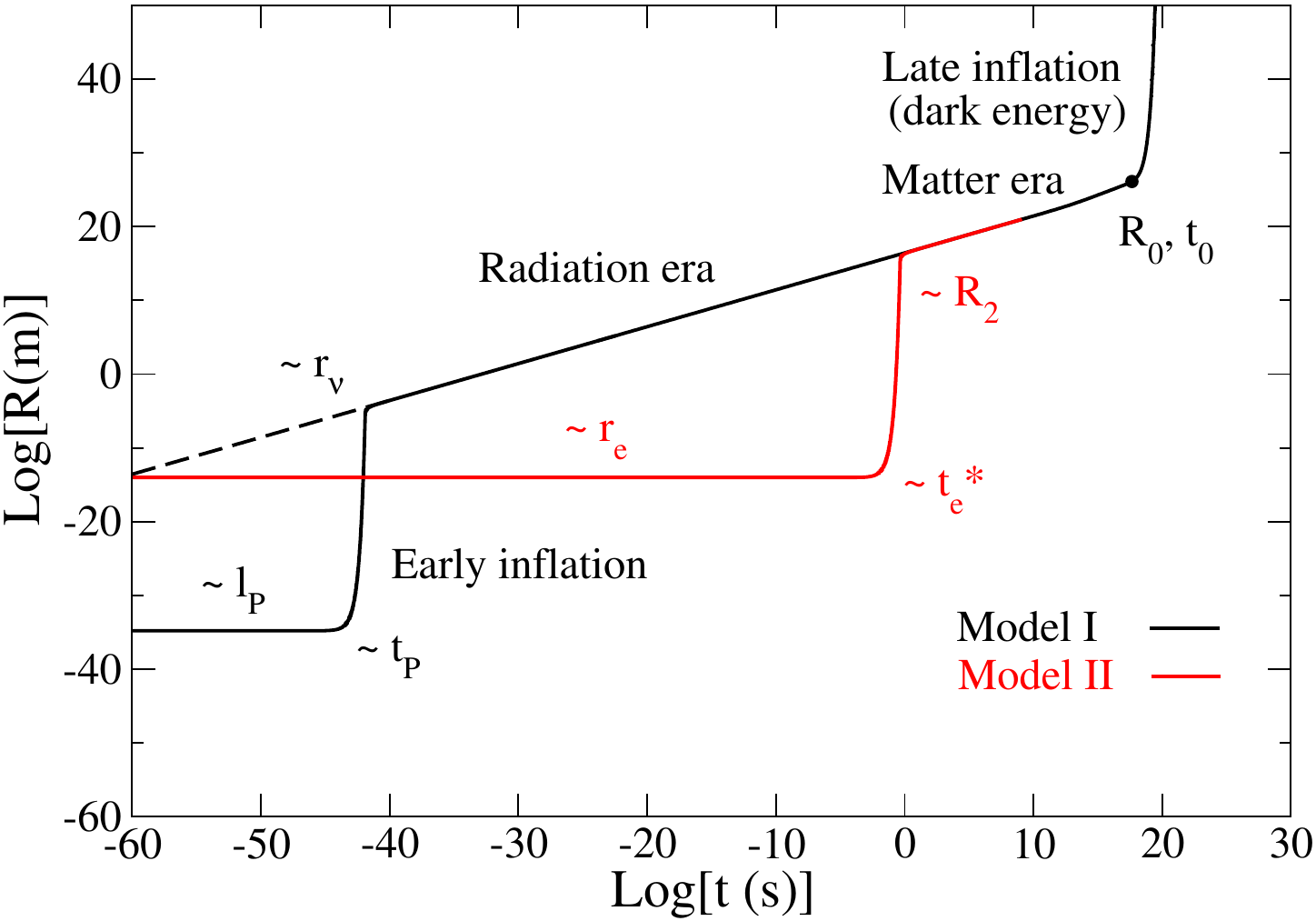}
\caption{Evolution of the scale factor (radius) of the Universe from the early inflation to the late accelerating expansion (dark energy). In between, the Universe undergoes a radiation era and a matter era. This model is based on a quadratic equation of state  \cite{aip,universe,vacuumon} or, equivalently, on nonlinear electrodynamics \cite{gbi}. In Model I, the initial density is identified with the Planck density $\rho_P$. In Model II, the initial density is identified with the electron density $\rho_e$. The final density corresponds to the cosmological density $\rho_\Lambda$.}
\label{cosmo}
\end{center}
\end{figure}

In a recent paper \cite{gbi} we have given a physical 
argument justifying why the radius of the electron represents the minimum
length. We have developed a model of magnetic Universe based on nonlinear
electrodynamics. The basic idea is that in the primordial Universe the magnetic
fields are so intense that linear (Maxwell) electrodynamics is not valid. We
must use a nonlinear model of electrodynamics. The model proposed in \cite{gbi}
naturally leads to a phase of inflation with a constant density $\rho_*$ (de
Sitter era)  instead of the big bang singularity yielding an infinite density at
$t=0$. The question is to know what the primordial density $\rho_*$ is. We have
identified the maximum density of the primordial Universe during the phase of
inflation with the electron density   $\rho_e=m_e/r_e^3=4.07\times 10^{16}\,
{\rm g\, m^{-3}}$, as a consequence of nonlinear electrodynamics, instead of the
more conventional Planck density $\rho_P=M_P/l_P^3=c^5/G^2\hbar=5.16\times
10^{99}\, {\rm g\, m^{-3}}$ usually advocated in inflationary scenarios. This
choice was justified by applying the same Lagrangian of nonlinear
electrodynamics both to the primordial magnetic Universe and  to the electron
(in the spirit of a generalized Born-Infeld model). The adopted
Lagrangian
\begin{equation}
{\cal L}=\frac{-{\cal F}}{1+\frac{\cal F}{\rho_* c^2}},
\end{equation}
where ${\cal F}=(1/4\mu_0)F_{\mu\nu}F^{\mu\nu}$ is the electromagnetic
invariant, depends on a characteristic length $r_*=(e^2/8\pi\rho_* c^2)^{1/4}$
which is zero for linear (Maxwell) electrodynamics but nonzero for nonlinear
electrodynamics. Nonlinear electrodynamics prevents the divergence of the
Coulombian potential at the origin, gives a finite electromagnetic energy ${\cal
E}=16 e^2/15 r_*$ for the electron, and allows us to define the mass $m_e$ of
the electron through the relation ${\cal E}=m_e c^2$. We then obtain
$r_*=16e^2/(15m_e c^2)=(16/15)r_e$ showing that $r_*$
represents the electron radius. As a result, the density
$\rho_*=e^2/(8\pi r_*^4c^2)$ associated with $r_*$ that occurs in the
electromagnetic Lagrangian is the electron density $\rho_e$, not the Planck
density $\rho_P$. When this Lagrangian is applied to a cosmological context
(magnetic Universe), we have shown that the Universe ``starts'' at $t=0$ with a
radius  $R(0)$ equal to the classical radius of the electron \cite{gbi}:
\begin{equation}
R(0)\sim r_e,
\label{rore}
\end{equation}
instead of being zero (big bang singularity) in the context of Maxwell's linear electrodynamics or instead of being equal to the Planck length $l_P$ in conventional inflationary scenarios (see Fig. \ref{cosmo}). Therefore, this theory provides a physical argument why the radius of the electron represents a minimum length scale as it corresponds to the original size of the Universe. It also confirms that quantum gravity, prevailing in the early Universe, is not a Planck scale phenomenon.

\subsection{Justification of the minimum length}

Combining the above results, we conclude that the minimum length $l_{\rm min}$ in the
Zeldovich relation leading to the correct value of the cosmological constant corresponds to:

(i) The classical radius of the electron [Eq. (\ref{el2})].

(ii) The limit on the measurability of spacetime distances [Eq. (\ref{ka2})]
in quantum gravity according to Karolyhazy \cite{karolyhazy}.

(iii) The radius of the Universe at the beginning of the inflation in a model of nonlinear electrodynamics [Eq. (\ref{rore})].  
This radius defines
an absolute ``minimum'' length $r_e=2.82\times 10^{-15}\, {\rm
m}$ which could  replace the Planck length $l_P=1.62\times
10^{-35}\, {\rm m}$ and yield the correct value of
the vacuum energy density when substituted into Eq. (\ref{zel3}).

If we identify the minimum length $l_{\rm min}$ in Zeldovich's relation (\ref{zel3}) with the classical radius
of the electron $r_e$,   we can express the cosmological constant in terms
of $r_e$, $\lambda_e$ or $m_e$ [see Eq. (\ref{rer1})]. This provides a refined Eddington relation \cite{preouf1,preouf2,ouf,gbi}. By using the empirical value of the mass of the electron, we can explain the measured value of $\Lambda$ or even predict its value very accurately (since the mass of the electron is known very precisely) if Eq. (\ref{rer1}) is exact. Alternatively, we can express the radius $r_e$, the Compton wavelength $\lambda_e$ and the mass $m_e$ of the electron as a function of the cosmological constant [see Eqs. (\ref{rer2}) and (\ref{rer3})]  and justify the mysterious Eddington and Weinberg relations \cite{preouf1,preouf2,ouf,gbi}.

{\it Remark:} Recently, by studying the early history of quantum mechanics \cite{kgdiss}, we came across the papers of Flint and Wilson  \cite{flint45,flint90,flintwilson} who also introduced the notion of a minimum length related to the ``size'' of the electron.  Flint \cite{flint45,flint90}  (see also March \cite{march1,march2}) showed that it is impossible that two electric charges $e$ lie closer together than a distance of the order $r_e=e^2/m_e c^2$, the classical electron radius. A charge $e$ can be regarded as surrounded by a region of linear dimension $r_e$ which is the exclusive property of the charge. It would have no meaning to distinguish between points lying within this region.  The electron radius in electromagnetism is a bit similar to the Schwarzschild radius in gravity.  Below that length, our customary conception of space and time breaks down.  Flint and Wilson \cite{flintwilson} argued that the length $x^5$ in the fifth dimension of the Kaluza-Klein theory can  only take the quantized values $nh/m_e c$ with $n=1,2,...$. Therefore, the classical electron radius $r_e=e^2/m_e c^2$ or the Compton wavelength of the electron $\lambda_e=h/m_e c$ may be identified with a fundamental minimum length. The corresponding time $t_e=r_e/c=e^2/m_e c^3=9.40\times 10^{-24}\, {\rm s}$ or $t_e=\lambda_e/c=h/m_e c^2$, which coincides with the inverse of the de Broglie proper pulsation $\omega_0=m_e c^2/\hbar$ \cite{debroglie1923b,dbphil,broglie25,brogliethese}, can be regarded as representing the smallest interval of time which has any significance in the physical world  \cite{flint45,flint90}. This is a quantum of time\footnote{This quantum of time was later called the ``chronon'' by Caldirola \cite{chronon}. This is the time it takes for a light wave to travel across the ``size'' of an electron. This timescale first appeared in the Abraham-Lorentz \cite{abraham,lorentz} theory of the extended electron when they tried to calculate the recoil force on an accelerated charged particle (e.g. an electron) caused by the particle emitting electromagnetic radiation. This is also the unit of time provided by the atomic constants that  Dirac used in his cosmological theory based on a large number hypothesis \cite{dirac1,dirac2,ouf}.  It can be written as $t_e=\alpha\hbar/(m_e c^2)$. This timescale is also connected to the flight time in the relativistic extension of Nelson's stochastic quantum mechanics \cite{nelson} developed by  Lehr and Park \cite{lp}.}  and $dt$ can only change by integral multiples of this quantity: $t\sim n t_e=n h/m_e c^2$. Similarly, the length is quantized by integral multiples of the Compton wavelength of the electron: $l\sim n\lambda_e=n h/m_e c$ \cite{flint630,flintrichardson} (see also \cite{ff}).  In this sense, the classical radius of the electron (possibly with some prefactor of order unity) may be interpreted as a quantum of length. Interestingly, this returns the fundamental minimum length $l_{\rm min}=r_e=2.82\times 10^{-15}\, {\rm m}$ obtained in \cite{ouf,gbi} from different arguments and allowing us to establish the refined Eddington relation (\ref{rer1}) between the cosmological constant $\Lambda$ and the mass  $m_e$ of the electron.

\section{Conclusion}

In a previous paper \cite{epjp}, we asked the question: ``Is the Universe logotropic?'' Although there is no definite answer to  that question for the moment, the present paper assembles several pieces of evidence towards the possibility that the logotropic distribution plays an important role in the Universe.

We know since the 19th century through the works of  Maxwell and Boltzmann that collisional systems of particles such as neutral gases achieve a universal equilibrium distribution: the isothermal (or Maxwell-Boltzmann) DF $f_B=A\, e^{-\beta m\epsilon}$. This is the statistical equilibrium state of a system resulting from a collisional evolution. It corresponds to the most probable macrostate, i.e., the one that is the most represented at the microscopic level. In this sense, it maximizes the Boltzmann entropy at fixed mass and energy. This result is confirmed by the Boltzmann kinetic equation that admits 
an $H$-theorem for the Boltzmann entropy: $\dot S_B\ge 0$ with an equality if, and only if, $f$ is the Maxwell-Boltzmann distribution. These results extend to Coulombian plasmas whose ``collisional'' evolution is described by the Landau equation (or more generally by the  Lenard-Balescu equation), which can be derived from the Boltzmann equation in the limit of weak deflections.\footnote{The statistical mechanics and kinetic theory of stellar systems are more complicated because of the spatial inhomogeneity of the system and the problems of evaporation and gravothermal catastrophe \cite{aakin}.}

For collisionless systems with long-range interactions the situation is much more complicated. The Boltzmann entropy (which is a particular Casimir functional) is conserved by the Vlasov equation. However, it can increase on a coarse-grained scale. A statistical theory of violent collisionless relaxation has been proposed by Lynden-Bell \cite{lb}. In the single level case, it leads to a Fermi-Dirac-like DF, which reduces to a Maxwell-Boltzmann-like DF in the dilute (nondegenerate) limit. In the multilevel case, the Lynden-Bell theory leads to a larger class of DFs $\overline{f}$, which has been investigated in \cite{superstat}.  Recent arguments \cite{ewart1,ewart2} suggest that a ``universal'' DF may be singled out of all these possible DFs. This DF decreases for large energies as $f\sim \epsilon^{-(d+2)/2}$. In this paper, we have noticed that this asymptotic behavior precisely corresponds to the decay law of  the logotropic distribution 
$f_L=A_*/(\epsilon_m+\epsilon)^{d/2+1}$ which is a degenerate polytrope of index
$n=-1$ \cite{cslogo}.\footnote{More generally, we have shown that the Tsallis
(polytropic) distributions $f_T=A(\epsilon_m\pm \epsilon)^{n-d/2}$ of index $n$
could be justified from the Lynden-Bell theory of violent relaxation if the
phase levels follow a 
$\chi$-squared distribution.} The same asymptotic behavior has been obtained in
\cite{banik1,banik2} from the SDD equation introduced in
\cite{epjp1,sdduniverse}. This suggests that the logotropic DF is, for
collisionless systems, the counterpart of the isothermal DF for collisional
systems. In this sense, 
the logotropic temperature $A$ is the counterpart of the ordinary temperature
$T$. We must be careful, however, that the universality of a DF for
collisionless systems on the coarse-grained scale is much less well-established
than the universality of the 
Maxwell-Boltzmann distribution for collisional systems. Indeed, it rests on many assumptions that may not always be fulfilled (see the many other possible DFs discussed in \cite{superstat}).  However, it is interesting to notice the emergence of logotropic (or logotropic-like) DFs in different areas of physics \cite{mlp,cslogo,epjp,lettre,jcap,preouf1,action,logosf,preouf2,ouf,gbi,ewart1,ewart2,banik1,banik2}, which have been discussed in this paper.

If the logotropic distribution is characteristic of collisionless systems with long-range interactions, then it is natural that the Universe (which is essentially collisionless) is logotropic. The logotropic DF could be the ``universal'' statistical equilibrium state of a collisionless systems (with the reserves made previously), just like the isothermal DF is the statistical equilibrium state of a collisional system. In this interpretation, the logotropic temperature $A$ would be a measure of the temperature of the Universe in a generalized sense. It could represent the temperature of vacuum. We suggest that the external fluctuations in the SDD equation leading to the logotropic distribution in a cosmological framework may correspond to vacuum fluctuations. This could explain why the logotropic temperature $A$ is related  to the cosmological constant $\Lambda$ (interpreted as the vacuum energy) [see Eq. (\ref{logomodel8})]  as we have found in our previous papers \cite{epjp,lettre,jcap,preouf1,action,logosf,preouf2,ouf,gbi}.

\begin{table*}[t]
\centering
\begin{tabular}{|c|c|c|c|c|c|c|c|c|}
\hline
$-\infty\le n<-1$ & $n=-1$ & $-1<n<0$ & $0\le n<3$ & $n=3$ & $3<n<5$ & $n=5$ & $5<n\le +\infty$ \\
\hline
$M=\infty$ &  $M=\infty$ &  $M=\infty$ & $M$ finite & $M$ finite & $M$ finite & $M$ finite & $M=\infty$ \\
%\hline
$R=\infty$ &  $R=\infty$ &  $R=\infty$ & $R$ finite & $R$ finite & $R$ finite & $R=\infty$& $R=\infty$\\
 &  & & stable & marginal & unstable & unstable  &\\
\hline
\end{tabular}
\label{table1}
\caption{Structure of polytropes in $d=3$ (we have indicated the stability of gaseous polytropes).}
\end{table*}
\begin{table*}[t]
\centering
\begin{tabular}{|c|c|c|c|c|c|c|c|c|}
\hline
$n=-\infty$ & $-\infty<n<-1$ & $n=-1$ & $-1<n<0$ & $0\le n<+\infty$ & $n=+\infty$ \\
\hline
$M$ finite & $M=\infty$ & $M=\infty$ & $M=\infty$  & $M$ finite & $M$ finite \\
$R=\infty$ &  $R=\infty$ & $R=\infty$ & $R=\infty$ &  $R$ finite & $R=\infty$ \\
marginal & & & & stable  & marginal\\
\hline
\end{tabular}
\label{table2}
\caption{Structure of polytropes in $d=2$ (we have indicated the stability of gaseous polytropes).}
\end{table*}
\begin{table*}[t]
\centering
\begin{tabular}{|c|c|c|c|c|c|c|c|c|}
\hline
$n=-\infty$ & $-\infty< n<-1$ & $n=-1$ & $-1<n<0$ & $0\le n< +\infty$ & $n=+\infty$\\
\hline
$M$ finite & $M$ finite & $M=\infty$ & $M=\infty$  & $M$ finite & $M$ finite \\
$R=\infty$ & $R=\infty$ & $R=\infty$ & $R=\infty$ &  $R$ finite &  $R=\infty$ \\
stable & stable & & & stable  & stable\\
\hline
\end{tabular}
\label{table3}
\caption{Structure of polytropes in $d=1$ (we have indicated the stability of gaseous polytropes).}
\end{table*}

The recent results obtained from statistical mechanics \cite{ewart1,ewart2} (based on the Lynden-Bell theory \cite{lb,superstat})  and from  kinetic theory \cite{banik1,banik2} (based on the SDD equation \cite{epjp1,sdduniverse}) give further support to our logotropic model \cite{epjp,lettre,jcap,preouf1,action,logosf,preouf2,ouf,gbi} (see its original justification in Appendix \ref{sec_just}). When applied to DM halos, DE, and cosmology, this model makes interesting predictions:

(i) It can account for the accelerated expansion of the Universe and is indistinguishable from the $\Lambda$CDM model at the present time. The two models will differ in about $27$ Gyrs when the logotropic model  becomes phantom while the $\Lambda$CDM model has a de Sitter behavior.

(ii)  It explains the universal surface density $\Sigma_0$ of DM halos and
predicts its value [see Eq. (\ref{logomodel5})]  in terms of the fundamental
constants of physics (including the cosmological constant) 
without adjustable parameter. From a related, but
different, viewpoint it accounts for the $r^{-1}$ NFW cusp and predicts the
value of the product $\rho_s r_s$ [see Eq. (\ref{logomodel10})]. These
predictions are in good agreement with the observations.

(iii) It leads to rotation curves of galaxies with a constant surface density
$\Sigma(r)$  or a constant gravitational acceleration $g(r)$ at large distances
and predicts their universal values $\Sigma_{\infty}$ and $g_{\infty}$ [see Eqs.
 (\ref{logomodel15}) and (\ref{aga5})] in agreement with the observations. A
constant surface density or a constant gravitational acceleration leading to a
circular velocity increasing as $v\propto r^{1/2}$, as predicted by
the logotropic model, may be more relevant than a
constant circular velocity predicted by the MOND theory (see Appendix 
\ref{sec_another}). Both theories (logotropic and MOND) lead to the Tully-Fisher relation. In the logotropic model, a test particle (star) feels the Newtonian gravitational force produced by a DM halo and the Tully-Fisher relation is valid at each point of the approximately flat rotation curve [see Eq. (\ref{ano8})]. In the MOND theory there is no DM. A test particle (star) feels the modified gravitational force produced by the baryonic mass. The rotation curve is exactly flat and the Tully-Fisher relation determines the constant value of the circular velocity $v_c$ for a given baryonic mass $M_b$ [see Eq. (\ref{ano3})]. Isothermal DM halos predicted by ordinary statistical mechanics also lead to flat rotation curves but they do not have a universal surface density and do not satisfy the Tully-Fisher relation unless the temperature changes from halo to halo in an appropriate manner (see Appendix 
\ref{sec_another}).

(iv) It leads to a refined relation [see Eq. (\ref{rer1})] 
between the cosmological constant $\Lambda\sim 1/R^2$ (where
$R$ is the size of the visible universe) and the mass $m_e$ of the electron,
which connects microphysics and macrophysics
\cite{ouf}. This relation expresses the universality of the surface density of
various objects including DM halos, the Universe, and the electron. This rather
mysterious universality is likely to be related to a form of holographic
principle \cite{ouf}, which favors the surface density over the volume density.

(v) The logotropic model also predicts that the present ratio between DE and DM is equal to the Euler number $\Omega_{\rm de,0}/\Omega_{\rm dm,0}=e=2.71828...$,\footnote{Of course, the occurrence of the Euler number $e$ is connected to the logarithmic nature of the logotropic equation of state.} which is in the error bars of observations, but this ``prediction''  has a bit to do with ``dark magic'' (see Sec. 3.6 of \cite{preouf2}).

At this day, the status of these intriguing results is still unclear. Either these results are correct, implying that the logotropic model is of fundamental interest in physics, astrophysics and cosmology,\footnote{We have given simple arguments in Appendix  \ref{sec_another} that the logotropic model is {\it inevitable} if the rotation curves of the galaxies are characterized by a constant (universal) gravitational acceleration (or surface density). This property of the rotation curves should therefore be checked very carefully.}  or these results are pure coincidences, and that makes a lot of coincidences!  Indeed, we found that  logotropes arise in different domains of physics and astrophysics: statistical mechanics (Lynden-Bell entropy) and kinetic theory (SDD equation) of collisionless systems with long-range interactions, generalized thermodynamics (Tsallis entropy), DM (rotation curves, NFW cusps, the universal surface density of DM halos, the universal value of the gravitational acceleration, the Tully-Fisher relation...), DE (accelerated expansion of the Universe, refined Eddington relation between the cosmological constant and the mass of the electron, holographic principle...) etc. They could provide an alternative to the MOND theory. We hope that our study will be a step towards a better understanding of our Universe and a partial answer to the question raised at the beginning of this section.

\appendix

\section{Summary of the structure of the polytropes depending on their index $n$ and on the dimension of space $d=1,2,3$}
\label{sec_sumex}

In Tables I-III, we summarize the structure (mass and radius) of the polytropes depending on their index $n$ and on the dimension of space $d=1,2,3$.  We also specify the stability of complete gaseous polytropes (complete stellar polytropes are always stable).

\section{Summary of the main relations between the different exponents}
\label{sec_summary}

In this Appendix, we summarize the main relations between the different exponents obtained in this paper and we discuss specific applications. We restrict ourselves to the dimension of space $d=3$.  

Case (i):  Gaseous polytropes exist for $n>0$ and stellar polytropes exist for $n\ge 3/2$. Regular polytropes (i.e. with a finite central density)  are complete (i.e. bounded) for $n<5$ and incomplete (i.e. unbounded) for $n\ge 5$. Polytropes with $n>5$ have an infinite mass. There is a singular power-law solution for $n>3$.

Case (ii): Gaseous polytropes exist for $n<0$ and stellar polytopes exist for $n<-1$ (logotropes correspond to $n=-1$). They have an infinite mass. There is a singular power-law solution for any $n<0$.

The DF of a stellar polytrope of index $n$ is related to the individual energy $\epsilon$ by  $f\propto (\epsilon_m\mp \epsilon)^{n-3/2}$. The upper sign corresponds to case (i) and the lower sign corresponds to case (ii). The density of a gaseous or stellar polytrope is related to the gravitational potential $\Phi$ by $\rho\propto (\epsilon_m\mp \Phi)^n$.  The singular density profile of a polytrope of index $n$ is $\rho\propto r^{-2n/(n-1)}$.

In the context of the SDD equation, we have shown in Appendix \ref{sec_sol} that the diffusion equation with a diffusion coefficient $D\propto v^{\alpha}$ produces a DF with a power-law tail $f\propto \epsilon^{-(1+\alpha)/2}$ provided that $\alpha>2$. This DF can be associated with a stellar polytrope of index $n=1-\alpha/2$ with $n\le -1$ provided that $\alpha\ge 4$. Conversely, a stellar polytrope of index $n\le -1$ can be obtained from the diffusion (or SDD) equation with a diffusion coefficient velocity exponent $\alpha=2(1-n)$. 

In the context of Lynden-Bell's theory of violent collisionless relaxation, we have seen in Sec. \ref{sec_ew} that a phase space hypervolume $\gamma(\eta)\sim \eta^k$, implying a distribution of phase levels $g(\eta)\sim \eta^{c-1}$ with $c=5/2+k$, produces a coarse-grained DF with a power-law tail $\overline{f}\sim \epsilon^{-(c+1)}$ when $c>0$ (i.e. $k>-5/2$). It can be associated with a stellar polytrope of index $n=1/2-c$ (i.e. $n=-2-k$) with $n\le -1$ provided that $c\ge 3/2$ (i.e. $k\ge -1$). Conversely,  a stellar polytrope of index $n\le -1$ can be obtained from a power-law (or $\chi$-squared) distribution of phase levels with $c=1/2-n$ (i.e. $k=-2-n$).

Let us now consider specific applications of these results.

A NFW cusp $\rho\sim r^{-1}$ corresponds to a polytropic index $n=-1$ (logotrope) yielding   $\rho\propto \Phi^{-1}$ and $f\propto \epsilon^{-5/2}$. It is obtained from the SDD equation with $\alpha=4$ or from the Lynden-Bell theory of violent relaxation with $c=3/2$ (i.e. $k=-1$).

A  NFW cusp $\rho\sim r^{-3/2}$ corresponds to a polytropic index $n=-3$ yielding  $\rho\propto \Phi^{-3}$ and $f\propto \epsilon^{-9/2}$. It is obtained from the SDD equation  with $\alpha=8$ or from the Lynden-Bell theory of violent relaxation with $c=7/2$ (i.e. $k=1$).

A NFW tail $\rho\sim r^{-3}$ corresponds to a polytropic index $n=3$ (critical) yielding  $\rho\propto (-\Phi)^{3}$ and $f\propto (-\epsilon)^{3/2}$.

\section{A new derivation of the logotropic equation of state}
\label{sec_another}

In this Appendix, we provide a new derivation of the logotropic equation of state. We first recall why the observation of (approximately) flat rotation curves and the Tully-Fisher relation force us either to (i) change the laws of physics (MOND theory) or (ii) invoke the existence of DM with very specific properties (universal constant surface density and universal gravitational acceleration). In the second case, we show that the logotropic equation of state emerges naturally and inevitably.

\subsection{Basics of MOND theory: Constant circular velocity}
\label{sec_mond}

Let us consider a test particle of mass $m$ submitted to a central force (e.g. a planet around the Sun or a star in a galaxy). We assume that the particle is on a circular orbit of radius $r$ and that it has a constant angular velocity $\omega$. Its circular velocity is $v=\omega r$ and its centripetal acceleration is 
\begin{eqnarray}
a=\omega^2 r=\frac{v^2}{r}.
\label{ano1}
\end{eqnarray}

According to the Newtonian theory, the fundamental equation of the dynamics is ${\bf F}=m{\bf a}$, where ${\bf F}$ is the force exerted on the particle. The gravitational force created by a possibly extended spherical object of mass $M$ (e.g. the Sun of mass $M_\odot$ or the baryonic mass $M_b$ of a galaxy) is $F=GM m/r^2$. The gravitational acceleration of the test particle and its circular velocity are then
\begin{eqnarray}
a=\frac{GM}{r^2},\qquad v=\left (\frac{GM}{r}\right )^{1/2}.
\label{ano2}
\end{eqnarray}
The circular velocity of the test particle decreases as $r^{-1/2}$ according to Kepler's law. 

This decay law is valid for the planets around the Sun, but not for the stars in a galaxy.

In order to account for this discrepancy at the galactic scale without introducing DM, Milgrom \cite{mond} has proposed a modified gravity theory. According to the MOND theory, the equation of the dynamics ${\bf F}=m{\bf a}$ is not valid for very small accelerations. Milgrom assumes that ${\bf F}=m{\bf a}$ if $a\gg a_0$ and ${\bf F}=m (a/a_0) {\bf a}$ if $a\ll a_0$, where the transition acceleration $a_0$ is a new constant of physics. The force is the same as in the Newtonian theory. The motion of the stars in a galaxy corresponds to the second (peculiar) regime.  As a result, the acceleration and the circular velocity of a star caused by the baryonic mass $M_b$  are
\begin{eqnarray}
\frac{a^2}{a_0}=\frac{GM_b}{r^2},\qquad v^4=GM_b a_0.
\label{ano3}
\end{eqnarray}
In the MOND theory, the circular velocity of a star is constant (implying exactly flat rotation curves) and it satisfies the baryonic Tully-Fisher relation $v\propto M_b^{1/4}$. Note that the acceleration of a star $a\propto 1/r$ is not constant. Using Eq. (\ref{ano3}), one obtains from the observations $a_0\simeq 1.2\times 10^{-10}\, {\rm m/s^2}$ .

\subsection{Logotropic model: Constant acceleration}
\label{sec_ac}

We now assume that the force experienced by the test particle (star) is caused by a DM halo with a density profile $\rho(r)$ and a mass profile $M(r)$. Using the usual Newtonian law ${\bf F}=m{\bf a}$, the acceleration and the circular velocity of the test particle are
\begin{eqnarray}
a=\frac{d\Phi}{dr}=\frac{GM(r)}{r^2},\qquad v^2=\frac{GM(r)}{r}.
\label{ano4}
\end{eqnarray}
In the main part of our paper we have assumed that the DM halos have a logotropic equation of state and we have found that a test particle submitted to the force created by a logotropic DM halo has a constant acceleration $g_{\infty}$ at large distances. The constancy of $a$ is in agreement with the observations \cite{gentile}. Here, we take the reverse viewpoint and assume from the start that a test particle submitted to the force created by a DM halo has a constant acceleration. We thus assume that
\begin{eqnarray}
a=g_{\infty},\qquad v^2=g_{\infty}r.
\label{ano5}
\end{eqnarray}
This means that the test particle experiences a constant force $F=m g_{\infty}$. We then find from Eq. (\ref{ano4}) that the mass profile of the DM halo and the gravitational potential that it creates are given by
\begin{eqnarray}
M(r)=\frac{g_{\infty}}{G}r^2,\qquad \Phi=g_{\infty}r.
\label{ano6}
\end{eqnarray}
Using $dM/dr=4\pi\rho r^2$, or the Poisson equation (\ref{g3}), we find that the DM halo has a density profile
\begin{eqnarray}
\rho=\frac{g_{\infty}}{2\pi Gr}.
\label{ano7}
\end{eqnarray}
On the other hand, according to Eqs. (\ref{ano5}) and (\ref{ano6}), we have the following relation
\begin{eqnarray}
v^4(r)=G g_{\infty}M(r)
\label{ano8}
\end{eqnarray}
between the circular velocity and the mass profile. In the present model, the 
circular velocity is not strictly constant (it increases as
$v\propto r^{1/2}$) but the Tully-Fisher relation is satisfied at each point.
Therefore, instead of imposing a constant acceleration [see Eq. (\ref{ano5})],
or the constancy of the surface density $\langle \Sigma\rangle(r)=M(r)/(\pi
r^2)=g(r)/(\pi G)=g_{\infty}/(\pi G)$ of the DM 
halo [see Eq. (\ref{aga5})], we could have imposed the validity of the
Tully-Fisher relation at each point of the approximately flat rotation curve
[see Eq. (\ref{ano8})]. All these equivalent relations are justified from
observations. It seems that a constant surface density 
at large distances implying $v\propto r^{1/2}$  (as predicted
by the logotropic model) is more relevant than a constant circular velocity (as
predicted by the MOND theory).

The previous equations are valid for a test particle (star) submitted only to the gravity of the DM halo. If we now consider a ``fluid'' particle of DM at equilibrium we must take into account the self-gravity of the DM halo and the pressure force acting on the fluid particle. The condition of hydrostatic equilibrium then reads
\begin{eqnarray}
-\frac{1}{\rho}\frac{dP}{dr}=a(r)=\frac{d\Phi}{dr}=\frac{GM(r)}{r^2}.
\label{ano9}
\end{eqnarray}
Using the previous results, we find
\begin{eqnarray}
\frac{dP}{dr}=-\frac{g_{\infty}^2}{2\pi Gr}\quad\Rightarrow \quad P=-\frac{g_{\infty}^2}{2\pi G}\ln r+C.
\label{ano10}
\end{eqnarray}
Combining Eqs. (\ref{ano7}) and (\ref{ano10}), we recover the logotropic equation of state
\begin{eqnarray}
P=\frac{g_{\infty}^2}{2\pi G}\ln\rho+C.
\label{ano}
\end{eqnarray}
It can be written as Eq. (\ref{logomodel1}) with $A=g_{\infty}^2/(2\pi G)$ [see Eq. (\ref{aga6})]. Note that the value of the constant $A$ (or $g_{\infty}$)  is not determined by the present derivation, unlike the model developed in \cite{epjp,lettre,jcap,preouf1,action,logosf,preouf2,ouf,gbi}. In a sense, this derivation clarifies the origin of the logotropic equation of state and shows that it is inevitable if we want to account  for a constant acceleration (equivalent to the universal surface density of DM halos or the Tully-Fisher relation) at large distances. However, our original approach \cite{epjp,lettre,jcap,preouf1,action,logosf,preouf2,ouf,gbi} starting directly from the logotropic equation of state is more general and more fundamental since we can apply it both to galaxies and cosmology and this is how we were able to determine the value of the fundamental constant $A$ (or $g_{\infty}$) [see Eqs. (\ref{logomodel8}) and (\ref{aga6})]. Furthermore, the logotropic equation of state describes in principle the whole DM halo (core + envelope) while the results of the present section are only asymptotically valid at large distances.

{\it Remark:} The logotropic model implies a constant gravitational acceleration $g_{\infty}=2.82\times
10^{-11}\, {\rm m \, s^{-2}}$ while the MOND theory involves a transition acceleration $a_0\simeq 1.2\times 10^{-10}\, {\rm m/s^2}$ between Newtonian dynamics and modified Newtonian dynamics. Interestingly, these two fundamental accelerations have a similar value (see Sec. \ref{sec_modtheo} for a more detailed comparison).

\subsection{Isothermal model: Constant circular velocity}
\label{sec_imccv}

It is interesting to compare the previous results with those obtained by assuming that DM halos are isothermal with a universal temperature $T$. Indeed, this is the prediction of standard thermodynamics.\footnote{The cosmological evolution of an isothermal universe is treated in \cite{cosmopoly1,cosmopoly2,stiff,csfpoly}.} Let us thus assume that DM halos are described by the isothermal equation of state 
\begin{eqnarray}
P=\rho\frac{k_B T}{m}.
\label{kam1}
\end{eqnarray}
By solving the Emden equation for isothermal self-gravitating spheres, it can be shown that the density profile decreases at large distances as \cite{chandrabook}
\begin{eqnarray}
\rho\sim \frac{k_B T}{2\pi Gm r^2}.
\label{kam2}
\end{eqnarray}
Using Eq. (\ref{ano4}), we obtain
\begin{eqnarray}
M(r)\sim  \frac{2k_B Tr}{Gm},\qquad v^2\rightarrow  \frac{2k_B T}{m}.
\label{kam3}
\end{eqnarray}
Therefore, isothermal DM halos lead to a constant circular velocity (like the MOND theory). Conversely, assuming that the circular velocity is constant, one is necessarily led to the isothermal equation of state (\ref{kam1}). However, the isothermal distribution does not explain the universality of the surface density of DM halos nor the Tully-Fisher relation. Indeed, the surface density of isothermal DM halos decreases as $\langle \Sigma\rangle (r)=M(r)/(\pi r^2)\sim 2k_B T/(\pi Gm r)$. In addition, a universal temperature does not imply a universal surface density (it gives $M_h/r_h\sim 1$ instead of $M_h/r_h^2\sim 1$). In order to account for the universal surface density (and the Tully-Fisher relation), the temperature $T$ must change from halo to halo and scale as $k_B T/m=0.954\, G\Sigma_0 r_h=0.719\, G\sqrt{\Sigma_0 M_h}$ \cite{modeldmB,modeldmF,frontiers}.

\subsection{Standard thermodynamics and generalized thermodynamics}
\label{sec_svsg}

The isothermal distribution is the universal distribution predicted by standard
thermodynamics for collisional systems. When applied to astrophysics, it leads
to isothermal DM halos and flat rotation curves. However, it does not explain
the universality of the surface density of DM halos, nor the Tully-Fisher
relation. The present work suggests that the  logotropic distribution is the
``universal'' distribution predicted by generalized thermodynamics for
collisionless systems with long-range interactions. Logotropic DM halos have a
constant surface density and a constant gravitational acceleration. The
logotropic model explains the universality of the surface density of DM halos
and the Tully-Fisher 
relation without adjustable parameter. If it is confirmed from observation that
a constant surface density (or a constant gravitational acceleration) is more
relevant than a constant circular velocity, that would favor the logotropic
model  over the isothermal model and generalized thermodynamics over standard
thermodynamics.

\section{Collisionless relaxation and generalized kinetic equations}
\label{sec_gfp}

In this Appendix, we review the kinetic equations which have been proposed to describe the process of collisionless relaxation discussed in Sec. \ref{sec_ew}.

\subsection{Kinetic equation for the local distribution of phase levels}
\label{sec_mlk}

A kinetic equation governing the evolution of the local distribution of phase levels $\rho({\bf r},{\bf v},\eta,t)$ in the context of Lynden-Bell's theory of violent relaxation has been introduced in \cite{sl} and further discussed in \cite{chavmnras,kingen,kinVR,ewart0}.  Making a local approximation, it reads\footnote{In the general case, $\partial/\partial t$ should be replaced by the Stokes material derivative $D/Dt=\partial/\partial t+{\bf v}\cdot \partial/\partial {\bf r}-\nabla\Phi\cdot \partial/\partial {\bf v}$.}
\begin{equation}
\frac{\partial \rho}{\partial t}=\frac{\partial}{\partial v_i}\int d{\bf v}' K_{ij} \left \lbrack f'_2\frac{\partial \rho}{\partial v_j}-\rho(\eta-\overline{f})\frac{\partial \overline{f}'}{\partial v'_j}\right \rbrack,
\label{gfp1}
\end{equation}
where $\overline{f}'=\overline{f}({\bf r},{\bf v}',t)$ [see Eq. (\ref{g94})], $f_2'=f_2({\bf r},{\bf v}',t)$ [see Eq. (\ref{g96cprime}) or Eq. (\ref{fdt2})], and $K_{ij}$ is a specific collision kernel of the Landau or Lenard-Balescu type that we do not write explicitly here (see, e.g.,  \cite{kinVR} for details). This equation conserves the normalization condition $\int\rho\, d\eta=1$, the Casimirs [Eq. (\ref{g106})] and the energy [Eq. (\ref{g61}) with the bar on $f$]. It satisfies an H-theorem for the mixing entropy [Eq. (\ref{mix})] and relaxes towards the Lynden-Bell distribution [Eq. (\ref{g92})] of statistical equilibrium. Another kinetic equation for $\rho({\bf r},{\bf v},\eta,t)$ of the Kramers type has been introduced in \cite{csr} under the form\footnote{Here and in the following, $D$ should be interpreted as a tensor $D_{ij}$ for the sake of generality.}
\begin{equation}
\frac{\partial \rho}{\partial t}=\frac{\partial}{\partial {\bf v}}\cdot \left\lbrace D\left \lbrack \frac{\partial \rho}{\partial {\bf v}}+\beta\rho(\eta-\overline{f}) {\bf v}\right \rbrack\right\rbrace.
\label{gfp2}
\end{equation}
This equation can be obtained heuristically from a maximum entropy production principle (MEPP). It can also be derived from Eq. (\ref{gfp1}) by making a sort of thermal bath approximation, i.e., by replacing $\overline{f}'$ by its equilibrium expression from Eq. (\ref{g95}). Then, by using Eq. (\ref{g96c}), we get $\partial\overline{f}'/\partial {\bf v}'=-\beta f'_2{\bf v}'$.  Using the property of the collision kernel $K_{ij}(v'_i-v_i)=0$ \cite{kinVR}, we obtain Eq. (\ref{gfp2}). This procedure also determines the diffusion tensor according to $D_{ij}=\int K_{ij}f_2'\, d{\bf v}'$, where $f_2$ can be calculated at each time $t$. Then, one can heuristically allow $\beta(t)$ to depend on time so as to satisfy the conservation of energy \cite{csr}.

{\it Remark:} We note that Eq. (\ref{gfp2}) -- and the following equations of the same type -- has the form of a generalized (fermionic-like) Fokker-Planck (or Kramers) equation involving a diffusion term and a friction term. The coefficients of diffusion and friction are connected by an Einstein-like relation of the form $\xi=D\beta\eta$, where $\xi$ is the friction coefficient \cite{csr}. This is a manifestation of the fluctuation-dissipation theorem. In the nondegenerate limit $\overline{f}\ll \eta$, Eqs. (\ref{gfp1}) and (\ref{gfp2}) reduce to the usual Landau and Kramers equations for a multi-species collisional stellar system (see Sec. 2 of \cite{superstat} and the Remark at the end of Sec. \ref{sec_fgalle}).

\subsection{Kinetic equation for the coarse-grained DF}
\label{sec_cgv}

According to Eqs. (\ref{gfp1}) and (\ref{gfp2}), the coarse-grained DF defined by Eq. (\ref{g94}) satisfies an equation of the form
\begin{equation}
\frac{\partial \overline{f}}{\partial t}=\frac{\partial}{\partial v_i}\int d{\bf v}' K_{ij} \left ( f'_2\frac{\partial \overline{f}}{\partial v_j}-f_2\frac{\partial \overline{f}'}{\partial v'_j}\right )
\label{gfp3}
\end{equation}
or
\begin{equation}
\frac{\partial \overline{f}}{\partial t}=\frac{\partial}{\partial {\bf v}}\cdot \left\lbrack D\left ( \frac{\partial \overline{f}}{\partial {\bf v}}+\beta f_2 {\bf v}\right )\right\rbrack.
\label{gfp4}
\end{equation}
More generally, we can write down a hierarchy of equations for the local moments $\overline{f^k}$ of the distribution. This hierarchy of equations is not closed except in the single level (+ vacuum) case $f\in \lbrace 0,\eta_0\rbrace$, where we exactly have $f_2=\overline{f}(\eta_0-\overline{f})$. Eq. (\ref{gfp3}) then reduces to the fermionic Landau equation first obtained in \cite{kp} and further discussed in \cite{chavmnras,kingen,kinVR,ewart0}. Similarly, Eq. (\ref{gfp4}) reduces to the fermionic Kramers equation \cite{csr,chavmnras,gen,kingen,nfp,kinVR}. 

\subsection{Generalized Landau equation}
\label{sec_gla}

In more general situations, we have proposed to close the hierarchy of equations for the local moments of the distribution by using the relation \cite{gen,kingen,jupiter,kinVR}
\begin{equation}
f_2({\bf r},{\bf v},t)=\frac{1}{C''[\overline{f}({\bf r},{\bf v},t)]}.
\label{impt}
\end{equation}
Substituting this closure relation into Eq. (\ref{gfp3}), we obtain the generalized Landau equation \cite{kingen,kinVR}
\begin{equation}
\frac{\partial f}{\partial t}=\frac{\partial}{\partial v_i}\int d{\bf v}' K_{ij} \left \lbrack \frac{1}{C''(f')}\frac{\partial f}{\partial v_j}- \frac{1}{C''(f)}\frac{\partial f'}{\partial v'_j}\right\rbrack.
\label{gfp5}
\end{equation}
It can also be written as
\begin{equation}
\frac{\partial f}{\partial t}=\frac{\partial}{\partial v_i}\int d{\bf v}' {\tilde K}_{ij} f f' \left \lbrack C''(f)\frac{\partial f}{\partial v_j}- C''(f')\frac{\partial f'}{\partial v'_j}\right\rbrack
\label{gfp6}
\end{equation}
by defining a new kernel ${\tilde K}_{ij}=K_{ij}/\lbrack ff' C''(f)C''(f')\rbrack$.\footnote{For the reasons explained below we do not put the bar on $f$ in these equations and in the following ones.}

The generalized Landau equation conserves the mass $M=\int f\, d{\bf r}d{\bf v}$ and the energy [Eq. (\ref{g61})]. It satisfies an H-theorem for the generalized entropy [Eq. (\ref{g63})] and relaxes towards the distribution from Eq. (\ref{g64b}). This corresponds to a microcanonical description.

We have given two justifications for the closure relation from Eq. (\ref{impt}). The relation from Eq.  (\ref{imp}) is exact at statistical equilibrium and it may remain approximately valid if we are sufficiently close to equilibrium with $\overline{f}({\bf r},{\bf v})$ replaced by $\overline{f}({\bf r},{\bf v},t)$ and $f_2({\bf r},{\bf v})$ replaced by $f_2({\bf r},{\bf v},t)$. The closure relation from Eq. (\ref{impt}) can also be justified from an out-of-equilibrium maximum entropy principle.\footnote{This method was initially presented for the 2D Euler equation in Appendix C of \cite{jupiter} but it can be extended straightforwardly to the Vlasov equation because of the analogy between these two equations \cite{csr,kinVR}.} We can compute $f_2({\bf r},{\bf v},t)$ with the distribution $\rho({\bf r},{\bf v},\eta,t)$ which maximizes the Lynden-Bell mixing entropy (\ref{mix}) at fixed normalization, Casimirs, {\it and} DF $\overline{f}({\bf r},{\bf v},t)$ (the conservation of energy $E$ and mass $M$ is then automatically satisfied through the kinetic equations (\ref{gfp5}) and (\ref{gfp6})). This procedure leads to Eq. (\ref{impt}) with a generalized entropy $C_t(\overline{f})$ that has to be calculated at each time $t$ in order to account for the conservation of the Casimirs. This leads to a very complicated kinetic equation of the form of Eq. (\ref{gfp5}) where $C_t$ changes with time. A simplification can be made by assuming that the generalized entropy $C$ is fixed. It can be taken equal to its equilibrium expression (obtained by solving the maximum entropy problem at equilibrium once and for all) or be specified {\it a priori}, i.e., being adapted to the situation contemplated. This second procedure may be particularly relevant when the system is not isolated so that the Casimirs (fragile invariants) are {\it not} conserved and the distribution of the phase levels results from a sort of equilibrium between forcing and dissipation (see \cite{gen,kingen,jupiter,kinVR} for more details). In this respect, the resulting kinetic equation may be connected to the SDD equation (\ref{ba1})-(\ref{ba3}) introduced in \cite{epjp1,sdduniverse}.

The generalized Landau equations (\ref{gfp5}) and (\ref{gfp6}) may also have applications in other domains of physics apart from collisionless relaxation. For example, they may describe quantum statistics (when one uses for $C(f)$ the Fermi-Dirac or Bose-Einstein entropy) or more exotic statistics connected to generalized thermodynamics (e.g. when one uses for $C(f)$ the Tsallis entropy). This is why we have not put the bar on $f$ in Eqs. (\ref{gfp5}) and (\ref{gfp6}) in order to be as general as possible. Some examples of generalized entropies are given below.

\subsection{Generalized Kramers equation}
\label{sec_gkr}

Substituting the closure relation from Eq. (\ref{impt}) into Eq. (\ref{gfp4}) we obtain the generalized Kramers equation \cite{gen,kingen,nfp,kinVR}
\begin{equation}
\frac{\partial f}{\partial t}=\frac{\partial}{\partial {\bf v}}\cdot \left\lbrace D\left \lbrack \frac{\partial f}{\partial {\bf v}}+\frac{\beta}{C''(f)}  {\bf v}\right\rbrack\right\rbrace.
\label{gfp7}
\end{equation}
It can also be written as
\begin{equation}
\frac{\partial f}{\partial t}=\frac{\partial}{\partial {\bf v}}\cdot \left\lbrace {\tilde D}\left \lbrack f C''(f)\frac{\partial f}{\partial {\bf v}}+\beta f {\bf v}\right\rbrack\right\rbrace
\label{gfp8}
\end{equation}
by defining a new diffusion coefficient ${\tilde D}=D/\lbrack f C''(f)\rbrack$. 

The generalized Kramers equations (\ref{gfp7}) and (\ref{gfp8}) can also be derived from the generalized Landau equations (\ref{gfp5}) and (\ref{gfp6}) by making a sort of thermal bath approximation, i.e., by replacing $f'$ by its equilibrium expression from Eq. (\ref{g64b}). Then, by using Eq. (\ref{g64ca}), we get $\partial f'/\partial {\bf v}'=-\beta{\bf v}'/C''(f')$. Using the property of the kernel $K_{ij}(v'_i-v_i)=0$ \cite{kinVR}, we obtain Eqs. (\ref{gfp7}) and (\ref{gfp8}). This procedure also determines the diffusion tensor according to $D_{ij}=\int [K_{ij}/C''(f')]\, d{\bf v}'$, where $f'$ can be calculated at each time $t$. 

The generalized Kramers equation conserves the mass  $M=\int f\, d{\bf r}d{\bf v}$. When $\beta$ is constant, it does not conserve the energy but it satisfies an H-theorem for the generalized free energy $F=E-TS$ (Legendre transform) constructed with the energy  [Eq. (\ref{g61})] and the generalized entropy [Eq. (\ref{g63})], and relaxes towards the distribution from Eq. (\ref{g64b}). This corresponds to a canonical description (thermal bath approximation) instead of a microcanonical one. If we allow $\beta(t)$ to depend on time so as to conserve the energy we can restore a microcanonical description. In that case, the generalized Kramers equation has the same properties as the generalized Landau equation of the previous section. 

By taking the moments of the generalized Kramers equation (\ref{gfp8}) and making a local thermodynamic equilibrium (LTE) approximation,  we obtain the generalized damped Euler equations \cite{gen,kingen,nfp}
\begin{equation}
\label{ch2q1}
\frac{\partial \rho}{\partial t}+\nabla\cdot (\rho{\bf u})=0,
\end{equation}
\begin{equation}
\label{ch2q2}
\frac{\partial {\bf u}}{\partial t}+({\bf u}\cdot \nabla){\bf
u}=-\frac{1}{\rho}\nabla P-\nabla\Phi-\xi{\bf
u}.
\end{equation}
In the strong friction limit $\xi\rightarrow +\infty$, they reduce to the generalized Smoluchowski equation \cite{gen,kingen,nfp}
\begin{equation}
\label{ch7}
\xi\frac{\partial\rho}{\partial t}=\nabla\cdot \left (\nabla
P+\rho\nabla\Phi\right ).
\end{equation}
The equation of state $P(\rho)$ is determined from the generalized entropy $C(f)$ by using a procedure similar to the one developed in Sec. \ref{sec_sta} at equilibrium (see \cite{gen,kingen,nfp} for details).

\subsection{Standard examples}
\label{sec_ex}

The foregoing equations can be made explicit for the Boltzmann, Tsallis, Fermi-Dirac, Bose-Einstein and other generalized entropies as detailed in \cite{gen,kingen,nfp}. For the Tsallis entropy $C(f)=\frac{1}{q-1}(f^q-f)$ associated with stellar polytropes we have $C'(f)=\frac{1}{q-1}(qf^{q-1}-1)$ and $C''(f)=qf^{q-2}$. The corresponding Landau equation reads
\begin{equation}
\frac{\partial f}{\partial t}=\frac{\partial}{\partial v_i}\int d{\bf v}' {\tilde K}_{ij} \left \lbrack f' \frac{\partial f^q}{\partial v_j}- f\frac{\partial {f'}^q}{\partial v'_j}\right\rbrack
\label{gfp9}
\end{equation}
and the corresponding Kramers equation reads
\begin{equation}
\frac{\partial f}{\partial t}=\frac{\partial}{\partial {\bf v}}\cdot \left\lbrace {\tilde D}\left \lbrack \frac{\partial f^q}{\partial {\bf v}}+\beta f {\bf v}\right\rbrack\right\rbrace.
\label{gfp10}
\end{equation}
The associated damped Euler and Smoluchowski equations are given by Eqs. (\ref{ch2q1})-(\ref{ch7}) with the polytropic equation of state from Eq. (\ref{g8}).

{\it Remark:} For spatially homogeneous systems, the generalized Kramers equation (\ref{gfp10}) has an analytical self-similar solution with an invariant Tsallis profile which relaxes towards the DF of stellar polytropes or stellar logotropes (see Ref. \cite{cslogo} for details). 

\subsection{Another example}
\label{sec_cun}

As another tractable example, let us consider the exponential distribution
\begin{equation}
g(\eta)=\frac{1}{b}e^{-\eta/b},
\label{cun1}
\end{equation}
corresponding to the $\chi$-squared distribution from Eq. (\ref{g97}) with
$c=1$. Using Eqs.
(\ref{g93}) and (\ref{g96}), we obtain 
\begin{equation}
Z=1+\frac{1}{1+b\beta\epsilon}
\label{cun2}
\end{equation}
and
\begin{equation}
\overline{f}=\frac{b}{(1+b\beta\epsilon)(2+b\beta\epsilon)}.
\label{cun3}
\end{equation}
Inverting this function, which is a second degree equation in $\beta\epsilon$, and using Eq. (\ref{g64}), we obtain
\begin{equation}
C'(\overline{f})=\frac{3}{2b}-\frac{1}{2b}\sqrt{1+\frac{4b}{\overline{f}}}.
\label{cun4}
\end{equation}
Integrating this relation with respect to $\overline{f}$, we recover Eq. (\ref{cun}). Taking the derivative of Eq. (\ref{cun4}) with respect to $\overline{f}$, we obtain
\begin{equation}
C''(\overline{f})=\frac{1}{\overline{f}^2\sqrt{1+\frac{4b}{\overline{f}}}}.
\label{cun5}
\end{equation}
This result can also be obtained from Eq. (\ref{g64ca}), or from Eqs. (\ref{g96c}) and (\ref{imp}), by taking the derivative of Eq. (\ref{cun3}) with respect to $\beta\epsilon$. We can now substitute Eq. (\ref{cun5}) into the kinetic equations listed previously. For example, the corresponding Landau equation reads
\begin{equation}
\frac{\partial f}{\partial t}=\frac{\partial}{\partial v_i}\int d{\bf v}' {\tilde K}_{ij} \left \lbrack \frac{f'}{f\sqrt{1+\frac{4b}{f}}} \frac{\partial f}{\partial v_j}-  \frac{f}{f'\sqrt{1+\frac{4b}{f'}}}\frac{\partial {f'}}{\partial v'_j}\right\rbrack
\label{cun6}
\end{equation}
and the corresponding Kramers equation reads
\begin{equation}
\frac{\partial f}{\partial t}=\frac{\partial}{\partial {\bf v}}\cdot \left\lbrace {\tilde D}\left \lbrack \frac{1}{f\sqrt{1+\frac{4b}{f}}}\frac{\partial f}{\partial {\bf v}}+\beta f {\bf v}\right\rbrack\right\rbrace.
\label{cun7}
\end{equation}

\subsection{Construction of generalized kinetic equations}
\label{sec_const}

Our approach provides a methodology for constructing  generalized kinetic
equations (see \cite{superstat,kinVR} for a more precise discussion). Basically,
we can assume that forcing and dissipation determine the distribution $g(\eta)$
of phase levels. If we specify this function (called a prior) depending on the
situation contemplated, one can use Eqs. (\ref{g93}) and (\ref{gensb}) to
determine the generalized entropy $C(\overline{f})$, which can then be
substituted into the kinetic equations listed
previously.\footnote{Alternatively, we can determine  $C(\overline{f})$ from the
Lynden-Bell equilibrium state for a given initial condition and then use it
out-of-equilibrium.}  A more general model\footnote{See Sec. 4.4 of
\cite{superstat} for similar ideas in 2D turbulence.} would be to couple these
kinetic equations to a stochastic equation for the phase levels
\cite{gen,nfplangevin,nfp} 
\begin{equation}
\frac{d\eta}{dt}=-\chi U'(\eta)+\sqrt{2Dg\left\lbrack \frac{{\cal C}(g)}{g}\right \rbrack'}\, \eta(t),
\label{const1}
\end{equation}
where $U(\eta)$ and ${\cal C}(g)$ are adjustable functions depending on the situation contemplated, and $\eta(t)$ is a Gaussian white noise. This equation determines at each time $t$ the function $g(\eta,t)$ -- hence the generalized entropy  $C_t(\overline{f})$ -- through the generalized Fokker-Planck equation \cite{gen,nfplangevin,nfp} 
\begin{equation}
\frac{\partial g}{\partial t}=\frac{\partial}{\partial\eta}\left\lbrack D g {\cal C}''(g)\frac{\partial g}{\partial \eta}+\chi g U'(\eta)\right\rbrack.
\label{const2}
\end{equation}
If the distribution $g(\eta,t)$ relaxes faster than the DF $\overline{f}({\bf r},{\bf v},t)$, we can use the equilibrium distribution $g(\eta)$ of Eq. (\ref{const2}), which is determined by \cite{gen,nfplangevin,nfp} 
\begin{equation}
{\cal C}'(g)=-\frac{\chi}{D}U(\eta)+{\rm Cst}.
\label{gfp10k}
\end{equation}

\section{Fluctuation-dissipation theorem for the Lynden-Bell theory}
\label{sec_fdt}

The local moments of the distribution of phase levels $\rho({\bf r},{\bf v},\eta,t)$ are defined by
\begin{equation}
\overline{f^k}=\int \rho \eta^k\, d\eta.
\label{fdt1}
\end{equation}
The first moment is the coarse-grained DF $\overline{f}$, the second moment is $\overline{f^2}$ etc. The local centered variance of the distribution is
\begin{equation}
f_2=\overline{f^2}-\overline{f}^2=\overline{(f-\overline{f})^2}=\int \rho (\eta-\overline{f})^2\, d\eta.
\label{fdt2}
\end{equation}

At equilibrium, substituting the Gibbs state from Eq. (\ref{g92}) into Eq. (\ref{fdt1}) with $k=1,2$, we obtain
\begin{equation}
\overline{f}=\frac{1}{Z(\epsilon)}\int_0^{+\infty} g(\eta)\eta e^{-\beta \eta\epsilon}\, d\eta=-\frac{1}{Z}\frac{\partial Z}{\partial (\beta \epsilon)},
\label{fdt3}
\end{equation}
\begin{equation}
\overline{f^2}=\frac{1}{Z(\epsilon)}\int_0^{+\infty} g(\eta)\eta^2 e^{-\beta \eta\epsilon}\, d\eta=\frac{1}{Z}\frac{\partial^2 Z}{\partial (\beta \epsilon)^2}.
\label{fdt4}
\end{equation}
We note that the moments $\overline{f^k}$ are functions of $\beta\epsilon$. From Eqs. (\ref{fdt2}), (\ref{fdt3}) and (\ref{fdt4}), we obtain 
\begin{equation}
f_2=\frac{1}{Z}\frac{\partial^2 Z}{\partial (\beta \epsilon)^2}-\left\lbrack \frac{1}{Z} \frac{\partial Z}{\partial(\beta\epsilon)}\right\rbrack^2.
\label{fdt5}
\end{equation}
On the other hand, taking the derivative of $\overline{f}(\beta\epsilon)$ with respect to $\beta\epsilon$ in Eq. (\ref{fdt3}), we get
\begin{equation}
\frac{d\overline{f}}{d(\beta\epsilon)}=-\frac{1}{Z}\frac{\partial^2 Z}{\partial (\beta \epsilon)^2}+\left\lbrack \frac{1}{Z} \frac{\partial Z}{\partial(\beta\epsilon)}
\right\rbrack^2.
\label{fdt6}
\end{equation}
Comparing Eqs. (\ref{fdt5}) and (\ref{fdt6}), we obtain the relation \cite{grand,kingen,superstat,assise,kinVR}
\begin{equation}
\frac{d\overline{f}}{d(\beta\epsilon)}=-f_2.
\label{fdt7}
\end{equation}
This relation is similar to the fluctuation-dissipation theorem (Gibbs-Einstein relation) $C\equiv d\langle E\rangle/dT=k_B\beta^2 \langle (\Delta E)^2\rangle$ in thermodynamics (see, e.g., \cite{n2d3}), where $\overline{f}'(\epsilon)$ plays the role of the specific heat $C=d\langle E\rangle/dT$ and $f_2$ plays the role of the fluctuations of energy $\langle E^2\rangle-\langle E\rangle^2$ in the canonical ensemble.

\section{Calculation of the phase space hypervolume $\gamma(\eta)$}
\label{sec_a}

For a given DF $f({\bf r},{\bf v})$, the phase space hypervolume occupied by the level $\eta$ is
\begin{equation}
\gamma(\eta)=\int \delta\left (\eta-f({\bf r},{\bf v})\right)\, d{\bf r}d{\bf v}.
\label{a1}
\end{equation}
It is conserved by the Vlasov-Poisson equations. We can write 
\begin{equation}
\gamma(\eta)=-\frac{d\Gamma}{d\eta},
\label{a2}
\end{equation} 
where
\begin{equation}
\Gamma(\eta)=\int_{f({\bf r},{\bf v})\ge \eta} \, d{\bf r}d{\bf v}
\label{a3}
\end{equation}
is the phase space hypervolume where the DF $f({\bf r},{\bf v})$ takes a value larger than $\eta$.

In the homogeneous case, the foregoing expressions reduce to
\begin{equation}
\gamma(\eta)=\int \delta\left (\eta-f({\bf v})\right)\, d{\bf v}
\label{a6}
\end{equation}
and
\begin{equation}
\Gamma(\eta)=\int_{f({\bf v})\ge \eta} \, d{\bf v},
\label{a4}
\end{equation}
where we have absorbed the domain volume in $\gamma(\eta)$. If $f({\bf v})=f(v^2/2)$ is a monotonically decreasing function of the energy $\epsilon=v^2/2$, the condition $f({\bf v})\ge \eta$ corresponds to $v\le v_{\rm max}(\eta)$ with $f(v_{\rm max}^2/2)=\eta$. As a result, $\Gamma(\eta)$ is just the volume of a $d$-dimensional sphere in velocity space:
\begin{equation}
\Gamma(\eta)=\frac{S_d}{d}v_{\rm max}^d(\eta).
\label{a5}
\end{equation}
The function $\gamma(\eta)$ can then be obtained from Eq. (\ref{a2}) giving
\begin{equation}
\gamma(\eta)=-S_d v_{\rm max}^{d-1}(\eta)v_{\rm max}'(\eta).
\label{a5g}
\end{equation}

We can proceed differently without using the function $\Gamma(\eta)$. In the homogeneous case, for an isotropic velocity distribution, Eq. (\ref{a6}) can be written as
\begin{equation}
\gamma(\eta)=\int_0^{+\infty} \delta\left \lbrack\eta-f(v^2/2)\right\rbrack S_d v^{d-1}\, dv.
\label{a7}
\end{equation}
Writing $\epsilon=v^2/2$, we obtain
\begin{equation}
\gamma(\eta)=\int_0^{+\infty} \delta\left (\eta-f(\epsilon)\right)S_d (2\epsilon)^{(d-2)/2}\, d\epsilon.
\label{a8}
\end{equation}
Using the identity
\begin{equation}
\delta\left (\eta-f(\epsilon)\right)=\frac{\delta(\epsilon-\epsilon_{\rm max})}{|f'(\epsilon_{\rm max})|},
\label{a9}
\end{equation}
where $\epsilon_{\rm max}(\eta)$ is the zero of the function $F(\epsilon)=\eta-f(\epsilon)$, i.e. the solution of $f(\epsilon_{\rm max})=\eta$, we find that
\begin{equation}
\gamma(\eta)=S_d \frac{1}{|f'(\epsilon_{\rm max})|}(2\epsilon_{\rm max})^{(d-2)/2}.
\label{a10}
\end{equation}
We can easily check that Eqs. (\ref{a5g}) and (\ref{a10}) with $\epsilon_{\rm max}(\eta)=v_{\rm max}(\eta)^2/2$ are equivalent.

{\it Remark:} The phase space hypervolume occupied by the level $\eta$ is similar to the density of states in thermodynamics which is defined by (see, e.g., \cite{n2d3}) 
\begin{equation}
g(E)=\int \delta\left (E-H({\bf r},{\bf v})\right)\, d{\bf r}d{\bf v},
\label{a11}
\end{equation}
where $H$ is the Hamiltonian. We can write
\begin{equation}
g(E)=\frac{d\Gamma}{dE},
\label{a12}
\end{equation} 
where
\begin{equation}
\Gamma(E)=\int_{H({\bf r},{\bf v})\le E} \, d{\bf r}d{\bf v}
\label{a13}
\end{equation}
is the phase space hypervolume where the Hamiltonian $H({\bf r},{\bf v})$ takes a value smaller than $E$, i.e., the phase space hypervolume occupied by states with energy less than $E$.

\subsection{Maxwell distribution}
\label{sec_sadm}

For the Maxwell distribution
\begin{equation}
f({\bf v})=\left (\frac{\beta}{2\pi}\right )^{d/2}\rho e^{-\beta v^2/2}
\label{a14}
\end{equation}
with variance
\begin{equation}
\langle v^2\rangle=dT=\frac{d}{\beta},
\label{a14b}
\end{equation}
we have
\begin{equation}
v_{\rm max}(\eta)=\sqrt{\frac{2}{\beta}\ln\left\lbrack \left (\frac{\beta}{2\pi}\right )^{d/2}\frac{\rho}{\eta}\right\rbrack}.
\label{a15}
\end{equation}
For $\eta\le \eta_{\rm max}=(\beta/2\pi)^{d/2}\rho$, using Eq. (\ref{a5}),  we obtain
\begin{equation}
\Gamma(\eta)=\frac{S_d}{d}\left\lbrace \frac{2}{\beta}\ln\left\lbrack \left (\frac{\beta}{2\pi}\right )^{d/2}\frac{\rho}{\eta}\right\rbrack\right\rbrace^{d/2}.
\label{a16}
\end{equation}
After differentiation, using Eq. (\ref{a2}), we find that
\begin{equation}
\gamma(\eta)=\frac{S_d}{\beta\eta}\left\lbrace \frac{2}{\beta}\ln\left\lbrack \left (\frac{\beta}{2\pi}\right )^{d/2}\frac{\rho}{\eta}\right\rbrack\right\rbrace^{d/2-1}.
\label{a17}
\end{equation}
Up to logarithmic corrections, we obtain
\begin{equation}
\gamma(\eta)\sim \eta^{-1}.
\label{a17mag}
\end{equation}
Using Eq. (\ref{g110}), this gives
\begin{equation}
g(\eta)\sim \eta^{d/2-1}.
\label{lemay}
\end{equation}
Comparing Eq. (\ref{lemay}) with Eq. (\ref{g97}), we see that this function selects the power-law exponent
\begin{equation}
c=d/2.
\end{equation}
According to the results of Sec. \ref{sec_ndg}, this leads to an equilibrium coarse-grained DF which is (at least asymptotically) a  polytropic DF  with an index [see Eq. (\ref{g105})] 
\begin{equation}
n_f=-1\qquad  {\it i.e.}\qquad  q_f=\frac{d}{d+2},
\end{equation}
that is to say a stellar logotrope. These results correspond to Eqs. (\ref{ck}) and (\ref{nk}) with $k=-1$.

\subsection{Polytropic (Tsallis) distribution}
\label{sec_sadt}

We now consider the two types of polytropic distributions discussed in Sec. \ref{sec_spol}.

(i) For the polytropic distribution 
\begin{equation}
f({\bf v})=A\left (\epsilon_m-\frac{v^2}{2}\right )_+^{n-d/2},
\label{a18}
\end{equation}
where $A$ is given in terms of $\rho$ by Eq. (\ref{g72}) with $\Phi=0$, we have
\begin{equation}
v_{\rm max}(\eta)=\sqrt{2\left\lbrack \epsilon_m-\left (\frac{\eta}{A}\right )^{2/(2n-d)}\right\rbrack}.
\label{a19}
\end{equation}
For $\eta\le\eta_{\rm max}=A\left (\epsilon_m\right )^{n-d/2}$, using Eq. (\ref{a5}), we obtain
\begin{equation}
\Gamma(\eta)=\frac{S_d}{d}\left\lbrace 2\left\lbrack \epsilon_m-\left (\frac{\eta}{A}\right )^{2/(2n-d)}\right\rbrack\right \rbrace^{d/2}.
\label{a20}
\end{equation}
After differentiation, using Eq. (\ref{a2}), we find that
\begin{eqnarray}
\gamma(\eta)&=&\frac{2S_d}{2n-d}\left\lbrace 2\left\lbrack \epsilon_m-\left (\frac{\eta}{A}\right )^{2/(2n-d)}\right\rbrack\right\rbrace^{d/2-1}\nonumber\\
&\times&\left (\frac{\eta}{A}\right )^{2/(2n-d)}\frac{1}{\eta}.
\label{a21}
\end{eqnarray}
Recalling that $n>d/2$, we obtain for $\eta\rightarrow 0$:
\begin{eqnarray}
\gamma(\eta)\sim \eta^{2/(2n-d)-1}.
\label{a22}
\end{eqnarray}
Using Eq. (\ref{g110}), this gives
\begin{eqnarray}
g(\eta)\sim \eta^{d/2+2/(2n-d)-1}.
\label{a23}
\end{eqnarray}
Comparing Eq. (\ref{a23}) with Eq. (\ref{g97}), we see that this function selects the power-law exponent
\begin{eqnarray}
c=\frac{d}{2}+\frac{2}{2n-d}.
\label{a24}
\end{eqnarray}
According to the results of Sec. \ref{sec_ndg},  this leads to an equilibrium coarse-grained DF which is (at least asymptotically) a  polytropic DF  with an index [see Eq. (\ref{g105})]
\begin{eqnarray}
n_f=-1-\frac{2}{2n-d}.
\label{a25}
\end{eqnarray}
These results correspond to Eqs. (\ref{ck}) and (\ref{nk}) with $k=2/(2n-d)-1$.
When $n\rightarrow +\infty$ (Maxwell distribution), we recover the logotropic DF characterized by the index $n_f=-1$ (see Appendix \ref{sec_sadm}). For $n=n_{5/2}=1+d/2$ (minimum enstrophy state) we get $n_f=-2$ ($k=0$).

(ii) For the polytropic distribution 
\begin{equation}
f({\bf v})=\frac{A}{\left (\epsilon_m+\frac{v^2}{2}\right )^{d/2-n}},
\label{a26}
\end{equation}
where $A$ is given in terms of $\rho$ by Eq. (\ref{g78}) with $\Phi=0$, we have
\begin{equation}
v_{\rm max}(\eta)=\sqrt{2\left\lbrack \left (\frac{A}{\eta}\right )^{2/(d-2n)}-\epsilon_m\right\rbrack}.
\label{a27}
\end{equation}
For $\eta\le\eta_{\rm max}=A/\left (\epsilon_m\right )^{d/2-n}$, using Eq. (\ref{a5}), we obtain
\begin{equation}
\Gamma(\eta)=\frac{S_d}{d}\left\lbrace 2\left\lbrack \left (\frac{A}{\eta}\right )^{2/(d-2n)}-\epsilon_m\right\rbrack\right \rbrace^{d/2}.
\label{a28}
\end{equation}
After differentiation, using Eq. (\ref{a2}), we find that
\begin{eqnarray}
\gamma(\eta)&=&\frac{2S_d}{d-2n}\left\lbrace 2\left\lbrack \left (\frac{A}{\eta}\right )^{2/(d-2n)}-\epsilon_m\right\rbrack\right\rbrace^{d/2-1}\nonumber\\
&\times&\left (\frac{A}{\eta}\right )^{2/(d-2n)}\frac{1}{\eta}.
\label{a29}
\end{eqnarray}
Recalling that $n<-1$, we obtain for $\eta\rightarrow 0$:
\begin{eqnarray}
\gamma(\eta)\sim \eta^{-d/(d-2n)-1}.
\label{a30}
\end{eqnarray}
Using Eq. (\ref{g110}), this gives
\begin{eqnarray}
g(\eta)\sim \eta^{d/2-d/(d-2n)-1}.
\label{a31}
\end{eqnarray}
Comparing Eq. (\ref{a31}) with Eq. (\ref{g97}) we see that this function selects the power-law exponent
\begin{eqnarray}
c=\frac{d}{2}-\frac{d}{d-2n}.
\label{a32}
\end{eqnarray}
According to the results of Sec. \ref{sec_ndg},  this leads to an equilibrium coarse-grained DF which is (at least asymptotically) a  polytropic DF  with an index [see Eq. (\ref{g105})]
\begin{eqnarray}
n_f=\frac{2n}{d-2n}.
\label{a33}
\end{eqnarray}
These results correspond to Eqs. (\ref{ck}) and (\ref{nk}) with $k=-d/(d-2n)-1$. When $n\rightarrow -\infty$ (Maxwell distribution), we recover the logotropic DF characterized by the index $n_f=-1$ (see Appendix \ref{sec_sadm}).

\section{Density of states and differential energy distribution}
\label{sec_dsg}

We consider a stellar system in phase space.
The individual energy (by unit of mass) of a star is $\epsilon=v^2/2+\Phi({\bf r})$. The density of states $g(\epsilon)$, which is the volume of phase space per unit energy,  is defined by
\begin{equation}
g(\epsilon)=\int \delta\left \lbrack \epsilon-\frac{v^2}{2}-\Phi({\bf r})\right\rbrack\, d{\bf r}d{\bf v}.
\label{dsg1}
\end{equation}
We can write 
\begin{equation}
g(\epsilon)=\frac{d\Gamma}{d\epsilon},
\label{dsg2}
\end{equation} 
where
\begin{equation}
\Gamma(\epsilon)=\int_{v^2/2+\Phi({\bf r})\le \epsilon} \, d{\bf r}d{\bf v}
\label{dsg3}
\end{equation}
is the phase space hypervolume where the individual energy $\epsilon({\bf r},{\bf v})=v^2/2+\Phi({\bf r})$ takes a value smaller than $\epsilon$, i.e., the phase space hypervolume occupied by states with energy less than $\epsilon$. It can be rewritten as
\begin{equation}
\Gamma(\epsilon)=\int d{\bf r} \int_{v\le \sqrt{2\left ( \epsilon-\Phi({\bf r})\right )}} \, d{\bf v}.
\label{dsg4}
\end{equation}
The second integral is just the volume of an hypersphere of ``radius'' $\sqrt{2\left ( \epsilon-\Phi({\bf r})\right )}$. Therefore,
\begin{equation}
\Gamma(\epsilon)=\frac{S_d}{d}\int  \left\lbrack 2\left ( \epsilon-\Phi({\bf r})\right )\right\rbrack^{d/2}\, d{\bf r}.
\label{dsg5}
\end{equation}
Taking its derivative with respect to $\epsilon$, we obtain
\begin{equation}
g(\epsilon)=S_d\int  \left\lbrack 2\left ( \epsilon-\Phi({\bf r})\right )\right\rbrack^{d/2-1}\, d{\bf r}.
\label{dsg6}
\end{equation}
For spherically symmetric systems, the density of states can be written as
\begin{equation}
g(\epsilon)=S_d\int_0^{r_m(\epsilon)} \left\lbrack 2\left ( \epsilon-\Phi({r})\right )\right\rbrack^{d/2-1}\, S_d r^{d-1}dr,\label{dsg7}
\end{equation}
where $r_m(\epsilon)$ is the apocenter radius such that $\epsilon=\Phi(r_m)$. 

For a spherically symmetric and isotropic stellar system, whose DF $f=f(\epsilon)$ depends only on the energy $\epsilon$, the differential energy distribution is
\begin{equation}
N(\epsilon)=g(\epsilon)f(\epsilon).
 \label{dsg9}
\end{equation}
$N(\epsilon)d\epsilon$ represents the fraction of the system's stars that have energies in the range $(\epsilon,\epsilon+d\epsilon)$. In general, $N(\epsilon)$ is an increasing function of the energy. The total number of stars is $N=\int N(\epsilon)\, d\epsilon$. In addition, we can write the density from Eq. (\ref{g59}) and the pressure from Eq. (\ref{g60}) as
\begin{equation}
\rho=S_d\int f(\epsilon)  \left\lbrack 2\left ( \epsilon-\Phi({\bf r})\right )\right\rbrack^{d/2-1}\, d\epsilon,
 \label{dsg10}
\end{equation}
\begin{equation}
P=\frac{1}{3}S_d\int f(\epsilon)  \left\lbrack 2\left ( \epsilon-\Phi({\bf r})\right )\right\rbrack^{d/2}\, d\epsilon.
 \label{dsg11}
\end{equation}

For spatially homogeneous systems ($\Phi=0$ and $\epsilon=v^2/2$), the density of states [see Eq. (\ref{dsg6})] reads
\begin{equation}
g(\epsilon)=S_d V  (2 \epsilon)^{d/2-1}.
 \label{dsg8}
\end{equation}
The DF of stellar polytropes of index $n<-1$ behaves for $\epsilon\rightarrow +\infty$ as $f\sim \epsilon^{-(d/2-n)}$ (see Sec. \ref{sec_spol}). Therefore, in the homogeneous case, using Eqs. (\ref{dsg9}) and (\ref{dsg8}), we find that the differential energy distribution scales as
\begin{equation}
N(\epsilon)\propto \epsilon^{-(1-n)},
 \label{dsg12}
\end{equation}
irrespective of the dimension of space. For stellar logotropes ($n=-1$) for which $f\sim \epsilon^{-(d/2+1)}$ (see Sec. \ref{sec_slog}), we find that 
\begin{equation}
N(\epsilon)\propto \epsilon^{-2}\qquad ({\rm logotropes}).
 \label{dsg13}
\end{equation}

\section{Original justification of the logotropic model}
\label{sec_just}

In this Appendix, we briefly explain how we came to introduce the logotropic model in \cite{epjp,lettre,jcap,preouf1,action,logosf,preouf2,ouf,gbi}.

Our initial goal was to find a simple equation of state that yields a universal surface density for all DM halos [see Eq. (\ref{univ})]. Naturally, as a first guess, we considered a polytropic equation of state [see Eq. (\ref{g8})]. Using the mass-radius relation from Eq. (\ref{completemr}) in $d=3$ we find that
\begin{eqnarray}
\label{just}
\Sigma\sim \frac{M}{R^2}\sim \frac{1}{R^{(n+1)/(n-1)}}.
\end{eqnarray}
A universal surface density (i.e. independent on $R$) seems to correspond to a polytropic index $n=-1$. However, for $n=-1$, the polytropic equation of state yields a constant pressure ($P={\rm cst}$) so there is no pressure gradient. Therefore, the standard polytropic equation of state with $n=-1$ cannot account for a situation of hydrostatic equilibrium. Actually, the index $n=-1$ turns out to be ``degenerate''. Indeed, we knew from our previous study on logotropes \cite{cslogo} that the index $n=-1$ was peculiar and that it had to be treated specifically.  We therefore came to the conclusion that a good candidate for the equation of state of DM halos is the logotropic equation of state from Eq. (\ref{logomodel1}). Since the logotropic density profile decreases as $\rho\sim r^{-1}$ [see Eq. (\ref{logomodel2})] it yields a constant surface density $\Sigma\sim \rho r\rightarrow {\rm cst}$ at large distances. Conversely, the arguments developed in Appendix \ref{sec_ac} show that the logotropic equation of state  is inevitable if the large distance surface density of the DM halos (rather than their circular velocity) is constant [see Eq. (\ref{aga5})]. A drawback of the logotropic equation of state is that it yields DM halos with an infinite mass. However, it should not be applied at arbitrarily large distances. It is valid only up to a few halo radii, after which the density profile decreases more rapidly that $r^{-1}$ (see Sec. \ref{sec_interpol}). The logotropic equation of state yields a universal surface density given by Eq. (\ref{logomodel3}). However, this equation of state depends on two constants $A$ and $\rho_P$ that are {\it a priori} unknown.\footnote{Only the constant $A$ plays a role for nonrelativistic systems since $\rho_P$ just yields a constant term in the pressure. However, the constant $\rho_P$ becomes important for relativistic systems, e.g., in cosmology.} Apparently, this prevents us from making any prediction...

However, we had the idea to apply the same equation of state in cosmology in a relativistic framework. We know that a UDME model equivalent to the $\Lambda$CDM model is provided by a single dark fluid with a constant equation of state $P=-\rho_{\Lambda}c^2$ \cite{cosmopoly1,cosmopoly2}. We also know that the $\Lambda$CDM model works well in cosmology (large scales). However, it suffers from drawbacks at the level of  DM halos (small scales). Indeed, a constant (or zero) pressure has a vanishing gradient and leads to gravitational collapse creating cusps instead of cores and generating many satellites that are not observed. Therefore, the idea is to  find an equation of state  that is close to  a constant $P=-\rho_{\Lambda}c^2$  but not exactly constant in order to provide a pressure gradient preventing gravitational collapse. The logotropic equation of state, involving a weakly varying logarithmic function, is a good candidate for that.\footnote{We have seen that the logotropic equation of state is closely related to the polytropic equation of state of index $\gamma=0$ (or $n=-1$) yielding a constant pressure $P=K$.  In this sense, the logotropic model  may be viewed as the simplest extension  of the $\Lambda$CDM model (corresponding to a polytropic equation of state $P=K\rho^{\gamma}$ with $\gamma=0$ and $K=-\rho_{\Lambda}c^2$ \cite{cosmopoly1,cosmopoly2}) in the framework of UDME models \cite{epjp}.}

By solving the Friedmann equations with the logotropic equation of state (\ref{logomodel1}) to determine the evolution of the energy density $\epsilon$ of the Universe as a function of the scale factor $a$, and by applying this equation in the present Universe ($a=1$) where we have observational data, we obtained the relation from Eq. (\ref{logomodel4}) between $A$ and $\rho_P$. By determining $A$ [using Eq. (\ref{logomodel3})]  from the empirical value of the universal surface density of DM halos provided by Eq. (\ref{univ}), we could then use  Eq. (\ref{logomodel4}) to estimate $\rho_P$. We found a gigantic value! After a moment of surprise, we realized that this value was of the order of the Planck density. Therefore, reversing the argument, we assumed that $\rho_P$ in Eq. (\ref{logomodel1}) {\it is} the Planck density (or a fraction of it) and we used the cosmological relation between $A$ and $\rho_P$ [Eq. (\ref{logomodel4})] to predict $A$, then $\Sigma_0$ [Eq. (\ref{logomodel3})], finding a good agreement with the observational value [compare Eqs. (\ref{logomodel5}) and (\ref{univ})]. 

The next step is to try to justify the logotropic equation of state (\ref{logomodel1}). This was the purpose of the present paper where we showed that logotropic DFs arise naturally in statistical mechanics and kinetic theory of collisionless systems with long-range interactions, like self-gravitating systems. They can be justified from generalized thermodynamics, from the Lynden-Bell theory of violent relaxation, or from the SDD equation. They have been recently observed in many physical systems \cite{ewart1,ewart2,banik1,banik2}. This may explain why the logotropic equation of state is relevant to DM halos and cosmology \cite{epjp,lettre,jcap,preouf1,action,logosf,preouf2,ouf,gbi}.

\section{Analytical solution of the diffusion equation with a power-law diffusion coefficient}
\label{sec_sol}

We consider a diffusion equation of the form
\begin{eqnarray}
\frac{\partial f}{\partial t}=\frac{1}{v^{d-1}}\frac{\partial}{\partial v}\left \lbrack v^{d-1}D(v)\frac{\partial f}{\partial v}\right \rbrack
\label{sol1}
\end{eqnarray}
with a power-law diffusion coefficient $D(v)=Av^{\alpha}$. By an appropriate change of variables, we can rewrite this diffusion equation as
\begin{eqnarray}
\frac{\partial f}{\partial t}=\frac{1}{x^{d-1}}\frac{\partial}{\partial x}\left ( x^{d-1+\alpha}\frac{\partial f}{\partial x}\right )
\label{sol2}
\end{eqnarray}
with the normalization condition $\int f\, d{\bf x}=1$. The variance of the distribution is $\sigma^2=\int f x^2\, d{\bf x}$.

This type of diffusion equation has been studied in Appendix F of \cite{kinQ} for the dimension $d=3$ and the exponent 
$\alpha=-3$. Here, we consider the general case of arbitrary $d$ and $\alpha$. The diffusion equation (\ref{sol2}) admits a self-similar solution of the form
\begin{eqnarray}
f(x,t)=t^{-a}F\left (\frac{x}{t^{\beta}}\right ).
\label{sol3}
\end{eqnarray}
The scaling $x\sim t^{1/(2-\alpha)}$ deduced from Eq. (\ref{sol2}) yields $\beta=1/(2-\alpha)$. On the other hand, the normalization condition $\int f\, d{\bf x}=1$ implies $t^{-a}t^{d\beta}\sim 1$, hence $a=d\beta=d/(2-\alpha)$. Therefore, we can rewrite Eq. (\ref{sol3}) as
\begin{eqnarray}
f(x,t)=t^{-d/(2-\alpha)}F\left \lbrack\frac{x}{t^{1/(2-\alpha)}}\right \rbrack
\label{sol4}
\end{eqnarray}
with $\int_0^{+\infty} F(X)S_d X^{d-1}\, dX=1$, where $X=x/t^{1/(2-\alpha)}$ is the scaled variable. Substituting Eq. (\ref{sol4}) into Eq. (\ref{sol2}), we find that the invariant profile $F(X)$ satisfies the differential equation
\begin{eqnarray}
\frac{1}{X^{d-1}}\frac{d}{dX}\left (X^{d-1+\alpha}\frac{dF}{dX}\right )+\beta X \frac{dF}{dX}+aF=0
\label{sol5}
\end{eqnarray}
or, equivalently,
\begin{equation}
\frac{d^2F}{dX^2}+\left\lbrack (d-1+\alpha)\frac{1}{X}+\beta X^{1-\alpha}\right\rbrack \frac{dF}{dX}+aX^{-\alpha}F=0,
\label{sol6}
\end{equation}
where $a$ and $\beta$ are defined above. If we make the change of variables 
\begin{eqnarray}
F(X)=e^{-\frac{X^{2-\alpha}}{(2-\alpha)^2}}V(X),
\label{sol7}
\end{eqnarray}
we find that the new function $V(X)$ satisfies
\begin{equation}
\frac{d^2V}{dX^2}+\left\lbrack (d-1+\alpha)\frac{1}{X}-\frac{1}{2-\alpha} X^{1-\alpha}\right\rbrack \frac{dV}{dX}=0.
\label{sol6b}
\end{equation}
This is a first order differential equation for $V'(X)$ whose solution is
\begin{equation}
V'(X)=A\, \frac{e^{\frac{X^{2-\alpha}}{(2-\alpha)^2}}}{X^{d-1+\alpha}}.
\label{sol6c}
\end{equation}
Therefore, the general solution of Eq. (\ref{sol6}) is
\begin{equation}
F(X)=A e^{-\frac{X^{2-\alpha}}{(2-\alpha)^2}}\int_{0}^{X}e^{\frac{w^{2-\alpha}}{(2-\alpha)^2}}\frac{dw}{w^{d-1+\alpha}}+Be^{-\frac{X^{2-\alpha}}{(2-\alpha)^2}},
\label{sol7b}
\end{equation}
where $A$ and $B$ are integration constants.

(i) When $\alpha<2$, the function $F(X)$ diverges for $X\rightarrow +\infty$ unless $A=0$. Therefore, the physical (normalizable) solution of Eq. (\ref{sol6}) is
\begin{equation}
F(X)=Be^{-\frac{X^{2-\alpha}}{(2-\alpha)^2}}.
\label{sol7bh}
\end{equation}
We note that the exponential term is
independent of the dimension of space. The constant $B$ is determined by the
normalization condition  $\int_0^{+\infty} F(X)S_d X^{d-1}\, dX=1$ yielding
\begin{eqnarray}
B=\frac{1}{S_d (2-\alpha)^{\frac{2d-2+\alpha}{2-\alpha}}\Gamma\left (\frac{d}{2-\alpha}\right )},
\label{sol8}
\end{eqnarray}
where $\Gamma(x)$ is the gamma function. Therefore,
\begin{equation}
f(x,t)=\frac{1}{t^{\frac{d}{2-\alpha}}}\frac{1}{S_d (2-\alpha)^{\frac{2d-2+\alpha}{2-\alpha}}\Gamma\left (\frac{d}{2-\alpha}\right )}e^{-\frac{x^{2-\alpha}}{(2-\alpha)^2t}}.
\label{sol8tot}
\end{equation}
The velocity dispersion is given by
\begin{eqnarray}
\sigma^2=t^{2\beta}\int_0^{+\infty} F(X)S_d X^{d+1}\, dX.
\label{sol9}
\end{eqnarray}
With the results from Eqs. (\ref{sol7}) and (\ref{sol8}) we obtain
\begin{eqnarray}
\sigma^2=t^{2/(2-\alpha)}(2-\alpha)^{4/(2-\alpha)}\frac{\Gamma\left (\frac{d+2}{2-\alpha}\right )}{\Gamma\left (\frac{d}{2-\alpha}\right )}.
\label{sol10}
\end{eqnarray}

(ii) When $\alpha>2$, the physical (normalizable) solution of Eq. (\ref{sol6}) is
\begin{equation}
F(X)=C e^{-\frac{X^{2-\alpha}}{(2-\alpha)^2}}\int_{X}^{+\infty}e^{\frac{w^{2-\alpha}}{(2-\alpha)^2}}\frac{dw}{w^{d-1+\alpha}}.
\label{sol7c}
\end{equation}
For $X\rightarrow 0$:
\begin{equation}
F(X)\sim C\frac{\alpha-2}{X^d}.
\label{sol7d}
\end{equation}
For $X\rightarrow +\infty$: 
\begin{equation}
F(X)\sim \frac{C}{d+\alpha-2}\frac{1}{X^{d+\alpha-2}}.
\label{sol7e}
\end{equation}
We note that the integral $\int_0^{+\infty} F(X)S_d X^{d-1}\, dX$ diverges logarithmically when $X\rightarrow 0$, so the distribution is not normalizable at small $X$. However, the foregoing results assume that $D=A v^{\alpha}$ for all velocities, which may not be strictly valid in practice for small $v$ (we may have $D=A v^{\alpha}$ only for $v>v_{\rm min}$). Therefore, the divergence at small $v$ is not a serious problem. If we consider only the behavior of $F(X)$ for $X\rightarrow +\infty$ the distribution is normalizable for $\alpha>2$ and the variance exists for $\alpha>4$. Coming back to the original variables, we find that the DF decreases as
\begin{equation}
f\propto v^{-(d-2+\alpha)},\qquad f\propto \epsilon^{-(d-2+\alpha)/2}.
\label{dai}
\end{equation}
This asymptotic behavior is equivalent to the one obtained in Eq. (\ref{dait}) by looking for a stationary solution of Eq. (\ref{sol1}). Comparing Eq. (\ref{dai}) with Eq. (\ref{g75}), we see that the DF has the same power-law decay as a  stellar polytrope of index $n=1-\alpha/2$ with $n<-1$ (assuming $\alpha>4$). For $\alpha=4$, we recover the case of stellar logotropes (see Sec. \ref{sec_slog}).

(iii) When $\alpha=2$ we expect that the solution $f(x,t)$ involves logarithmic terms.

\section{Justification of the Maxwellian fine-grained distribution}
\label{sec_jumax}

We have seen in Sec. \ref{sec_udf} that the ``universal'' $\epsilon^{-(d+2)/2}$ tail \cite{ewart1}  of the coarse-grained Lynden-Bell DF $\overline{f}(\epsilon)$, resembling the DF of a stellar logotrope, is due to the fact that the fine-grained DF $f$ is Maxwellian.\footnote{Actually, Ewart {\it et al.} \cite{ewart1} show that the $\epsilon^{-(d+2)/2}$ tail arises in more general circumstances.} How can we justify a Maxwellian fine-grained DF? We give below several answers.

(i) If the initial DF $f_0({\bf r},{\bf v})$ is Maxwellian, then the results of Sec. \ref{sec_udf} are exact in the framework of the ordinary Lynden-Bell theory, which assumes that the distribution of  phase levels $\gamma(\eta)$ is conserved by the dynamics. In that case,  the coarse-grained Lynden-Bell DF $\overline{f}(\epsilon)$ presents an $\epsilon^{-(d+2)/2}$ tail. However, in theory, this power-law tail should not be observed for other initial conditions (i.e. different from the Maxwellian) such as  the single level $\eta_0$ ($+$ vacuum) case for which the strict Lynden-Bell prediction is  a Fermi-Dirac-like or a Maxwell-Boltzmann-like distribution (see \ref{sec_fdl}).

(ii) In practice, the phase space hypervolume $\gamma(\eta,t)$ occupied by the level $\eta$ is not strictly conserved but changes with time (see Appendix \ref{sec_kramers}). Indeed, it is well-known that the Casimirs or the moments $M^{\rm f.g.}_{n>2}=\int f^n\, d{\bf r}d{\bf v}$ of the fine-grained DF are altered by ``collisions'' (or by viscosity in 2D turbulence) \cite{grand,superstat,ncd}. For that reason, the moments  $M^{\rm f.g.}_{n>2}=\int f^n\, d{\bf r}d{\bf v}$ are called ``fragile constraints'' because they are dissipated in the presence of weak  collisions (or a small viscosity  in 2D turbulence).  By contrast, the mass $M=\int f\, d{\bf r}d{\bf v}$ and the energy $E=\frac{1}{2}\int f v^2\, d{\bf r}d{\bf v}+\frac{1}{2}\int\rho\Phi\, d{\bf r}$ are called ``robust constraints'' because they are relatively well-conserved under the same conditions. Therefore, because of nonideal (dissipative) effects, the system progressively loses the memory of the initial condition. This is what Ewart {\it et al.} \cite{ewart2} recently called ``turbulent amnesia'' (this property had been known for a long time in the literature \cite{grand,superstat,ncd}). As a result, we should not compute the Lynden-Bell equilibrium DF from the initial  fine-grained DF $f_0({\bf r},{\bf v})$ but rather from the time-dependent fine-grained DF $f({\bf r},{\bf v},t)$, which is affected by the collisions. This time-dependent fine-grained DF can become Maxwellian even if the initial condition is not.\footnote{As a simple illustration we may assume that the fine-grained  DF $f({\bf r},{\bf v},t)$ is governed by the mean field  Kramers equation (see Appendix \ref{sec_kramers}), which relaxes ultimately towards the mean field Boltzmann distribution from Eq. (\ref{kramers2bol}). In the spatially homogeneous case, there are particular cases where the solution of the Kramers equation is a Gaussian distribution at any time (see Appendix \ref{sec_kramershom}). In that case, the evolution of  the phase space hypervolume $\gamma(\eta,t)$ is given by Eq. (\ref{a17}) with a time-dependent inverse temperature $\beta(t)=d/(m\langle v^2\rangle(t))$ related to the variance $\langle v^2\rangle(t)$ calculated in Appendix \ref{sec_kramershom}. These are simple examples where $\gamma(\eta,t)\propto \eta^{-1}$ at any time.} This is the case in the two-stream instability situation investigated by Ewart {\it et al.} \cite{ewart2}. The initial condition is characterized by a single level $\eta_0$  ($+$ vacuum) so that the strict Lynden-Bell prediction is a Fermi-Dirac-like or a Maxwell-Boltzmann-like distribution (see Sec. \ref{sec_fdl}). This is indeed the DF that is observed for short times in the numerical simulations. However, because of intrinsic collisions, the fine-grained DF $f({\bf r},{\bf v},t)$ rapidly becomes Maxwellian and, if we compute the Lynden-Bell distribution  with this ``new'' initial condition, we get a coarse-grained distribution with a $\overline{f}\sim \epsilon^{-(d+2)/2}$ tail (see Sec. \ref{sec_udf}) as observed at later times  in the numerical simulations.

(iii) Actually, it may be difficult in practice to disentangle the fine-grained
and coarse-grained DFs $f$ and $\overline{f}$. Indeed, it is difficult to know
if the alteration of the DF is due to collisions which affect the fine-grained
DF or if it is due to coarse-graining, which is inherent to any measurement.
Therefore, it may be relevant to evaluate the Lynden-Bell DF at late times not
with the initial fine-grained DF $f_0({\bf r},{\bf v})$ but rather with the
ulterior coarse-grained DF $\overline{f}({\bf r},{\bf v},t)$. This can be done
iteratively. From the initial condition $f_0$ we get a first Lynden-Bell DF
$\overline{f}_1$.  Then, we take $\overline{f}_1$ as a new fine-grained initial
condition $f_1$ and compute a second Lynden-Bell DF $\overline{f}_2$ and so on.
We may wonder if this iterative process leads to a relevant asymptotic DF.  In
the case considered by Ewart {\it et al.} \cite{ewart2}, this iterative process
leads to results consistent with the numerical simulations. Indeed, $f_0$ is a
step function,  $\overline{f}_1$ is a Maxwell-Boltzmann-like DF, and
$\overline{f}_2$ presents an $\epsilon^{-(d+2)/2}$ tail, as observed 
in the numerical simulations.

\section{Mean field Kramers equation}
\label{sec_kramers}

In this Appendix, we study how intrinsic collisions among the particles affect
the conservation of the Casimirs (see also \cite{superstat,capelpasmanter,ncd} for the
effect of viscosity in 2D turbulence). For simplicity, we model the collisions
by the mean field Kramers (or Vlasov-Kramers) equation
\begin{eqnarray}
\frac{\partial f}{\partial t}+{\bf v}\cdot \frac{\partial f}{\partial {\bf r}}-\nabla\Phi\cdot \frac{\partial f}{\partial {\bf v}}=\frac{\partial}{\partial {\bf v}}\cdot\left (D\frac{\partial f}{\partial {\bf v}}+\xi f {\bf v}\right ),
\label{kramers1}
\end{eqnarray}
\begin{eqnarray}
\Phi({\bf r},t)=\int u({\bf r}-{\bf r}')\rho({\bf r}',t)\, d{\bf r}'.
\label{kramers2}
\end{eqnarray}
Here, $\Phi({\bf r},t)$ could be the gravitational potential or, more
generally, 
any long-range potential of interaction. The diffusion and friction coefficients
$D$ and $\xi$ are related to each other by the Einstein relation
\begin{eqnarray}
D=\frac{\xi k_B T}{m},
\label{kramers2ein}
\end{eqnarray}
where $T$ is the temperature and $m$ the mass of the particles. This relation
guarantees that the mean field Kramers equation  relaxes towards the mean field
Boltzmann distribution 
\begin{eqnarray}
f({\bf r},{\bf v})=Ae^{-\beta m[\frac{v^2}{2}+\Phi({\bf r})]}
\label{kramers2bol}
\end{eqnarray}
at statistical equilibrium. Indeed, the mean field Kramers equation satisfies an $H$-theorem $\dot F_B\le 0$ for the Boltzmann free energy $F_B=E-TS_B$ constructed with the Boltzmann entropy $S=-k_B\int \frac{f}{m} \ln f \, d{\bf r}d{\bf v}$.  It can be rewritten as
\begin{eqnarray}
\frac{\partial f}{\partial t}+{\bf v}\cdot \frac{\partial f}{\partial {\bf r}}-\nabla\Phi\cdot \frac{\partial f}{\partial {\bf v}}=\xi\frac{\partial}{\partial {\bf v}}\cdot\left (\frac{k_B T}{m}\frac{\partial f}{\partial {\bf v}}+f {\bf v}\right ).\quad
\label{kramers1k}
\end{eqnarray}
We note that the mean field Kramers equation does not provide a perfect description of the collisional process because it does not conserve the energy (it rather assumes that the system is in contact with a thermal bath fixing the temperature). A better kinetic equation would be the Vlasov-Landau equation \cite{aakin} but it is more complicated for a simple illustration. 

\subsection{Evolution of the phase space hypervolume}

The  phase space hypervolume $\gamma(\eta,t)$ occupied by the level $\eta$ is defined by 
\begin{equation}
\gamma(\eta,t)=\int \delta\left (\eta-f({\bf r},{\bf v},t)\right)\, d{\bf r}d{\bf v}.
\label{a1t}
\end{equation}
Taking its time derivative,  using Eq. (\ref{kramers1}), and recalling that the advection (Vlasov)  term conserves the  phase space hypervolumes (or Casimirs) \cite{kinVR}, we get
\begin{eqnarray}
\frac{\partial \gamma}{\partial t}(\eta,t)&=&-\int d{\bf r}d{\bf v}\, \delta'(\eta-f)\frac{\partial f}{\partial t}\nonumber\\
&=&-\int d{\bf r}d{\bf v}\, \delta'(\eta-f)\frac{\partial}{\partial {\bf v}}\cdot\left (D\frac{\partial f}{\partial {\bf v}}+\xi f {\bf v}\right )\nonumber\\
&=&-\frac{\partial}{\partial\eta}\int d{\bf r}d{\bf v}\,\delta(\eta-f)\frac{\partial}{\partial {\bf v}}\cdot\left (D\frac{\partial f}{\partial {\bf v}}+\xi f {\bf v}\right )\nonumber\\
&=&-\frac{\partial}{\partial\eta}\int d{\bf r}d{\bf v}\,\delta'(\eta-f)\left (D\frac{\partial f}{\partial {\bf v}}+\xi f {\bf v}\right )\cdot \frac{\partial f}{\partial {\bf v}}\nonumber\\
&=&-\frac{\partial^2}{\partial\eta^2}\int d{\bf r}d{\bf v}\,\delta(\eta-f)\left (D\frac{\partial f}{\partial {\bf v}}+\xi f {\bf v}\right )\cdot \frac{\partial f}{\partial {\bf v}},\nonumber\\
\label{kramers3}
\end{eqnarray}
where we have made an integration by parts to obtain the third line. If we define the diffusion coefficient in the space of phase levels by
\begin{equation}
{\cal D}(\eta,t)=-\frac{1}{\gamma(\eta,t)}\int d{\bf r}d{\bf v}\,\delta(\eta-f)\left (D\frac{\partial f}{\partial {\bf v}}+\xi f {\bf v}\right )\cdot \frac{\partial f}{\partial {\bf v}},
\label{kramers4}
\end{equation}
we can rewrite Eq. (\ref{kramers3}) as
\begin{eqnarray}
\frac{\partial \gamma}{\partial t}=\frac{\partial^2}{\partial\eta^2}({\cal D}\gamma).
\label{kramers5}
\end{eqnarray}
In the absence of friction, i.e., in the purely diffusive case, the diffusion coefficient is negative:
\begin{equation}
{\cal D}(\eta,t)=-\frac{1}{\gamma(\eta,t)}\int d{\bf r}d{\bf v}\,\delta(\eta-f) D\left (\frac{\partial f}{\partial {\bf v}}\right )^2\le 0.
\label{kramers4b}
\end{equation}
In that case, Eq. (\ref{kramers5}) can be seen as an anti-diffusion equation in the space of phase levels.

{\it Remark:} These calculations can be extended to the generalized kinetic equations presented in Appendix \ref{sec_gfp} with the justification given in Appendix \ref{sec_jumax}.

\subsection{Moments of the fine-grained DF}

The moments of the fine-grained DF are defined by
\begin{eqnarray}
M_n^{\rm f.g.}=\int f^n\, d{\bf r}d{\bf v}=\int \gamma(\eta,t)\eta^n\, d\eta.
\label{kramers6}
\end{eqnarray}
Taking their derivatives with respect to time, using Eq. (\ref{kramers5}), or starting directly from Eq. (\ref{kramers1}), and integrating by parts, we find that
\begin{eqnarray}
\frac{dM_n^{\rm f.g.}}{dt}=n(n-1)\int d\eta\, \eta^{n-2}{\cal D}(\eta,t)\gamma(\eta,t)\nonumber\\
=-n(n-1)\int d{\bf r}d{\bf v}\, f^{n-2}\left (D\frac{\partial f}{\partial {\bf v}}+\xi f {\bf v}\right )\cdot \frac{\partial f}{\partial {\bf v}}.\quad
\label{kramers7}
\end{eqnarray}
In the purely diffusive case ($\xi=0$), the foregoing equation reduces to
\begin{eqnarray}
\frac{dM_n^{\rm f.g.}}{dt}=-n(n-1)\int d{\bf r}d{\bf v}\, f^{n-2}D\left (\frac{\partial f}{\partial {\bf v}}\right )^2\le 0.\quad
\label{kramers8}
\end{eqnarray}
This relation shows that the fragile moments $M_{n>1}^{\rm f.g.}$ are dissipated under the effect of collisions (even when $\xi\rightarrow 0$ we have $\dot M_{n>1}^{\rm f.g.}$ finite because of strong velocity gradients $(\partial f/\partial {\bf v})^2\rightarrow +\infty$) \cite{superstat}.

{\it Remark:} For the momentum ${\bf P}=\int f {\bf v}\, d{\bf r}d{\bf v}$ and the energy $E$, we obtain after straightforward integrations by parts
\begin{eqnarray}
\dot{\bf P}=-\xi {\bf P}, \quad \dot E=dDM-\xi \int f v^2\, d{\bf r}d{\bf v}.
\label{lib}
\end{eqnarray}
When $\xi\rightarrow 0$ we have $\dot {\bf P}\rightarrow {\bf 0}$ and $\dot E\rightarrow 0$.

\subsection{Solution of the spatially homogeneous Kramers equation}
\label{sec_kramershom}

The solution of the spatially homogeneous Kramers equation 
\begin{eqnarray}
\frac{\partial f}{\partial t}=\frac{\partial}{\partial {\bf v}}\cdot\left (D\frac{\partial f}{\partial {\bf v}}+\xi f {\bf v}\right )
\label{kramers1b}
\end{eqnarray}
with the initial condition $f({\bf v},t=0)=\rho\delta({\bf v}-{\bf v}_0)$ is the Gaussian
\begin{equation}
f({\bf v},t)=\rho\left\lbrack \frac{m}{2\pi k_B T\left (1-e^{-2\xi t}\right )}\right\rbrack^{d/2}e^{-\frac{m\left ({\bf v}-e^{-\xi t}{\bf v}_0\right )^2}{2k_B T\left (1-e^{-2\xi t}\right)}}.
\label{kramers9}
\end{equation}
Its variance is
\begin{equation}
\langle v^2\rangle(t)=e^{-2\xi t}v_0^2+\frac{d k_B T}{m}\left (1-e^{-2\xi t}\right).
\label{kramers9a}
\end{equation}
It relaxes towards $\langle v^2\rangle=dk_B T/m$. The solution of the spatially homogeneous diffusion equation (Kramers equation
without friction) 
\begin{eqnarray}
\frac{\partial f}{\partial t}=D\frac{\partial^2f}{\partial {\bf v}^2}
\label{kramers1c}
\end{eqnarray}
with the initial condition $f({\bf v},t=0)=\rho\delta({\bf v}-{\bf v}_0)$ is the Gaussian
\begin{equation}
f({\bf v},t)=\frac{\rho}{(4\pi Dt)^{d/2}} e^{-\frac{\left ({\bf v}-{\bf v}_0\right )^2}{4Dt}}.
\label{kramers10}
\end{equation}
Its variance is
\begin{equation}
\langle v^2\rangle(t)=v_0^2+2dDt.
\label{kramers10b}
\end{equation}
These solutions (divided by $\rho$) are the Green functions $W({\bf v},t;{\bf v}_0,0)$ of the Kramers equation (\ref{kramers1b}) and of the diffusion equation (\ref{kramers1c}) respectively. The solutions of these equations with an arbitrary initial condition $f({\bf v},0)$ are then given by the relation
\begin{equation}
f({\bf v},t)=\int W({\bf v},t;{\bf v}_0,0)f({\bf v}_0,0)\, d{\bf v}_0.
\label{birdy1}
\end{equation}
If the initial DF is the Maxwellian 
\begin{equation}
f({\bf v})=\left (\frac{\beta_0 m}{2\pi}\right )^{d/2}\rho e^{-\beta_0 m v^2/2}
\label{a14bis}
\end{equation}
with a temperature $T_0$, then the solution of the Kramers equation (\ref{kramers1b})  is the Gaussian
\begin{eqnarray}
f({\bf v},t)=\frac{\rho}{\left\lbrack \frac{2\pi k_B T}{m}\left (1-e^{-2\xi t}\right )+\frac{2\pi k_B T_0}{m}
e^{-2\xi t}\right\rbrack^{d/2}}\nonumber\\
\times e^{-\frac{mv^2}{2k_B T \left (1-e^{-2\xi t}\right )+2k_B T_0 e^{-2\xi t}}}.
\label{birdy2}
\end{eqnarray}
Its variance is
\begin{eqnarray}
\langle v^2\rangle(t)=\frac{dk_B T}{m} \left (1-e^{-2\xi t}\right )+\frac{dk_B T_0}{m} e^{-2\xi t}.
\label{birdy3}
\end{eqnarray}
For the same initial condition, the solution of the diffusion equation (\ref{kramers1c}) is
\begin{eqnarray}
f({\bf v},t)=\frac{\rho}{\left ( 4\pi Dt+\frac{2\pi k_B T_0}{m}\right )^{d/2}}e^{-\frac{v^2}{4Dt+\frac{2k_B T_0}{m}}}.
\label{birdy4}
\end{eqnarray}
Its variance is
\begin{eqnarray}
\langle v^2\rangle(t)=2dDt+\frac{dk_B T_0}{m}.
\label{birdy5}
\end{eqnarray}

{\it Remark:} The expressions for the average velocity and its variance can also be obtained by integrating the relations $\dot {\bf P}=-\xi {\bf P}$ and $\dot E=dDM-2\xi  E$ giving ${\bf P}={\bf P}_0 e^{-\xi t}$ and $E=(E_0-{dDM}/{2\xi})e^{-2\xi t}+{dDM}/{2\xi}$. More generally, if we write the Gaussian distribution as
\begin{eqnarray}
\label{toi1}
f({\bf v})=\frac{\rho}{\left (\frac{2\pi}{d}v_2\right )^{d/2}}e^{-\frac{({\bf v}-\langle {\bf v}\rangle)^2}{\frac{2}{d}v_2}}
\end{eqnarray}
with $v_2=\langle ({\bf v}-\langle {\bf v}\rangle)^2\rangle=\langle {v^2}\rangle-\langle {\bf v}\rangle^2$, we obtain
\begin{eqnarray}
\label{toi3}
M_{n}^{\rm f.g.}=\frac{M\rho^{n-1}}{n^{d/2}\left (\frac{2\pi}{d}v_2\right )^{\frac{d}{2}(n-1)}},
\end{eqnarray}
\begin{eqnarray}
E=\frac{1}{2}M\langle v^2\rangle=\frac{1}{2}Mv_2+\frac{1}{2}M\langle {\bf v}\rangle^2,
\label{toi3b}
\end{eqnarray}
\begin{eqnarray}
\label{toi4}
S_B=\frac{d}{2}Nk_B-Nk_B \ln \left\lbrack \frac{\rho}{\left (\frac{2\pi}{d}v_2\right )^{d/2}}\right\rbrack.
\end{eqnarray}
The Boltzmann free energy is $F_B=E-TS_B$. In the diffusive case, we have $M_{n}^{\rm f.g.}\propto t^{-\frac{d}{2}(n-1)}$, $E\propto t$ and $S_B\propto \ln t$.

\subsection{Diffusion coefficient in the space of phase levels}

In the homogeneous case, in order to determine the diffusion coefficient in the space of phase levels from Eq. (\ref{kramers4}), we have to compute an integral of the form
\begin{equation}
{\cal I}(\eta,t)=\int d{\bf v}\,\delta(\eta-f)\left (D\frac{\partial f}{\partial {\bf v}}+\xi f {\bf v}\right )\cdot \frac{\partial f}{\partial {\bf v}}.
\label{kramers11}
\end{equation}
If $f=f(\epsilon)$ with $\epsilon=v^2/2$ (for simplicity we do not write the time variable explicity), it can be rewritten as
\begin{equation}
{\cal I}(\eta,t)=\int \delta(\eta-f)\left (D f'(\epsilon)+\xi f \right )f'(\epsilon)v^2 S_d v^{d-1}\, dv
\label{kramers12}
\end{equation}
or as
\begin{equation}
{\cal I}(\eta,t)=\int \delta(\eta-f)\left (D f'(\epsilon)+\xi f \right )f'(\epsilon) S_d (2\epsilon)^{d/2}\, d\epsilon.
\label{kramers13}
\end{equation}
Using the identity from Eq. (\ref{a9}),  and introducing the function $\epsilon_{\rm max}(\eta)$ defined by $f(\epsilon_{\rm max})=\eta$ (see Appendix \ref{sec_a}), we get
\begin{equation}
{\cal I}(\eta,t)=-\left (D f'(\epsilon_{\rm max})+\xi \eta \right ) S_d (2\epsilon_{\rm max})^{d/2},
\label{kramers14}
\end{equation}
where we have assumed $f'(\epsilon)<0$. Recalling Eq. (\ref{a10}) we find that the diffusion coefficient in the space of phase levels defined by Eq. (\ref{kramers4}) is given by
\begin{equation}
{\cal D}(\eta,t)=-\left (D f'(\epsilon_{\rm max})+\xi \eta \right ) f'(\epsilon_{\rm max}) 2\epsilon_{\rm max}.
\label{kramers15}
\end{equation}

For the Maxwellian DF from Eq. (\ref{a14}) with a time-dependent inverse temperature $\beta(t)$, we get
\begin{equation}
{\cal D}(\eta,t)=\left (-D\beta(t)+\xi \right ) \beta(t) \eta^2 2\epsilon_{\rm max}(\eta).
\label{kramers16}
\end{equation}
Recalling Eq. (\ref{a15}) and $\epsilon_{\rm max}(\eta)=v_{\rm max}^2(\eta)/2$, we obtain
\begin{equation}
{\cal D}(\eta,t)=2\left (-D\beta(t)+\xi \right )  \eta^2 \ln\left\lbrack \left (\frac{\beta(t)}{2\pi}\right )^{d/2}\frac{\rho}{\eta}\right\rbrack.
\label{kramers16b}
\end{equation}
In particular, for the  diffusion equation (\ref{kramers1c}) corresponding to $\xi=0$, using Eq. (\ref{kramers10}) with ${\bf v}_0={\bf 0}$, the time-dependent inverse temperature reads $\beta(t)=1/(2Dt)$, and we get
\begin{equation}
{\cal D}(\eta,t)=-\frac{\eta^2}{t}  \ln\left\lbrack \left (\frac{1}{4\pi Dt}\right )^{d/2}\frac{\rho}{\eta}\right\rbrack.
\label{kramers18}
\end{equation}
The other cases can be treated similarly by using Eqs. (\ref{kramers9a}), (\ref{birdy3}) and (\ref{birdy5}) with Eq. (\ref{a14b}).

\subsection{Analogy with 2D turbulence}

The previous results are similar to those obtained in 2D turbulence \cite{superstat,ncd,capelpasmanter} (see \cite{houches} for the analogy between stellar systems, plasmas, and 2D vortices). In the presence of a small dissipation (collisions), the enstrophy $M_2=\int f^2\, d{\bf r}d{\bf v}$ (similar to $\Gamma_2=\int \omega^2\, d{\bf r}$)  is dissipated while the mass $M=\int f\, d{\bf r}d{\bf v}$ (similar to the circulation $\Gamma=\int \omega\, d{\bf r}$) and the energy $E=\frac{1}{2}\int f v^2\, d{\bf r}d{\bf v}+\frac{1}{2}\int\rho\Phi\, d{\bf r}$ (similar to the energy $E=\frac{1}{2}\int\omega\psi\, d{\bf r}$ or the angular momentum $L=\int \omega r^2\, d{\bf r}$) are approximately conserved. This is the inverse cascade process. We may therefore expect that the system will reach a minimum enstrophy state corresponding to a polytrope of index $n=5/2$ (see footnote 30). However, this state is not always achieved in practice \cite{superstat,brands} because other fragile moments $M_n=\int f^n\, d{\bf r}d{\bf v}$ in addition to the enstrophy are dissipated depending on the strength of the velocity gradient [see Eq. (\ref{kramers8})]. We also recall that, ultimately, the DF relaxes towards the Boltzmann distribution because of intrinsic collisions.

\section{Polytropes and logotropes with a negative squared speed of sound}
\label{sec_cs2neg}

In this Appendix, for completeness, we discuss the case of gaseous polytropes with a negative squared speed of sound: $c_s^2=P'(\rho)=K\gamma\rho^{\gamma-1}<0$. Therefore, we assume $K\gamma<0$ or, equivalently,  $K(n+1)/n<0$. The pressure is positive ($P>0$) when $-1<n<0$ and negative ($P<0$) when $n>0$ and $n<-1$. 

{\it Remark:} Since $\Phi(r)$ increases with the distance [$d\Phi/dr=GM(r)/r^2\ge 0$], the condition of hydrostatic equilibrium from Eq. (\ref{g4})  implies that $P(r)$ decreases with the distance. Therefore, when $c_s^2<0$, the density increases with the distance, which is possible but not very realistic. Furthermore,  self-gravitating systems with $c_s^2<0$ are dynamically unstable because they are energy maxima ($\delta^2E_{\rm tot}<0$) instead of energy minima \cite{wdsD}.
Indeed, $c_s^2<0$ is a sufficient condition of instability.

\subsection{Polytropes $n>0$ and $K<0$}

We first consider polytropes of index $n>0$. In that case, $K<0$ so the pressure is negative. The relation $\rho(\Phi)$ is given by Eq. (\ref{g11b}). These polytropes are described by the Lane-Emden equation  (\ref{g14}) with  $\epsilon=-1$. The function $\theta(\xi)$ increases with the distance and tends to $+\infty$, the density $\rho\propto \theta^n$  increases with the distance and tends to $+\infty$,  the pressure $P\propto -\theta^{n+1}<0$   decreases with the distance and tends to $-\infty$. The divergence may occur at a finite distance or at infinity. The total mass is infinite in all the cases that we have investigated (see below). 

The $d$-dimensional Schuster solution of Sec. \ref{sec_schuster} is valid in dimensions $d>2$ and the Schuster index $n_5$ satisfies $1<n_5<+\infty$. The function $\theta_5$ diverges at a finite normalized radius $\xi_*=\sqrt{d(d-2)}$ and behaves as $\theta_5\sim \lbrack 2(1-\xi/\xi_*)\rbrack^{-(d-2)/2}$ when $\xi\rightarrow \xi_*$. The total mass is infinite.

In $d=1$ dimension, we can use the results of Sec. \ref{sec_dimun}. For $n>1$,  the function $\theta$ diverges at a finite normalized radius
\begin{equation}
\label{melua1}
\xi_*=-\left (\frac{n+1}{2}\right )^{1/2}\sqrt{\pi}\frac{\Gamma\left (\frac{1}{2}-\frac{1}{1+n}\right )}{\Gamma\left (-\frac{1}{1+n}\right )}
\end{equation}
and behaves as 
\begin{equation}
\label{melua2}
\theta\sim \left\lbrack \frac{\sqrt{2(n+1)}}{n-1}\frac{1}{\xi_*-\xi}\right\rbrack^{2/(n-1)}
\end{equation}
when $\xi\rightarrow \xi_*$. The total mass is infinite. For $0<n<1$,  the function $\theta$ behaves as
\begin{equation}
\label{melua3}
\theta\sim \left\lbrack \frac{1-n}{\sqrt{2(n+1)}}\xi\right\rbrack^{2/(1-n)}
\end{equation}
when $\xi\rightarrow +\infty$. For $n=1$, the solution of the Lane-Emden equation is $\theta=\cosh\xi$, so that $\theta\sim e^{\xi}/2$ when $\xi\rightarrow +\infty$. For $0<n\le 1$, the density tends to $+\infty$ at infinity so that the mass is infinite.

{\it Remark:} There is a  singular solution of the form of Eq. (\ref{g17}) for $0<n<n_3=d/(d-2)$ in dimensions $d>2$ and for any $n>0$ in dimensions $d\le 2$ (with $n\neq 1$ in all cases). The corresponding singular density profile is $\rho_s\propto r^{-2n/(n-1)}$. The density increases with the distance when $n<1$ and decreases with the distance when $n>1$. In the second case it is integrable for $r\rightarrow +\infty$ but not for $r\rightarrow 0$.  We have not investigated this case further. 

\subsection{Polytropes $n<-1$ and $K<0$ }

We now consider polytropes of index $n<-1$. In that case, $K<0$ so the pressure is negative. The relation $\rho(\Phi)$ is given by Eq. (\ref{g11a}). These polytropes are described by the Lane-Emden equation  (\ref{g14}) with  $\epsilon=+1$. The function $\theta(\xi)$ decreases with the distance and vanishes at a finite normalized radius $\xi_1$,\footnote{The distance where $\theta(\xi)$ vanishes is finite in all the cases that we have investigated but this property may not be general.} the density $\rho\propto \theta^n$  increases with the distance and tends to $+\infty$ at $R=r_0\xi_1$,  the pressure $P\propto -\theta^{n+1}<0$ decreases with the distance and tends to $-\infty$ at $R$. Note that Eqs. (\ref{complete})-(\ref{completemr}) are not valid in the present situation because $\theta'_1\rightarrow -\infty$. The total mass is infinite in all the cases that we have investigated (see below).

The $d$-dimensional Schuster solution of Sec. \ref{sec_schuster} is valid in dimensions $d<2$ and the Schuster index $n_5$ satisfies $-\infty<n_5<-1$. The function $\theta_5$ vanishes at a finite normalized radius $\xi_1=\sqrt{d(2-d)}$ and behaves as $\theta_5\propto \lbrack 2(1-\xi/\xi_1)\rbrack^{(2-d)/2}$ when $\xi\rightarrow \xi_1$. The total mass is infinite.

In $d=1$ dimension, we can use the results of Sec. \ref{sec_dimun}. The function $\theta$ vanishes at a finite normalized radius $\xi_1$ given by Eq. (\ref{g32}) and behaves as 
\begin{equation}
\label{melua4}
\theta\sim \left\lbrack \frac{1-n}{\sqrt{-2(n+1)}}(\xi_1-\xi)\right\rbrack^{2/(1-n)}
\end{equation}
when $\xi\rightarrow \xi_1$. For $n_5=-3$ we recover the Schuster solution. The total mass is infinite.

{\it Remark:} There is a  singular solution of the form of Eq. (\ref{g16}) for $n<n_3=d/(d-2)$ in dimensions $d<2$.  The corresponding singular density profile $\rho_s\propto r^{-2n/(n-1)}$ is integrable for $r\rightarrow +\infty$ but not for $r\rightarrow 0$. We have not investigated this case further.

\subsection{Polytropes $-1<n<0$ and $K>0$ }

Finally, we consider polytropes of index $-1<n<0$. In that case, $K>0$ so the pressure is positive. The relation $\rho(\Phi)$ is given by Eq. (\ref{g11a}). These polytropes are described by the Lane-Emden equation  (\ref{g14}) with  $\epsilon=+1$.  The function $\theta(\xi)$ decreases with the distance and vanishes at a finite normalized radius $\xi_1$, the density $\rho\propto \theta^n$  increases with the distance and tends to $+\infty$ at $R=r_0\xi_1$,  the pressure $P\propto \theta^{n+1}>0$ decreases with the distance and vanishes at $R$.  Eqs. (\ref{complete})-(\ref{completemr}) remain valid in the present situation. The total mass is finite in all the cases that we have investigated (see below).

There is no Schuster solution with index $-1<n<0$ (more generally with index $-1<n<1$).

In $d=1$ dimension, we can use the results of Sec. \ref{sec_dimun} with $\epsilon=+1$. The function $\theta$ vanishes at a finite normalized radius $\xi_1$ given by Eq. (\ref{g32}).

{\it Remark:} There is a  singular solution of the form of Eq. (\ref{g16}) for $-1<n<n_3=d/(d-2)$ in dimensions $d<1$.  The corresponding singular density profile $\rho_s\propto r^{-2n/(n-1)}$ is integrable for $r\rightarrow +\infty$ but not for $r\rightarrow 0$. We have not investigated this case further. 

\subsection{Logotropes with $A<0$}

To complete our analysis, we treat the case of gaseous logotropes with a negative squared speed of 
sound: $c_s^2=P'(\rho)=A/\rho<0$. Therefore, we assume $A<0$. The pressure is positive when $\rho<\rho_*$ and negative when $\rho>\rho_*$. The relation $\rho(\Phi)$ is given by Eq. (\ref{g49}). Starting from Eq. (\ref{g50}), assuming that the system is spherically symmetric, and making the change of variables
\begin{equation}
\label{ng51}
\theta=\frac{\rho_0}{\rho},\qquad \xi=\left (\frac{S_d
G\rho_0^2}{|A|}\right )^{1/2}r=r/r_0,
\end{equation}
where $\rho_0$ is the central density and $r_0=({|A|}/{S_d
G\rho_0^2})^{1/2}$ is the logotropic core radius, we find that Eq.
(\ref{g50})
reduces to the Lane-Emden equation of index $n=-1$ and $\epsilon=+1$: 
\begin{equation}
\label{ng52}
\frac{1}{\xi^{d-1}}\frac{d}{d\xi}\left (\xi^{d-1}
\frac{d\theta}{d\xi}\right )=-\frac{1}{\theta},
\end{equation}
with the boundary conditions $\theta=1$ and $\theta'=0$ at $\xi=0$.  The
function $\theta(\xi)$ decreases and vanishes at a finite distance $\xi_1$. The
density $\rho\propto 1/\theta$  increases with the distance and tends to
$+\infty$ at $\xi_1$ while the pressure $P\propto \ln\theta+C$ decreases with the
distance and tends to $-\infty$ at $\xi_1$. The total mass is infinite in all
the cases that we have investigated (see below).

In $d=1$ dimension, the Lane-Emden equation (\ref{ng52}) reduces to Eq. (\ref{g27}) with $n=-1$ and $\epsilon=+1$. It is similar to the equation of motion of a fictive particle of unit mass in a potential
\begin{equation}
\label{mg36}
V(\theta)=\ln\theta.
\end{equation}
The first integral of motion takes the form
\begin{equation}
\label{mg37}
E=\frac{1}{2}\left (\frac{d\theta}{d\xi}\right )^2+\ln\theta,
\end{equation}
where $E$ is a constant. It is determined by the initial condition giving $E=0$. The solution of Eq. (\ref{g27}) with $n=-1$ and $\epsilon=+1$ is given in reversed form by
\begin{equation}
\label{bg38}
\xi=\int_{\theta}^1 \frac{dx}{\sqrt{-2\ln x}}.
\end{equation}
Making the change of variables $-\ln x=y^2$, we obtain
\begin{equation}
\label{bg39}
\xi=\sqrt{\frac{\pi}{2}}{\rm erf}(\sqrt{-\ln\theta}),
\end{equation}
where
\begin{equation}
\label{bg40}
{\rm erf}(x)=\frac{2}{\sqrt{\pi}}\int_0^x e^{-y^2}\, dy
\end{equation}
is the error function. The function $\theta(\xi)$ vanishes at 
\begin{equation}
\xi_1=\sqrt{\frac{\pi}{2}},
\label{bg40b}
\end{equation}
with an infinite derivative ($\theta'_1=-\infty$). For $\xi\rightarrow\xi_1$, we have
\begin{equation}
\label{bg40x}
\theta\sim \sqrt{-2\ln(\xi_1-\xi)}\, (\xi_1-\xi).
\end{equation}
The logotropic density profile is $\rho(r)=\rho_0/\theta(\xi)$ with $\xi=(2G\rho_0^2/|A|)^{1/2}r$. It diverges as $\rho\sim \rho_0/[\sqrt{-2\ln(R-r)}(R-r)]$ when $r\rightarrow R=\xi_1 r_0$.  The integrated density behaves as $M(r)\propto \sqrt{-\ln(R-r)}$ when $r\rightarrow R$, so the mass of the configuration is infinite.

{\it Remark:} In dimensions $d<1$, there exists an exact
analytical solution 
\begin{equation}
\label{ng53}
\theta_s=\frac{\xi}{\sqrt{1-d}},
\end{equation}
corresponding to the density profile (\ref{g54}) called the singular logotropic sphere.  This singular density profile is integrable for $r\rightarrow +\infty$ but not for $r\rightarrow 0$. We have not investigated this case further.

\end{document}